\documentclass[a4paper,11pt, table]{article}
\pdfoutput=1

\usepackage{jheppub}

\usepackage{epsf,amsmath,amssymb,graphicx,dcolumn}
\usepackage{scalefnt}
\usepackage{hyperref}
\usepackage{cleveref,etoolbox,color,soul}
\usepackage{listings,booktabs}
\usepackage[utf8]{inputenc}
\usepackage{epsfig}
\usepackage{graphicx}
\usepackage[colorinlistoftodos]{todonotes}
\usepackage{multirow}
\usepackage{bbold}

\usepackage{amssymb}
\usepackage{amsmath}

\newcommand{\dmatm}{\Delta m^2_{31}}
\newcommand{\dmsol}{\Delta m^2_{21}}

\newcommand{\U}{\mathbf{U}}

\newcommand{\Y}{\mathbf{Y}}

\newcommand{\BR}{{\rm BR}}
\newcommand{\CR}{{\rm CR}}

\def\EW{$\mathrm{SU(2)}_L \otimes \mathrm{U(1)}_Y$}

\def\Y{\mathbf{Y}}

\def\Yf{\mathbf{Y}_f}

\def\Mnu{\mathbf{M}_\nu}
\def\Mnuh{\widehat{\mathbf{M}}_\nu}

\def\U{\mathbf{U}}

\def\dmatm{\Delta m^2_{31}}
\def\dmsol{\Delta m^2_{21}}

\newcolumntype{K}[1]{>{\centering\arraybackslash}m{#1}}
\def\gsim{\raise0.3ex\hbox{$\;>$\kern-0.75em\raise-1.1ex\hbox{$\sim\;$}}}
\def\lsim{\raise0.3ex\hbox{$\;<$\kern-0.75em\raise-1.1ex\hbox{$\sim\;$}}}

\definecolor{linkcolor}{rgb}{0,0,0.5}

\def\Y{\mathbf{Y}}

\begin{document}

\title{Flavour and dark matter in a scoto/type-II seesaw model}

\author[a]{D.~M. Barreiros,}
\author[a]{H.~B. C\^amara,}
\author[a]{and F.R.~Joaquim}

\affiliation[a]{Departamento de F\'{\i}sica and CFTP, Instituto Superior T\'ecnico, Universidade de Lisboa, Av. Rovisco Pais 1, 1049-001 Lisboa, Portugal}

\emailAdd{debora.barreiros@tecnico.ulisboa.pt}
\emailAdd{henrique.b.camara@tecnico.ulisboa.pt}
\emailAdd{filipe.joaquim@tecnico.ulisboa.pt}

\abstract{The neutrino mass and dark matter (DM) problems are addressed in a Standard Model extension where the type-II seesaw and scotogenic mechanisms coexist. The model features a flavour $\mathcal{Z}_8$ discrete symmetry which is broken down to a $\mathcal{Z}_2$, stabilising the (scalar or fermion) DM particle. Spontaneous CP violation is implemented through the complex vacuum expectation value of a singlet scalar field, inducing observable CP-violating effects in the lepton sector. The structure of the effective neutrino mass matrix leads to constraints on the low-energy neutrino observables, namely the atmospheric neutrino mixing angle $\theta_{23}$, the Dirac CP-violating phase $\delta$ and the absolute neutrino mass scale $m_{\rm lightest}$. In particular, in most cases, the model selects one $\theta_{23}$ octant with $\delta \simeq 3\pi/2$. Moreover, the obtained lower bounds on $m_{\rm lightest}$ are typically in the range probed by cosmology. We also analyse the constraints imposed on the model by current experimental limits on charged lepton flavour violating (cLFV) processes, as well as future projected sensitivities. It is shown that the Higgs triplet and scotogenic contributions to cLFV never overlap and that the interplay among Yukawa couplings, \textit{dark} charged scalar masses and mixing leads to a wide parameter-space region compatible with current experimental bounds. We investigate the scalar and fermion DM parameter space of our model by considering relic density, direct-detection~(DD) and collider constraints. For scalar DM the mass interval  $68 \ \text{GeV} \lesssim m_{\text{DM}} \lesssim 90 \ \text{GeV}$ is viable and will be probed by future DD searches. In the fermion DM case, correct relic density is always obtained for $m_{\text{DM}} \gsim 45$ GeV thanks to \textit{dark} fermion-scalar coannihilation channels.}

\maketitle

\section{Introduction}
\label{sec:intro}

Two major puzzles for which the Standard Model (SM) of particle physics does not provide an explanation are the existence of neutrino masses and of dark matter (DM). The discovery of neutrino oscillations \cite{McDonald:2016ixn,Kajita:2016cak} and the observation of a DM matter relic energy density~\cite{Planck:2018vyg} thus call for physics beyond the SM. Over the past decades, the growing worldwide neutrino experimental programme has been remarkably improving the determination of neutrino oscillation parameters, namely the mass-squared differences, mixing angles and Dirac CP phase~\cite{deSalas:2020pgw,Esteban:2020cvm,Capozzi:2021fjo}. In spite of these experimental achievements, we still lack knowledge on which is the mechanism responsible for generating neutrino masses. From a theoretical perspective, and interpreting the SM as an effective theory, it is natural to consider that neutrinos are Majorana particles. The most popular framework where this is accounted for is based on the seesaw mechanism~\cite{Minkowski:1977sc,Gell-Mann:1979vob,Yanagida:1979as,Glashow:1979nm,Mohapatra:1979ia,Konetschny:1977bn,Cheng:1980qt,Lazarides:1980nt,Schechter:1980gr,Mohapatra:1980yp,Magg:1980ut} where naturally small neutrino masses are generated through the (tree-level) exchange of new degrees of freedom which are typically very heavy. A possible realization of this mechanism relies on extending the SM scalar sector with a triplet, triggering the so-called type-II seesaw~\cite{Konetschny:1977bn,Cheng:1980qt,Lazarides:1980nt,Schechter:1980gr,Mohapatra:1980yp,Magg:1980ut}. One of the attractive features of this mechanism is that the neutrino mass and mixing pattern is directly related to the dimensionless lepton-triplet couplings (apart from possible renormalization effects due to the gap between the triplet-decoupling and electroweak scale). This may have interesting phenomenological consequences as, for instance, in relating charged-lepton flavour violating (cLFV) processes for distinct flavours~\cite{Rossi:2002zb,Joaquim:2006uz,Joaquim:2006mn,Raidal:2008jk,Brignole:2010nh,Joaquim:2009vp}.  

The existence of DM has been established through indirect astrophysical measurements and cosmological observations which indicate that about $27 \%$ of the total matter in the Universe is DM~\cite{Planck:2018vyg}. Numerous candidate DM particles have been postulated~\cite{Bertone:2016nfn,Arbey:2021gdg}, among which the weakly interacting massive particles (WIMP) paradigm stands out. Although the neutrino mass and DM problem may be uncorrelated, it is interesting to explore scenarios in which both share the same solution. In this sense, the reason neutrinos are massive particles is somehow connected to the DM nature. This is exactly what happens in the scotogenic model~\cite{Ma:2006km} where neutrino masses arise at the quantum level via loops involving {\em dark} messengers which may be suitable DM particles. The most economical scotogenic model requires a couple of fermionic singlets and an inert scalar doublet. This {\em dark} sector is odd under a discrete symmetry (typically a $\mathcal{Z}_2$) stabilising the lightest odd particle and providing either a fermion or scalar DM particle. In this framework, it was shown that both Majorana and Dirac fermionic DM can reproduce the observed DM relic density (see, for instance, refs.~\cite{Vicente:2014wga} and \cite{Guo:2020qin} for a recent analysis on the phenomenology of Majorana and Dirac DM, respectively). On the other hand, the phenomenology of scalar DM in the canonical scotogenic model resembles that of the broadly studied inert two Higgs doublet model~\cite{Deshpande:1977rw,Barbieri:2006dq}, and the correct relic density can be readily obtained via scalar and gauge interactions in the early Universe. A recent analysis of scalar DM phenomenology in the simplest scotogenic model can be found in ref.~\cite{Avila:2021mwg}. Several scoto-inspired models have been explored relying on different kinds of fermionic and/or scalar extensions, as well as on imposing additional symmetries (see, e.g., refs.~\cite{Hirsch:2013ola,Fraser:2014yha,Ahriche:2016cio,vonderPahlen:2016cbw,Hagedorn:2018spx,Bonilla:2019ipe,Nomura:2019lnr,Avila:2019hhv,CentellesChulia:2019gic,Han:2019lux,Jana:2019mez,Escribano:2020iqq,Beniwal:2020hjc,Escribano:2021ymx,DeRomeri:2021yjo,deAnda:2021jzc}). Typically, these models enlarge the viable parameter space for DM and neutrino mass generation due to the presence of new interactions, opening up the possibility for exotic collider signatures.

It could also be that neutrino masses receive important contributions both at the tree and loop level via different mechanisms, as it is the case in the scoto-seesaw scenario where type-I seesaw and the scotogenic mechanisms are responsible for generating the solar and atmospheric neutrino mass scale independently~\cite{Rojas:2018wym,Mandal:2021yph}. Interestingly, such coexistence of two different neutrino mass mechanisms allows for the possibility of vacuum-induced CP-violating effects in the neutrino sector, as shown in ref.~\cite{Barreiros:2020gxu} for a model with a $\mathcal{Z}_8$ flavour symmetry which is broken down to a DM stabilising $\mathcal{Z}_2$. This is achieved by the complex vacuum expectation value (VEV) of a scalar singlet responsible for breaking CP spontaneously. The fermion-scalar couplings exhibit texture-zero patterns which lead to constraints of the low-energy neutrino mass and mixing parameters, namely the atmospheric mixing angle, the Dirac CP violating phase and the absolute neutrino mass scale.  

Motivated by the scoto-seesaw idea, in this work we investigate the flavour and DM phenomenology of a model where the interplay is between the type-II seesaw and scotogenic mechanisms. As in ref.~\cite{Barreiros:2020gxu}, a flavour $\mathcal{Z}_8$ symmetry is broken to a \textit{dark} $\mathcal{Z}_2$ and spontaneous CP violation (SCPV) is possible due to the complex VEV of a singlet scalar field. The paper is organised as follows. In section~\ref{sec:model} we present our model, namely its field content and symmetries, and we discuss the interplay between the type-II seesaw and scotogenic mechanisms for neutrino masses and lepton mixing. The compatibility with neutrino oscillation data is studied in section~\ref{sec:neutrinomass}, where the predictions from the two-texture zero effective neutrino mass matrix are presented for the different scenarios allowed by the $\mathcal{Z}_8$ symmetry. We also study leptonic CP violation (LCPV)~\cite{Branco:2011zb}, neutrinoless double-beta decay ($0\nu\beta\beta$) and the link between low-energy observables and the SCPV phase. In section~\ref{sec:cLFVmain} we analyse the implications for cLFV taking into consideration the current constraints and projected sensitivities from experimental searches. DM phenomenology is investigated in section~\ref{sec:DM} for the two types of DM candidates, namely scalar and fermion. We consider relic-density and collider constraints, as well as prospects for DM direct-detection (DD) in the context of our model. Finally, our concluding remarks are presented in section~\ref{sec:concl}. Details of the scalar sector and the DM (co)annihilation diagrams are presented in the appendices.

\section{A scoto/type-II seesaw model with a $\mathcal{Z}_{8}$ flavour symmetry}
\label{sec:model}

In this work we consider a SM extension with one neutral fermion $f$ and four complex scalar multiplets: two doublets $\eta_i$ ($i=1,2$), one triplet $\Delta$ and one singlet $\sigma$. In addition, we impose an Abelian discrete flavour symmetry $\mathcal{Z}_{8}$ which, as will be later seen, forbids some Yukawa couplings and, consequently, leads to low-energy predictions for neutrino mass and mixing parameters. After spontaneous symmetry breaking~(SSB), the $\mathcal{Z}_{8}$ symmetry breaks down to a {\em dark} $\mathcal{Z}_{2}$ symmetry, under which $f$ and $\eta_{1,2}$ are odd (these fields will be often referred as {\em dark}). The particle content of the model, together with the charge assignments under the original $\mathcal{Z}_{8}$ and remnant $\mathcal{Z}_{2}$ symmetries, is summarised in table~\ref{tab:part&sym}. The three cases $\mathcal{Z}_{8}^{e-\mu}$, $\mathcal{Z}_{8}^{e-\tau}$ and $\mathcal{Z}_{8}^{\mu-\tau}$ differ from each other by the $\mathcal{Z}_{8}$ charges of the SM lepton fields and, as discussed later, will lead to different predictions for the low-energy neutrino parameters.
\begin{table}[t!]
\renewcommand{\arraystretch}{1.15}
	\centering
	\begin{tabular}{| K{1.8cm} | K{1.5cm} | K{3.0cm} |  K{2.0cm} | K{2.0cm} | K{2.0cm} | }
		\hline 
&Fields&\EW&  $\mathcal{Z}_{8}^{e-\mu} \rightarrow \mathcal{Z}_2$ &  $\mathcal{Z}_{8}^{e-\tau} \rightarrow \mathcal{Z}_2$  &  $\mathcal{Z}_{8}^{\mu-\tau} \rightarrow \mathcal{Z}_2$ \\
		\hline 
		\multirow{4}{*}{Fermions} 
&$\ell_{e L} , e_R$&($\mathbf{2}, {-1/2}),(\mathbf{1}, {-1}$)& {$1$}   $\to$  $+$ & {$1$}   $\to$  $+$ & {$\omega^2$}   $\to$  $+$  \\
&$\ell_{\mu L} , \mu_R$&($\mathbf{2}, {-1/2}),(\mathbf{1}, {-1}$)& {$\omega^6$}   $\to$  $+$ & {$\omega^2$}   $\to$  $+$ & {$1$}   $\to$  $+$  \\
&$\ell_{\tau L} , \tau_R$&($\mathbf{2}, {-1/2}),(\mathbf{1}, {-1}$)& {$\omega^2$}   $\to$  $+$ & {$\omega^6$}   $\to$  $+$ & {$\omega^6$}   $\to$  $+$  \\
&$f$&($\mathbf{1}, {0}$)& {$\omega^3$}   $\to$  $-$ & {$\omega^3$}   $\to$  $-$ & {$\omega^3$}   $\to$  $-$ \\
		\hline 
		\multirow{5}{*}{Scalars}
&$\Phi$&($\mathbf{2}, {1/2}$)&\multicolumn{3}{c|}{{$1$}    $\to$ $+$}\\
&$\Delta$&($\mathbf{3}, {1}$)&\multicolumn{3}{c|}{{$1$}    $\to$ $+$} \\
&$\sigma$&($\mathbf{1}, {0}$)&\multicolumn{3}{c|}{{$\omega^2$}  $\to$  $+$}\\
&$\eta_{1}$&($\mathbf{2}, {1/2}$)&\multicolumn{3}{c|}{{$\omega^3$}    $\to$ $-$}\\
&$\eta_{2}$&($\mathbf{2}, {1/2}$)&\multicolumn{3}{c|}{{$\omega^5$}    $\to$ $-$}\\
\hline
	\end{tabular}
	\caption{Field content of the model and corresponding transformation properties under the \EW~gauge group . For the $\mathcal{Z}_8$ symmetry we have $\omega^k = e^{ik\pi/4}$.}
	\label{tab:part&sym} 
\end{table}

Taking into account the field content and the symmetries of our model, the most general Yukawa Lagrangian is
 \begin{equation}
- \mathcal{L}_{\text{Yuk.}} = \overline{\ell_L} \Y_{\ell}\, \Phi \,e_R + \overline{\ell_L} \Y_f^1 \tilde{\eta}_1 f  + \overline{\ell_L} \Y_f^2 \tilde{\eta}_2 f  + \overline{\ell_L^c} \Y_{\Delta} i \tau_2 \Delta \ell_L+ \frac{1}{2}  y_f \,\sigma \overline{f^c} f+\text{H.c.} \;,
\label{eq:LYuk}
\end{equation}
where $\ell_L=(\nu_L\; e_L)^T$ and $e_R$ denote the SM left-handed doublet and right-handed singlet charged-lepton fields, respectively. In a usual notation, the scalars $\Phi$, $\eta_i$ and $\Delta$ are defined as:
\begin{align}
    \Phi=\begin{pmatrix}
    \phi^+\\
    \phi^0
    \end{pmatrix}, \;
    \eta_i=\begin{pmatrix}
    \eta_i^+\\
    \eta_i^0
    \end{pmatrix},\;
    \Delta=\begin{pmatrix}
    \Delta^+/\sqrt{2} &\Delta^{++}\\
    \Delta^{0} &-\Delta^+/\sqrt{2}
    \end{pmatrix},
    \label{eq:scalardef}
\end{align}
with $\tilde{\Phi}=i\tau_2 \Phi^\ast$ and $\tilde{\eta_i}=i\tau_2 \eta_i^\ast$ ($i=1,2$), being $\tau_2$ the complex Pauli matrix. In this work we impose CP invariance at the Langrangian level, which implies real Yukawa couplings $\Y_\ell$, $\Y^{1,2}_f$, $\Y_\Delta$ and $y_f$. Notice that since the $\sigma f f$ coupling is allowed by the gauge and $\mathcal{Z}_{8}$ symmetries, LCPV can, in principle, be successfully transmitted to the neutrino sector as long as $\sigma$ acquires a complex VEV~\footnote{The scalar sector of our model is discussed in appendix~\ref{sec:scalar} where it is shown that SCPV can be realised in our framework, and the scalar mass spectrum is discussed.}
\begin{align}
\langle\sigma\rangle=u\,e^{i\theta}/\sqrt{2}\,,
\end{align}
where $u$ and $\theta$ are real. If this is so, LCPV probed in neutrino oscillation experiments originates dynamically from the vacuum. Notice also that all fermions become massive after SSB. As already mentioned, the three different $\mathcal{Z}_8$ lepton charge assignments given in table~\ref{tab:part&sym} will lead to different Yukawa couplings $\Y_\ell$, $\Y^{1,2}_f$ and $ \Y_\Delta$. For the specific $\mathcal{Z}_8^{e-\mu}$ case, we have
\begin{align}
    \Y_\ell=\begin{pmatrix}
    \times&0&0\\
    0&\times&0\\
    0&0&\times
    \end{pmatrix},\;
    \Y^1_f=\begin{pmatrix}
    \times\\
    0\\
    0
    \end{pmatrix},\;
    \Y^2_f=\begin{pmatrix}
    0\\
    \times\\
    0
    \end{pmatrix},\;
    \Y_\Delta=\begin{pmatrix}
    \times&0&0\\
    .&0&\times\\
    .&.&0
    \end{pmatrix},
    \label{eq:yukawastructures}
\end{align}
where the $\times$'s denote real entries (due to CP conservation) not determined by the symmetries of the model, and the dots in $\Y_\Delta$ reflect its symmetric nature. The corresponding matrices for $\mathcal{Z}_8^{e-\tau}$ and $\mathcal{Z}_8^{\mu-\tau}$ can be obtained from the above by performing row permutations in $\Y^i_f$, and column and row permutations in $\Y_\Delta$. For all cases, $\Y_\ell$ is diagonal and, thus, nontrivial lepton mixing originates solely from the neutrino sector.

\begin{figure}[t!]
    \centering
    \includegraphics[scale=0.9,trim={2.5cm 19.3cm 2cm 1.5cm},clip]{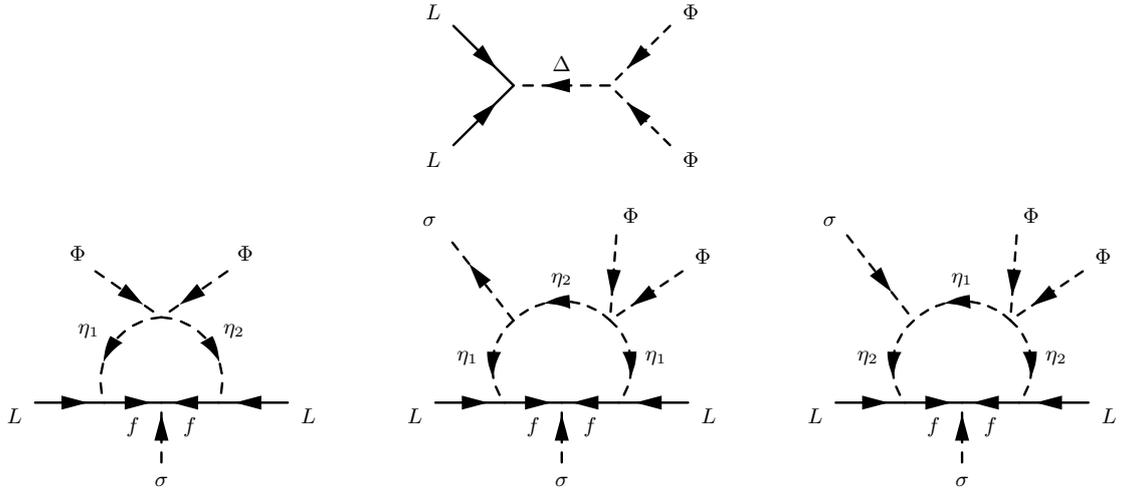}
    \caption{First row: type-II seesaw contribution to the dimension-5 effective mass operator $LL\Phi\Phi$. Second row: (scotogenic) one-loop contributions to the dimension-6 operator $LL\Phi\Phi\sigma$ (left diagram) and dimension-7 operators $LL\Phi\Phi\sigma\sigma^*$ and $LL\Phi\Phi\sigma\sigma$ (centre and right diagrams, respectively).}
    \label{fig:neutrinomassdiagrams}
\end{figure}

\begin{figure}[t!]
    \centering
    \includegraphics[scale=0.9,trim={2.5cm 11.5cm 2cm 1.5cm},clip]{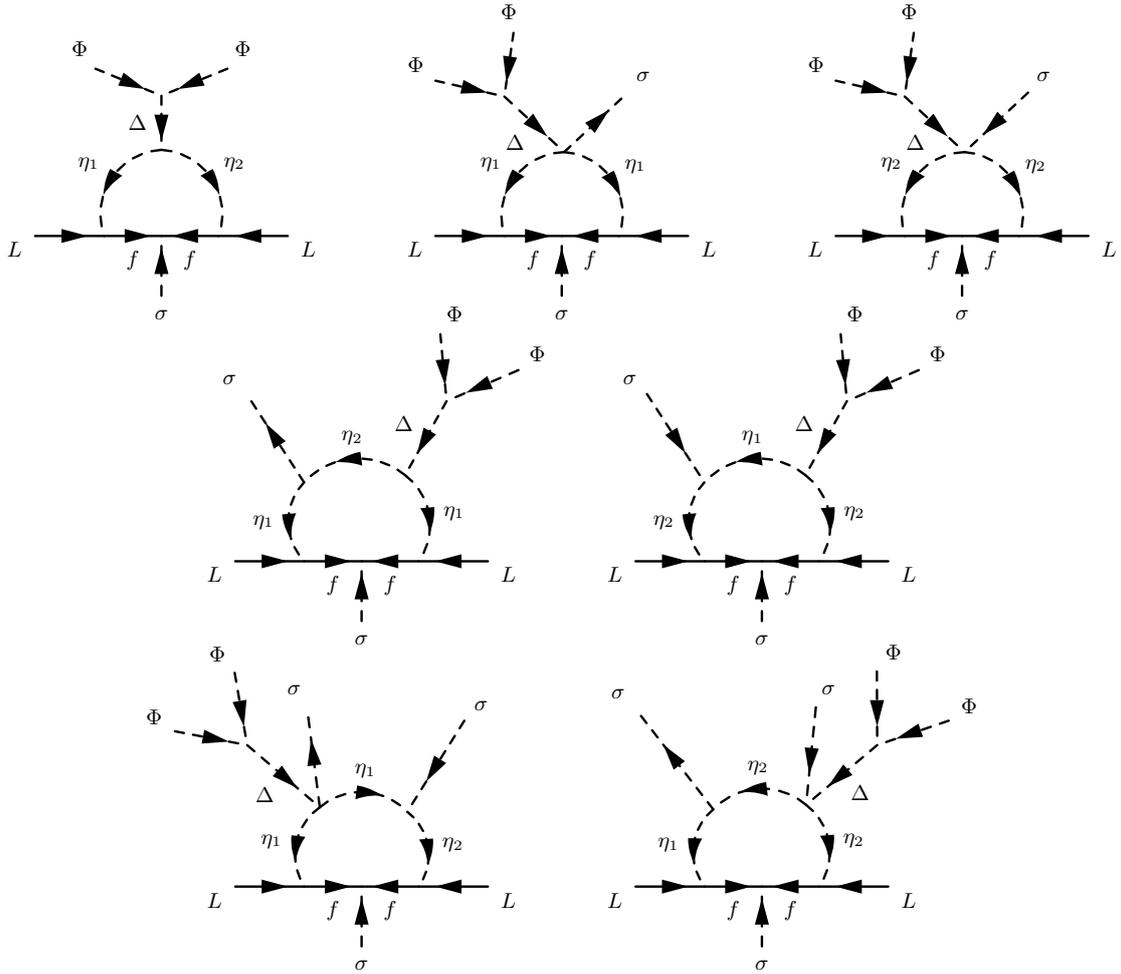}
    \caption{One-loop contributions to the dimension-6 operator $LL\Phi\Phi\sigma$ (first row diagrams), to the dimension-7 operators $LL\Phi\Phi\sigma\sigma^*$ and $LL\Phi\Phi\sigma\sigma$ (second row diagrams), and to the dimension-8 operator $LL\Phi\Phi\sigma\sigma\sigma^*$ (third row diagrams), involving the triplet $\Delta$.}
    \label{fig:neutrinomassdiagramsdelta}
\end{figure}

In our model, neutrino masses receive contributions both from the tree-level type-II seesaw~\cite{Konetschny:1977bn,Cheng:1980qt,Lazarides:1980nt,Schechter:1980gr,Mohapatra:1980yp,Magg:1980ut} and the radiative scotogenic mechanisms~\cite{Ma:2006km}. The resulting effective neutrino mass matrix is given by,
\begin{align}
    \Mnu=&\;\sqrt{2}w\Y_\Delta+\mathcal{F}_{11}(M_f,m_{S_k})M_f\Y^1_f\Y_f^{1T}+\mathcal{F}_{22}(M_f,m_{S_k})M_f\Y^2_f\Y_f^{2T}\nonumber\\
    &+\mathcal{F}_{12}(M_f,m_{S_k})M_f \left( \Y^1_f\Y_f^{2T}+\Y_f^2\Y_f^{1T}\right).
    \label{eq:neutrinomassmatrix}
\end{align}
The first term corresponds to the type-II seesaw (diagram in the first row of fig.~\ref{fig:neutrinomassdiagrams}) where $w/\sqrt{2}$ is the VEV of $\Delta^0$, while the remaining terms account for the one-loop scotogenic contributions (diagrams in the second row in fig.~\ref{fig:neutrinomassdiagrams} and diagrams in fig.~\ref{fig:neutrinomassdiagramsdelta}). The loop factors $\mathcal{F}_{11}(M_f,m_{S_k})$, $\mathcal{F}_{12}(M_f,m_{S_k})$ and $\mathcal{F}_{22}(M_f,m_{S_k})$ depend on the $f$ mass $M_f=u\, y_f/\sqrt{2}$ generated after SSB, and on the \textit{dark} neutral-scalar masses $m_{S_k}$ and mixing resulting from the neutral components of $\eta_1$ and $\eta_2$. The complete expressions for $\mathcal{F}_{11}$, $\mathcal{F}_{12}$ and $\mathcal{F}_{22}$ are given in  eqs.~\eqref{eq:F11loop}, \eqref{eq:F12loop} and \eqref{eq:F22loop} of appendix~\ref{sec:scalar}, respectively. 

A comment is in order regarding the diagrams in fig.~\ref{fig:neutrinomassdiagramsdelta}. It is natural to consider that the couplings $\mu_\Delta$ and $\lambda_5$ from the scalar potential terms $\mu_{\Delta} \Phi^{\dagger} \Delta i \tau_2 \Phi^{*}$ and $\lambda_{5} \eta_1^{\dagger} \Phi \eta_2^{\dagger} \Phi$ given in eqs.~\eqref{eq:VPhiDelta} and \eqref{eq:VetaPhi}, are small. In fact, by setting them to zero, it is straightforward to show that a lepton number symmetry is recovered. Therefore, those couplings are naturally small in the t'Hooft sense~\cite{tHooft:1979rat}. Furthermore, $\lambda_5$ and $\mu_\Delta$ are of special interest, due to their role in neutrino mass generation. Namely, for $\lambda_5=0$ the scotogenic diagrams in fig.~\ref{fig:neutrinomassdiagrams} vanish. On the other hand, by setting $\mu_\Delta=0$ there is no type-II seesaw contribution to neutrino mass generation (see fig.~\ref{fig:neutrinomassdiagrams}) and all one-loop diagrams of fig.~\ref{fig:neutrinomassdiagramsdelta} are identically zero. Notice that the quartic coupling $\lambda_5$ is not related to the VEVs of the neutral components of the Higgs triplet, doublet and singlet. However, as shown in eq.~\eqref{eq:VEVDelta} of appendix~\ref{sec:scalar}, the coupling $\mu_\Delta$ controls directly the scalar triplet VEV $w$, making the hierarchy $w \ll v \lsim u$ natural.~\footnote{In section~\ref{sec:cLFVmain}, we mention that electroweak precision measurements impose $w \lesssim 8$ GeV~\cite{Kanemura:2012rj,Arhrib:2011uy,Arhrib:2011vc,Aoki:2012yt,Aoki:2012jj}, making this VEV hierarchy choice phenomenologically suitable.} Following this reasoning, in this limit the $\Delta$ contribution to \textit{dark} neutral scalar masses is negligible [see eqs.~\eqref{eq:eta1eta2MassMatrix},~\eqref{eq:S12mass} and~\eqref{eq:S34mass}], and consequently the diagrams in fig.~\ref{fig:neutrinomassdiagramsdelta} are subdominant compared to those in fig.~\ref{fig:neutrinomassdiagrams}.

Once $\sigma$ acquires a VEV, and after performing the rephasings $f,\ell_L\rightarrow e^{-i\theta/2}\; f,\ell_L$, the $f$ mass $M_f$ becomes real and the phase $\theta$ is transmitted to $\Y_\Delta$. Thus, for the specific case of the  $\mathcal{Z}_8^{e-\mu}$ charge assignment, the Yukawa matrices read
\begin{align}
    \Y_f^1=\begin{pmatrix}
    y_e\\
    0\\
    0
    \end{pmatrix},\;
    \Y_f^2=\begin{pmatrix}
    0\\
    y_\mu\\
    0
    \end{pmatrix},\;
    \Y_\Delta=\begin{pmatrix}
    y_1&0&0\\
    0&0&y_2\\
    0&y_2&0
    \end{pmatrix}\;e^{-i\theta},
    \label{eq:rephasedyukawas}
\end{align}
which combined with eq.~\eqref{eq:neutrinomassmatrix} lead to
\begin{align}
\Mnu=\begin{pmatrix}
\mathcal{F}_{11} M_f\, y_e^2+\sqrt{2}w \,y_{1}\,e^{-i\theta} & &\mathcal{F}_{12} M_f\, y_e y_\mu& &0\\
\cdot & &\mathcal{F}_{22}M_f\, y_\mu^2 && \sqrt{2}w \,y_2e^{-i\theta}\\
\cdot & &\cdot & &0\\
\end{pmatrix}\, ,
\label{eq:finalmnu}
\end{align}
where the $\mathcal{F}_{ij}$ loop factors are those of eq.~\eqref{eq:neutrinomassmatrix}. The two texture-zero conditions $(\Mnu)_{13}=(\Mnu)_{33}=0$ stem from the considered flavour symmetry. It is worth stressing that the interplay among the type-II seesaw and the scotogenic terms is crucial to ensure compatibility with neutrino oscillation data and for LCPV to be possible. In fact, if one of the couplings $y_{e,\mu}$ or $y_{2}$ vanishes, compatibility with neutrino data is lost since $\Mnu$ would feature an additional vanishing entry, leading to two massless neutrinos and vanishing mixing angles. Alternatively, if $y_{1}=0$, there is no CP violation in the lepton sector. For each of the three possible charge assignments, $\mathcal{Z}_8^{e-\mu}$, $\mathcal{Z}_8^{e-\tau}$ and $\mathcal{Z}_8^{\mu-\tau}$, the obtained effective neutrino mass matrices exhibit the following texture-zero structures:
\begin{align}
   \mathcal{Z}_8^{e-\mu} \rightarrow \text{B}_4:\begin{pmatrix}
    \times&\times&0\\
    .&\times&\times\\
    .&.&0
    \end{pmatrix},\;
    \mathcal{Z}_8^{e-\tau}\rightarrow \text{B}_3:\begin{pmatrix}
    \times&0&\times\\
    .&0&\times\\
    .&.&\times
    \end{pmatrix},\;
    \mathcal{Z}_8^{\mu-\tau}\rightarrow\text{A}_1:\begin{pmatrix}
    0&0&\times\\
    .&\times&\times\\
    .&.&\times
    \end{pmatrix}\,,
    \label{eq:neutrinomasstextures}
\end{align}
where the nomenclature $\text{B}_{3,4}$ and $\text{A}_1$ follows that of ref.~\cite{Alcaide:2018vni}. In the next section, we will test our model against current neutrino data and analyse further constraints to which it leads to.

\section{Compatibility with neutrino data}
\label{sec:neutrinomass}

The compatibility of the neutrino mass matrices in eq.~\eqref{eq:neutrinomasstextures} with neutrino oscillation data has been previously analysed in ref.~\cite{Alcaide:2018vni}. In this work, we update the analysis in light of the latest global fits of neutrino data, the current bounds on $0\nu\beta\beta$ and on the absolute neutrino mass scale. In order to find the low-energy constraints imposed by textures A$_1$, B$_3$ and B$_4$ on the (physical) neutrino-mass and mixing parameters, we first express the effective neutrino mass matrix $\widehat{\mathbf{M}}_\nu$ as
\begin{align}
    \Mnuh=\U^\ast\, \text{diag}(m_1,m_2,m_3)\,\U^\dagger,
    \label{eq:mnureconstruction}
\end{align}
being $m_i$ the neutrino masses and $\U$ the lepton mixing matrix, parameterised through~\cite{Rodejohann:2011vc}
\begin{align}
\U=\begin{pmatrix}
c_{12}c_{13}&s_{12}c_{13}&s_{13}\\
-s_{12}c_{23}-c_{12}s_{23}s_{13}e^{i\delta}&c_{12}c_{23}-s_{12}s_{23}s_{13}e^{i\delta}&s_{23}c_{13}e^{i\delta}\\
s_{12}s_{23}-c_{12}c_{23}s_{13}e^{i\delta}&-c_{12}s_{23}-s_{12}c_{23}s_{13}e^{i\delta}&c_{23}c_{13}e^{i\delta}
\end{pmatrix} \begin{pmatrix}
1&0&0\\
0&e^{i\frac{\alpha_{21}}{2}}&0\\
0&0&e^{i\frac{\alpha_{31}}{2}}
\end{pmatrix}\;,
\label{eq:Uparam}
\end{align}
where $\theta_{ij}$ ($i<j=1,2,3$) are the lepton mixing angles (with $s_{ij}\equiv\sin\theta_{ij}$, $c_{ij}\equiv\cos\theta_{ij}$), $\delta$ is the Dirac CP-violating phase and $\alpha_{21,31}$ are Majorana phases. Since we are working in the mass basis of charged leptons [see eq.~\eqref{eq:yukawastructures}], the matrix $\U$ corresponds to the one which diagonalises $\Mnu$. For normal and inverted neutrino-mass ordering (NO and IO, respectively), two of the three neutrino masses may be expressed in terms of the lightest neutrino mass (corresponding to $m_1$ and $m_3$ for NO and IO, respectively) and the measured neutrino mass-squared differences $\dmsol=m_2^2-m_1^2$ and $\dmatm=m_3^2-m_1^2$ as
\begin{align}
   \text{NO:}&\quad m_2=\sqrt{m_1^2+\dmsol},\quad m_3=\sqrt{m_1^2+\dmatm}\;,\\
   \text{IO:} &\quad m_1=\sqrt{m_3^2+|\dmatm|},\quad m_2=\sqrt{m_3^2+\dmsol+|\dmatm|}\;.
\end{align}
The current allowed intervals for the lepton mixing angles, neutrino mass-squared differences and the Dirac phase $\delta$ obtained from global fits of neutrino oscillation data are shown in table~\ref{tab:dataref}~\cite{deSalas:2020pgw,Esteban:2020cvm,Capozzi:2021fjo}.
\begin{table}[!t]
\renewcommand{\arraystretch}{1.2}
\centering
\setlength{\tabcolsep}{10pt}
\begin{tabular}{|l|c|c|}  
\hline
Parameter  & Best Fit $\pm 1 \sigma$ & $3\sigma$ range \\ \hline
$\theta_{12} (^\circ)$ & $34.3\pm1.0$ &  $31.4 \rightarrow 37.4$ \\
$\theta_{23} (^\circ) [\text{NO}]$ & $49.26\pm0.79$ &  $ 41.20 \rightarrow 51.33 $ \\
$\theta_{23} (^\circ) [\text{IO}]$ & $49.46^{+0.60}_{-0.97}$  &  $ 41.16 \rightarrow 51.25$ \\
$\theta_{13} (^\circ) [\text{NO}]$ & $8.53^{+0.13}_{-0.12}$ &  $8.13 \rightarrow 8.92$\\
$\theta_{13} (^\circ) [\text{IO}]$ & $8.58^{+0.12}_{-0.14}$ &  $8.17 \rightarrow 8.96$ \\
$\delta  (^\circ) [\text{NO}]$ & $194^{+24}_{-22}$ & $128 \rightarrow 359 $ \\
$\delta  (^\circ) [\text{IO}]$ & $284^{+26}_{-28}$ &  $200 \rightarrow 353 $ \\
$\Delta m_{21}^2 \left(\times 10^{-5} \ \text{eV}^2\right)$ & $7.50^{+0.22}_{-0.20}$ &  $ 6.94 \rightarrow 8.14$ \\
$\left|\Delta m_{31}^2\right| \left(\times 10^{-3}  \ \text{eV}^2\right) [\text{NO}]$ & $2.55^{+0.02}_{-0.03}$  & $2.47 \rightarrow 2.63 $ \\
$\left|\Delta m_{31}^2\right| \left(\times 10^{-3} \ \text{eV}^2\right) [\text{IO}]$ & $2.45^{+0.02}_{-0.03}$ & $2.37 \rightarrow 2.53$\\
\hline
\end{tabular}
\caption{Current allowed intervals for the lepton mixing angles, neutrino mass-squared differences and the Dirac phase $\delta$ obtained from the global fit of neutrino oscillation data performed in ref.~\cite{deSalas:2020pgw}. See also refs.~\cite{Esteban:2020cvm} and~\cite{Capozzi:2021fjo}.}
\label{tab:dataref}
\end{table} 

The effective neutrino mass matrix $\Mnu$ of eq.~\eqref{eq:neutrinomassmatrix}, obtained considering the symmetries of the model, must be matched with $\Mnuh$ in eq.~\eqref{eq:mnureconstruction} by performing proper rephasings of the lepton fields and imposing the texture-zero conditions in eq.~\eqref{eq:neutrinomasstextures}. This allows checking whether those conditions are compatible with current neutrino data and possibly select regions of the physical neutrino-parameter space preferred by the model. It is worth stressing that the parameters in $\Mnu$, i.e. the Yukawa couplings in $\Yf^1$, $\Yf^2$ and $\mathbf{Y}_{\Delta}$ and the SCPV phase~$\theta$, can be fully determined once the masses of~$f$, $S_k$ and the VEV $w$ are fixed. In the $\mathcal{Z}_8^{e-\mu}$ case, and taking into account eq.~\eqref{eq:finalmnu}, we get:
\begin{align}
&y_e = \dfrac{\left|(\Mnuh)_{12}\right|}{\mathcal{F}_{12}}\sqrt{\dfrac{\mathcal{F}_{22}}{M_f\,\left|(\Mnuh)_{22}\right|}}, \; y_\mu  = \sqrt{\dfrac{\left|(\Mnuh)_{22}\right|}{M_f\,\mathcal{F}_{22}}}, \label{eq:yscoto}\\
&y_1  = \frac{1}{\sqrt{2}w} \left[\left|(\Mnuh)_{11}\right|^2 + \frac{\left|(\Mnuh)_{12}\right|^4}{\left|(\Mnuh)_{22}\right|^2} \mathcal{R}^2 - 2 \cos\xi \frac{\left|(\Mnuh)_{11}\right| \left|(\Mnuh)_{12}\right|^2}{\left|(\Mnuh)_{22}\right|} \mathcal{R} \right]^{\frac{1}{2}}, \nonumber \\
& y_2  = \frac{\left|(\Mnuh)_{23}\right|}{\sqrt{2}w}, \label{eq:yDelta} \\
&\text{cotan }\theta  = -\dfrac{1}{\sin\xi}\left[\cos\xi- \dfrac{ \left|(\Mnuh)_{12}\right|^2}{\left|(\Mnuh)_{11}\right|\left|(\Mnuh)_{22}\right|}\mathcal{R}\right],
\label{eq:theta}
\end{align}
where the phase $\xi$
reads
\begin{align}
    \xi  = \arg\left[(\Mnuh)_{11}\right] - 2 \arg\left[(\Mnuh)_{12}\right] + \arg\left[(\Mnuh)_{22}\right],
\end{align}
and the loop-function ratio $\mathcal{R}$ is defined as
\begin{align}
    \mathcal{R}=\dfrac{\mathcal{F}_{11}(M_f,m_{S_k})\mathcal{F}_{22}(M_f,m_{S_k})}{\mathcal{F}_{12}^2(M_f,m_{S_k})}.
    \label{eq:Rratio}
\end{align}
In the above equations, the elements $(\Mnuh)_{ij}\neq 0$ are evaluated for values of physical parameters which fulfil the B$_4$ texture-zero condition given in \eqref{eq:neutrinomasstextures}. Interestingly, the $\sigma$ VEV phase  $\theta$ is fully determined by the neutrino mass and mixing parameters and the loop-function ratio $\mathcal{R}$, as seen from eq.~\eqref{eq:theta} (further consequences of this will be addressed at the end of this section). Expressions similar to those in~\eqref{eq:yscoto}-\eqref{eq:Rratio} can be readily obtained for cases $\mathcal{Z}_8^{e-\tau}$ and $\mathcal{Z}_8^{\mu-\tau}$ by performing the index replacements in $(\Mnuh)_{ij}$ and Yukawa parameters ($22 \rightarrow 33$, $12 \rightarrow 13$, $\mu \rightarrow \tau$) and ($22 \rightarrow 33$, $12 \rightarrow 23$, $11 \rightarrow 22$, $23 \rightarrow 13$ , $e \rightarrow \mu$, $\mu \rightarrow \tau$), respectively. 

Let us proceed to the compatibility analysis of textures A$_1$, B$_3$ and B$_4$ with the data shown in table~\ref{tab:dataref}. The two texture-zero conditions of $\Mnu$ are tested performing a chi-squared analysis based on the minimisation of $\chi^2_\text{tot}$. This function is computed using the one-dimensional profiles $\chi^2(s^2_{ij})$ and $\chi^2(\Delta m^2_{ij})$, and the two-dimensional distribution  $\chi^2(\delta,s^2_{23})$ for $\delta$ and $\theta_{23}$ given in refs.~\cite{deSalas:2020pgw,10.5281/zenodo.4726908}. The constraints in $\Mnu$ are incorporated in $\chi^2_\text{tot}$, using the Lagrange multiplier method described in ref.~\cite{Alcaide:2018vni}. For a fixed pair of parameters, the 2D 1$\sigma$, 2$\sigma$ and 3$\sigma$ compatibility regions were obtained by requiring $\chi^2_\text{tot}\leq 2.30$, $6.18$ and $11.83$, respectively. 
\begin{figure}
    \centering
    \includegraphics[scale=0.36]{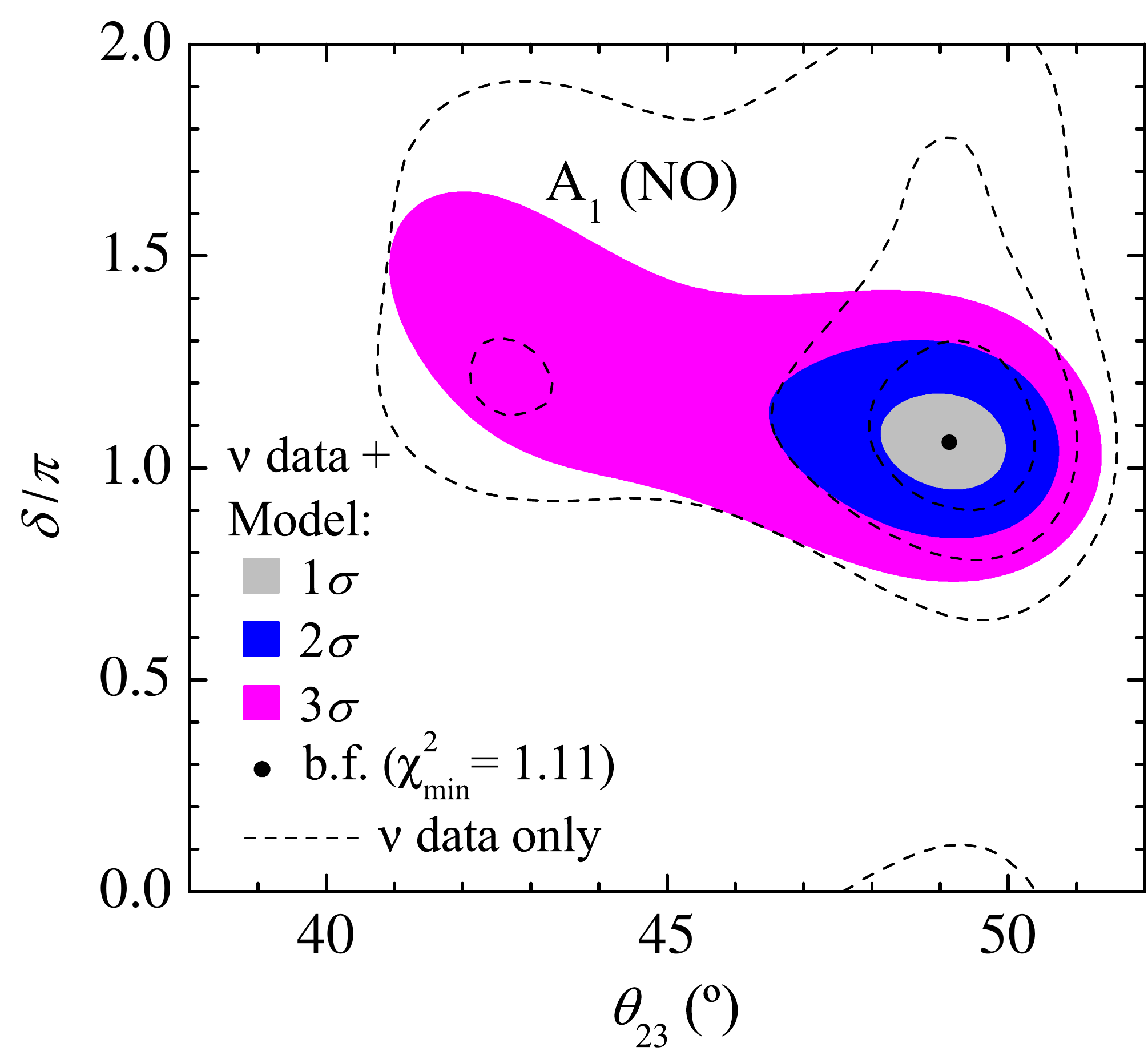}\hspace{0.2cm}
    \includegraphics[scale=0.36]{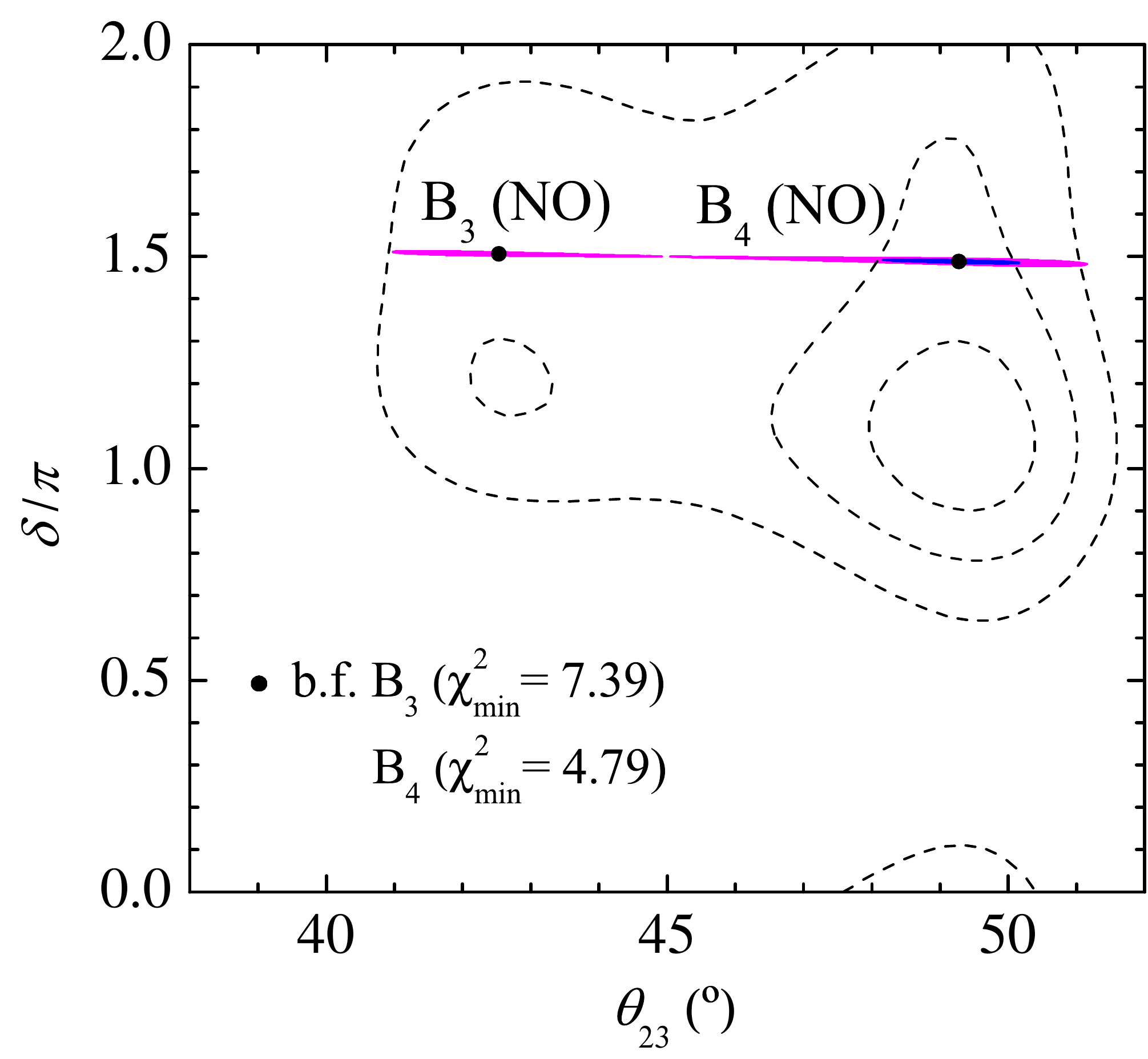}\\[0.2cm]
    \includegraphics[scale=0.36]{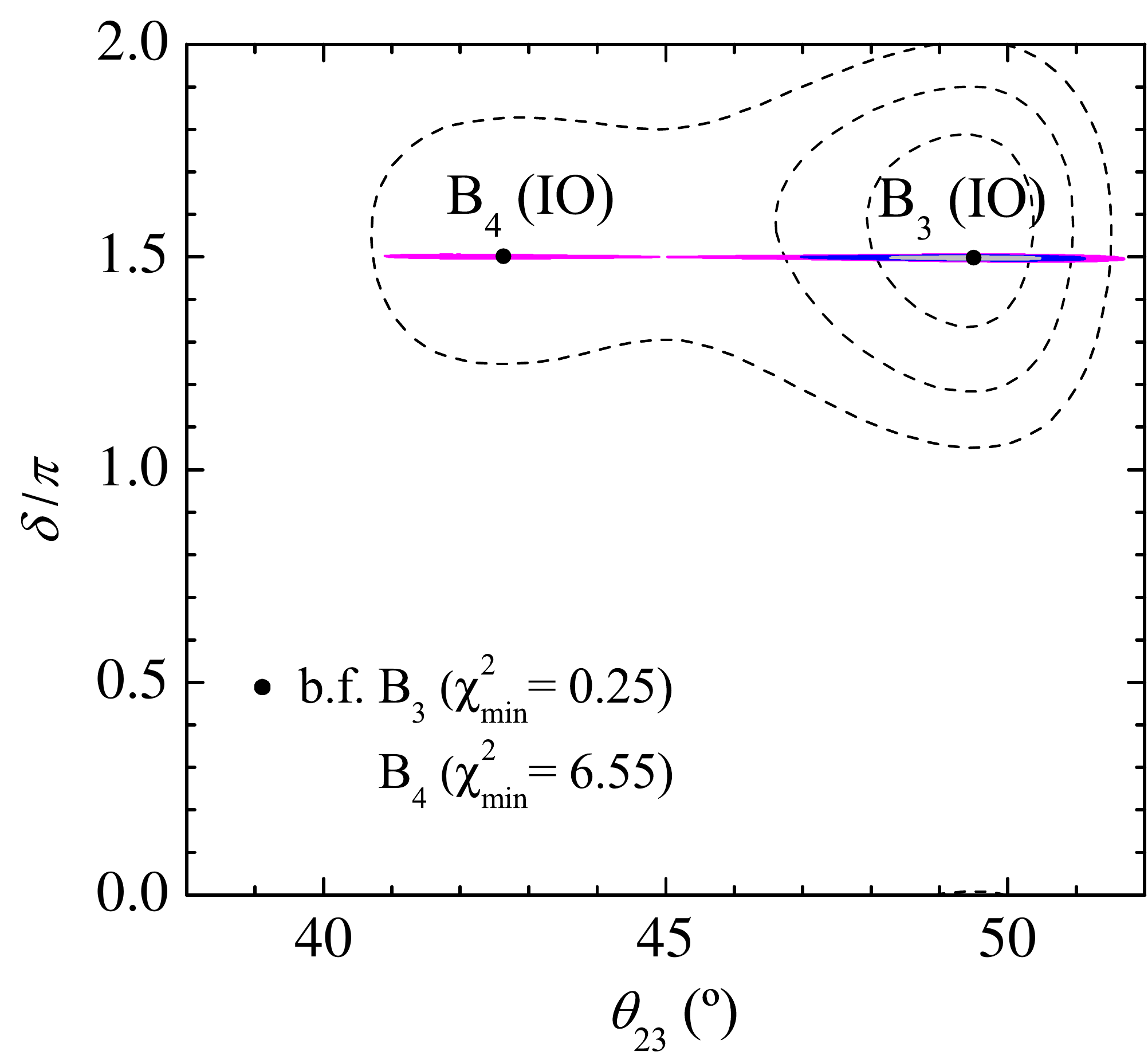}
    \caption{Allowed regions at the 1, 2 and 3$\sigma$ level (in grey, blue and magenta, respectively) in the plane ($\theta_{23}$,$\delta$), for case A$_1$ and NO (upper left plot), B$_3$/B$_4$ and NO (upper right plot), and B$_3$/B$_4$ and IO (lower plot). The black dots mark the best-fit value for each case, while the dashed contours correspond to the $\chi^2$ contours at 1, 2 and 3$\sigma$, allowed by the global fit of neutrino oscillation data.}
    \label{fig:predictionst23delta}
\end{figure}
\begin{figure}
    \centering
    \includegraphics[scale=0.37]{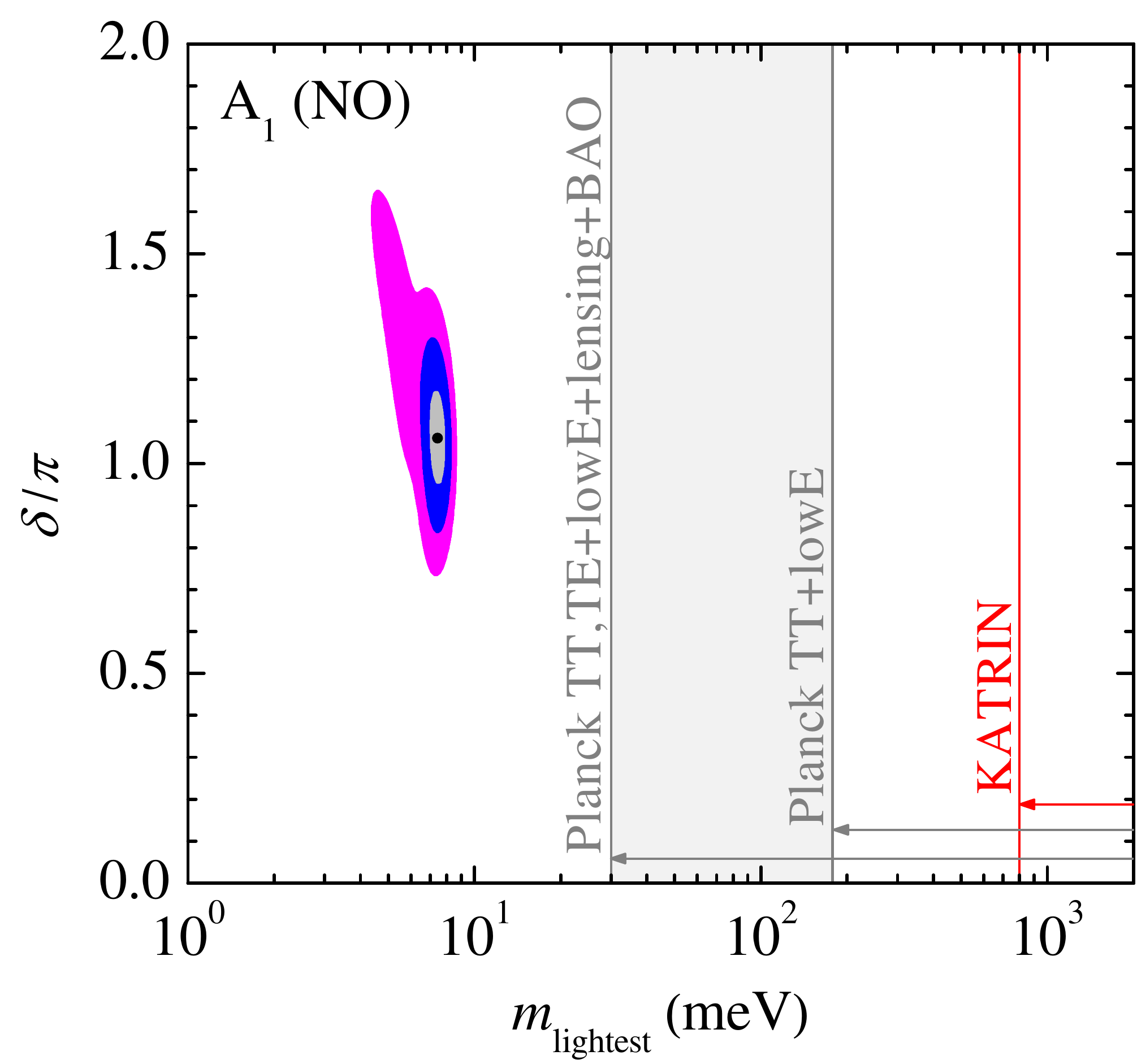}\\
    \includegraphics[scale=0.37]{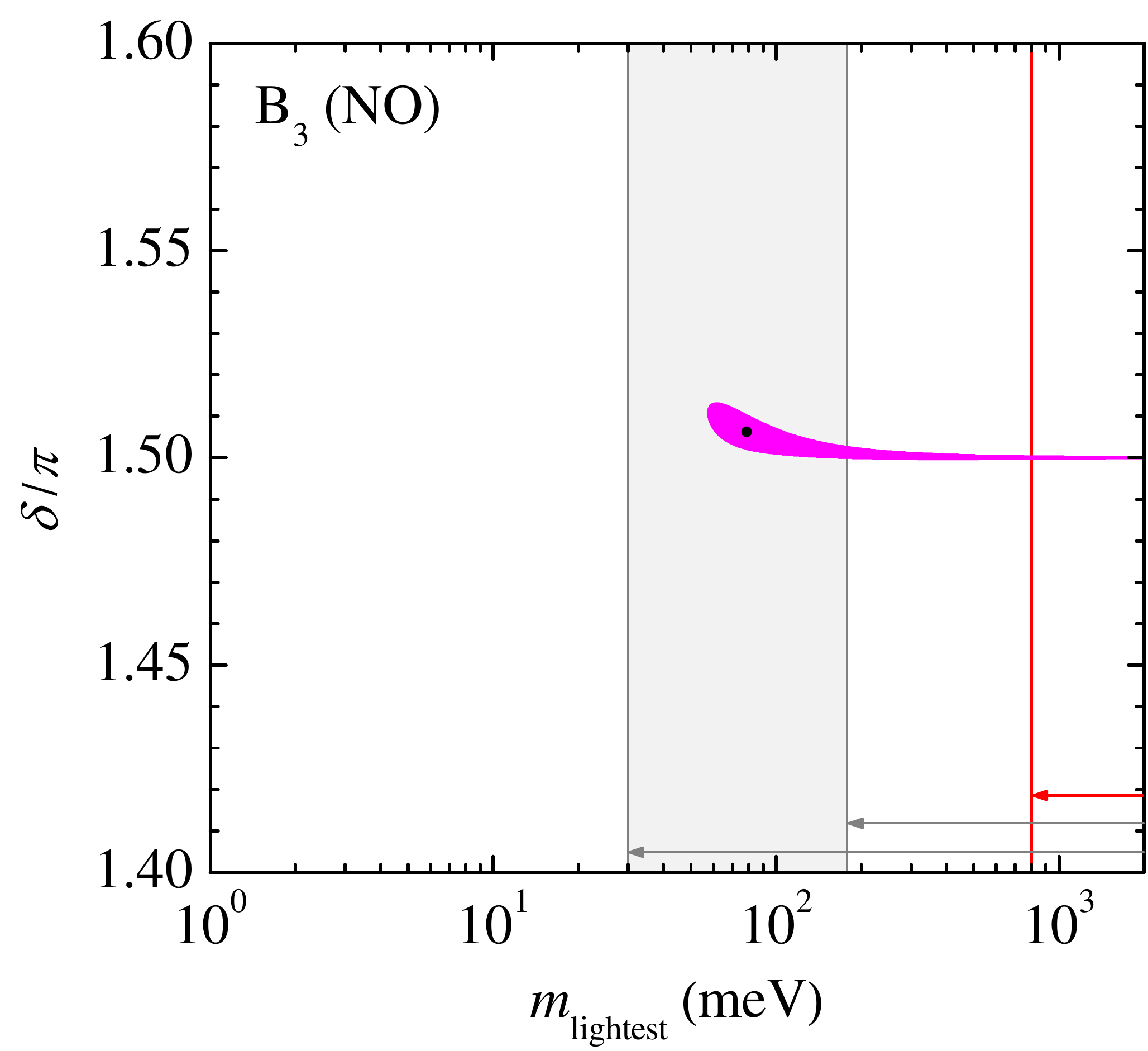}\hspace{0.1cm}
    \includegraphics[scale=0.37]{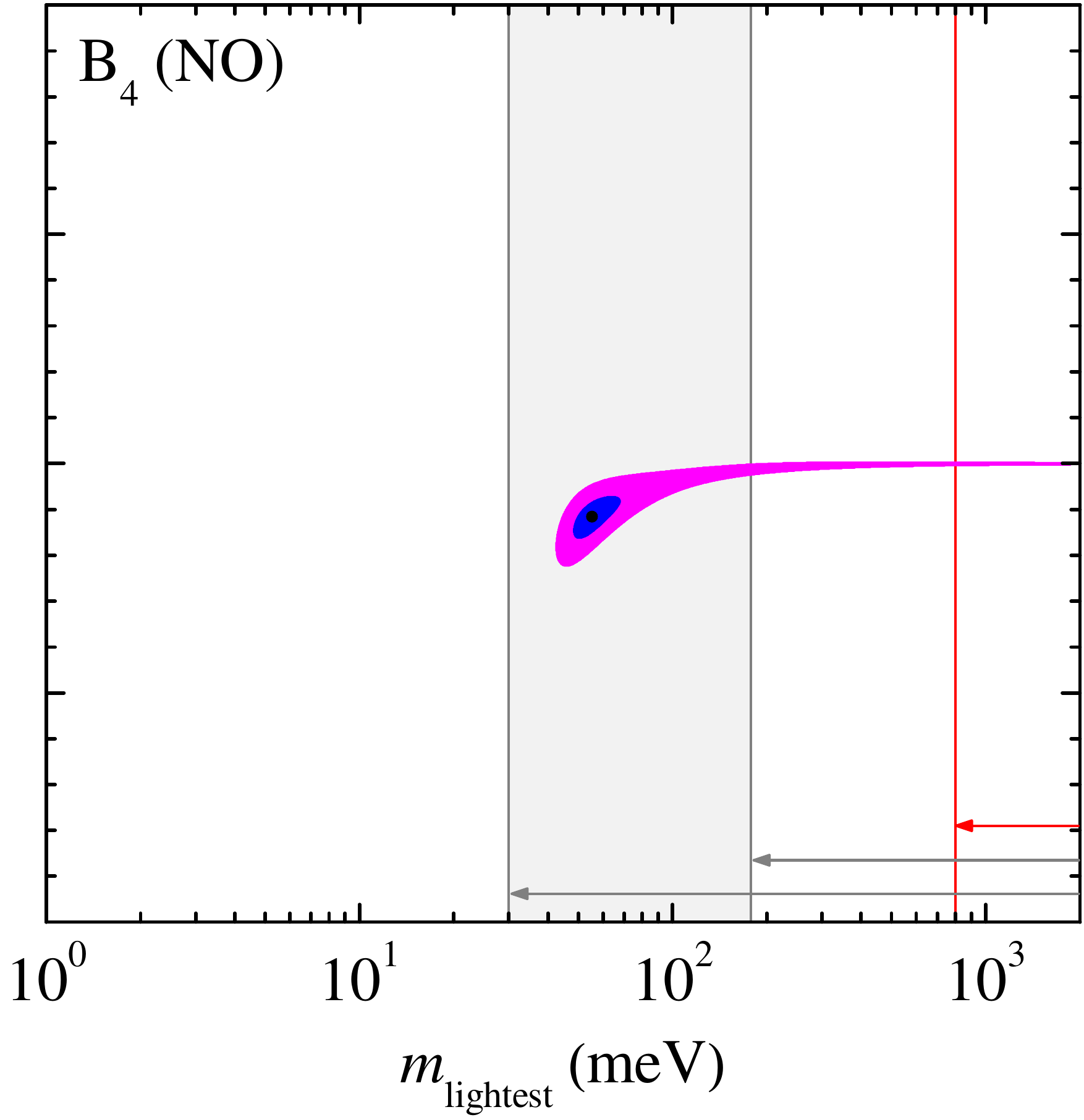}\\
    \includegraphics[scale=0.37]{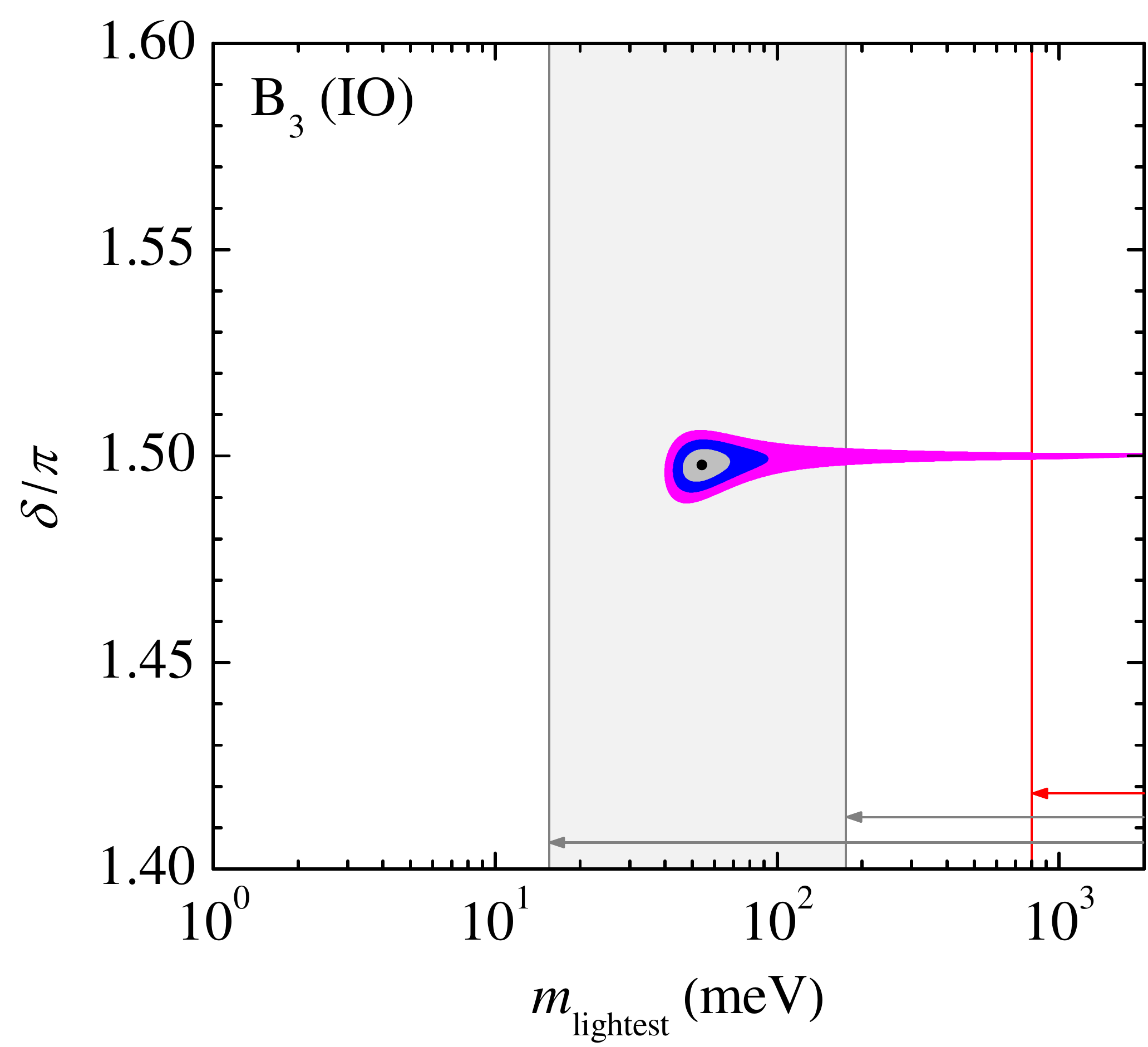}\hspace{0.1cm}
    \includegraphics[scale=0.37]{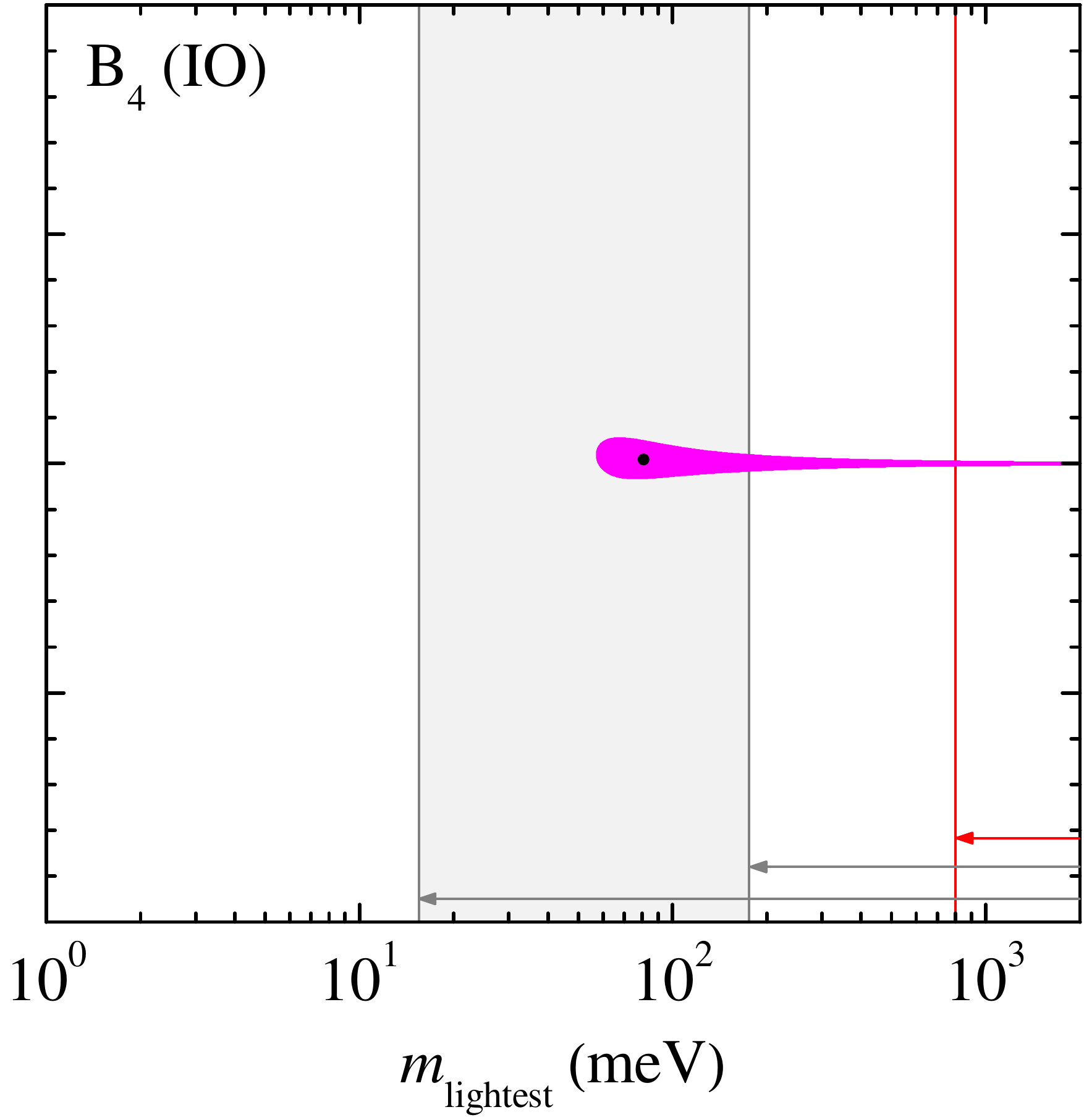}
    \caption{Allowed regions in the plane ($m_\text{lightest}$,$\delta$) for NO (top and middle plots) and IO (bottom plots). The colour scheme is the same as in fig.~\ref{fig:predictionst23delta}. The vertical red line corresponds to the upper limit for $m_\text{lightest}$ extracted from the KATRIN bound on $m_\beta$. The grey shaded vertical region shows the $m_\text{lightest}$ upper-limit interval obtained from the Planck cosmological bounds on $\sum_k m_k$.}
    \label{fig:predictionsmlightestdelta}
\end{figure}

In figs.~\ref{fig:predictionst23delta} and~\ref{fig:predictionsmlightestdelta} we show the 1$\sigma$, 2$\sigma$ and 3$\sigma$ allowed regions (in grey, blue and magenta, respectively) in the planes ($\theta_{23}$,$\delta$) and ($m_\text{lightest}$,$\delta$) for textures A$_1$, B$_3$ and B$_4$. For comparison, we show in fig.~\ref{fig:predictionst23delta} the 1$\sigma$, 2$\sigma$ and 3$\sigma$ contours of the regions allowed by neutrino oscillation data~\cite{deSalas:2020pgw,10.5281/zenodo.4726908} (dashed curves) together with the best-fit point for each case and corresponding $\chi^2_\text{min}$ values. The cases that fit the data best are B$_3$ for IO ($\chi^2_\text{min}=0.25$) and A$_1$ for NO ($\chi^2_\text{min}=1.11$), being both compatible at the 1$\sigma$ level. The case B$_4$ with NO ($\chi^2_\text{min}=4.79$) is viable at 2$\sigma$, while the cases B$_4$ with IO ($\chi^2_\text{min}=6.55$) and B$_3$ with NO ($\chi^2_\text{min}=7.39$) are only compatible with data at the 3$\sigma$ level. \footnote{Notice that we have not included in our chi-squared calculation the $2.5\sigma$ data preference for NO~\cite{deSalas:2020pgw}.} For A$_1$ we have $(\Mnu)_{11}=0$, which leads to a vanishing $0\nu\beta\beta$ decay rate. Therefore, this case is not allowed by neutrino oscillation data with IO neutrino masses and, for this reason, is not shown in our plots. Although we have not included $m_\text{lightest}$ bounds from cosmology or $\beta$-decay experiments in our $\chi^2_\text{tot}$ calculation, we present in fig.~\ref{fig:predictionsmlightestdelta} the upper limit on $m_\text{lightest}$ obtained from the KATRIN current bound $m_\beta<0.8$~eV (at 90\% CL)~\cite{KATRIN:2021uub} (red vertical line), and the upper-bound range on $m_\text{lightest}$ calculated from Planck limits on the sum of neutrino masses (grey shaded vertical band). The left (right) vertical dark-grey line refers to the less (most) conservative 95\% CL bound, corresponding to $\sum_k m_k<0.12$~eV ($0.54$~eV) where Planck TT,TE+lowE+lensing+BAO (Planck TT+lowE) data was used~\cite{Planck:2018vyg}.

From the plots in fig.~\ref{fig:predictionst23delta}, we conclude that all three textures lead to significant constraints on $\theta_{23}$ and $\delta$.  For NO, A$_1$ selects the second octant for $\theta_{23}$ at the 2$\sigma$ level and predicts a CP-violating phase $\delta$ in the interval $~[0.8,1.6]\pi$ at 3$\sigma$. On the other hand, cases B$_3$ and B$_4$ sharply predict $\delta\sim 3\pi/2$. Additionally, the model selects the first (second) $\theta_{23}$ octant for texture B$_3$ with NO and B$_4$ with IO (B$_4$ with NO and B$_3$ with IO). It is interesting to note that the $\theta_{23}$ octant preference of these textures, and also the $\delta$-phase predictions mentioned above, will be tested by improving the sensitivity on $\delta$ and $\theta_{23}$ with upcoming experiments like DUNE~\cite{DUNE:2015lol} or T2HK~\cite{Hyper-KamiokandeWorkingGroup:2014czz}.

The constraints of our model on the lightest neutrino mass are presented in fig.~\ref{fig:predictionsmlightestdelta}. One can see that, for all cases, the model predicts a 3$\sigma$ lower limit for $m_\text{lightest}$, while for A$_1$ with NO an upper bound is also established. These limits are a direct consequence of requiring compatibility of lepton mixing angles, mass-squared differences and $\delta$ with oscillation data. For A$_1$, $m_\text{lightest}$ is constrained to the range $\sim [4,9]$~meV (3$\sigma$), well below the current limits from cosmology and KATRIN. The remaining cases establish a lower limit on $m_\text{lightest}$ for values $\sim 40$~meV (3$\sigma$). This value is now being probed by cosmological observations, lying above (below) the less (most) conservative Planck bound. The case B$_4$ with NO (B$_3$ with IO) predicts as well an upper limit on $m_\text{lightest}$, but only at the 2$\sigma$ level for values around $60$~meV ($100$~meV). 

Lastly, we analyse the constraints on $0\nu\beta\beta$ decay rate by presenting the allowed regions for the effective neutrino mass parameter $m_{\beta\beta}$. Using the parameterisation given in eq.~\eqref{eq:Uparam}, $m_{\beta\beta}$ reads
\begin{align}
{\rm NO}: \;m_{\beta\beta}&=\left| c_{12}^2 c_{13}^2  \,m_{\rm lightest} + s_{12}^2 c_{13}^2  \,\sqrt{m_{\rm lightest}^2+\dmsol} \,e^{-i \alpha_{21}}\right. \nonumber \\
&\left.+s_{13}^2  \, \sqrt{m_{\rm lightest}^2+\dmatm}  \,e^{-i \alpha_{31}}\right| \,,\\
{\rm IO}: \;m_{\beta\beta}&=\left| c_{12}^2 c_{13}^2  \,\sqrt{m_{\rm lightest}^2+\lvert\dmatm\rvert}  + s_{12}^2 c_{13}^2  \,\sqrt{m_{\rm lightest}^2+\dmsol+|\dmatm|}  \,e^{-i \alpha_{21}}\right.\nonumber \\
&\left.+s_{13}^2  \, m_{\rm lightest}  \,e^{-i \alpha_{31}}
\right| \,,
\end{align}
for NO and IO, respectively.
\begin{figure}
    \centering
    \includegraphics[scale=0.56]{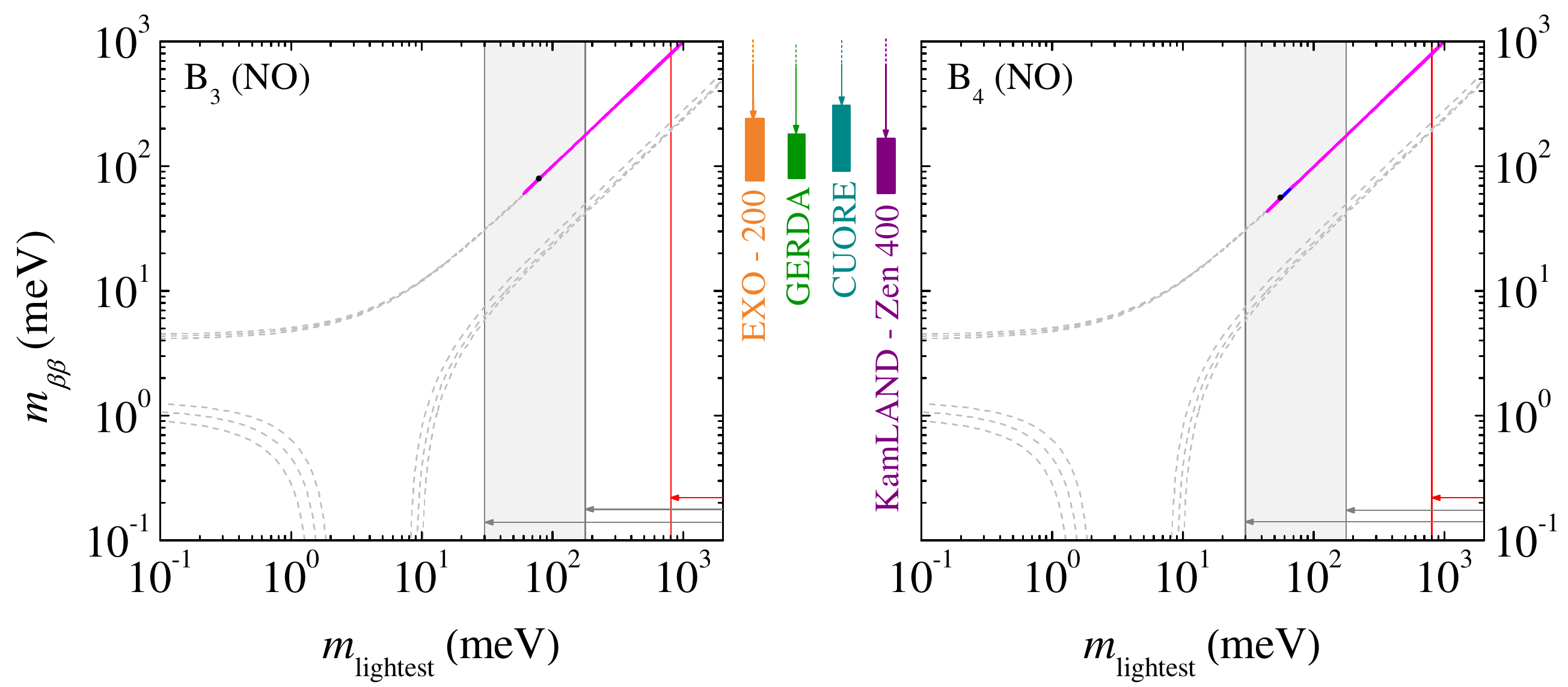}\\
    \includegraphics[scale=0.56]{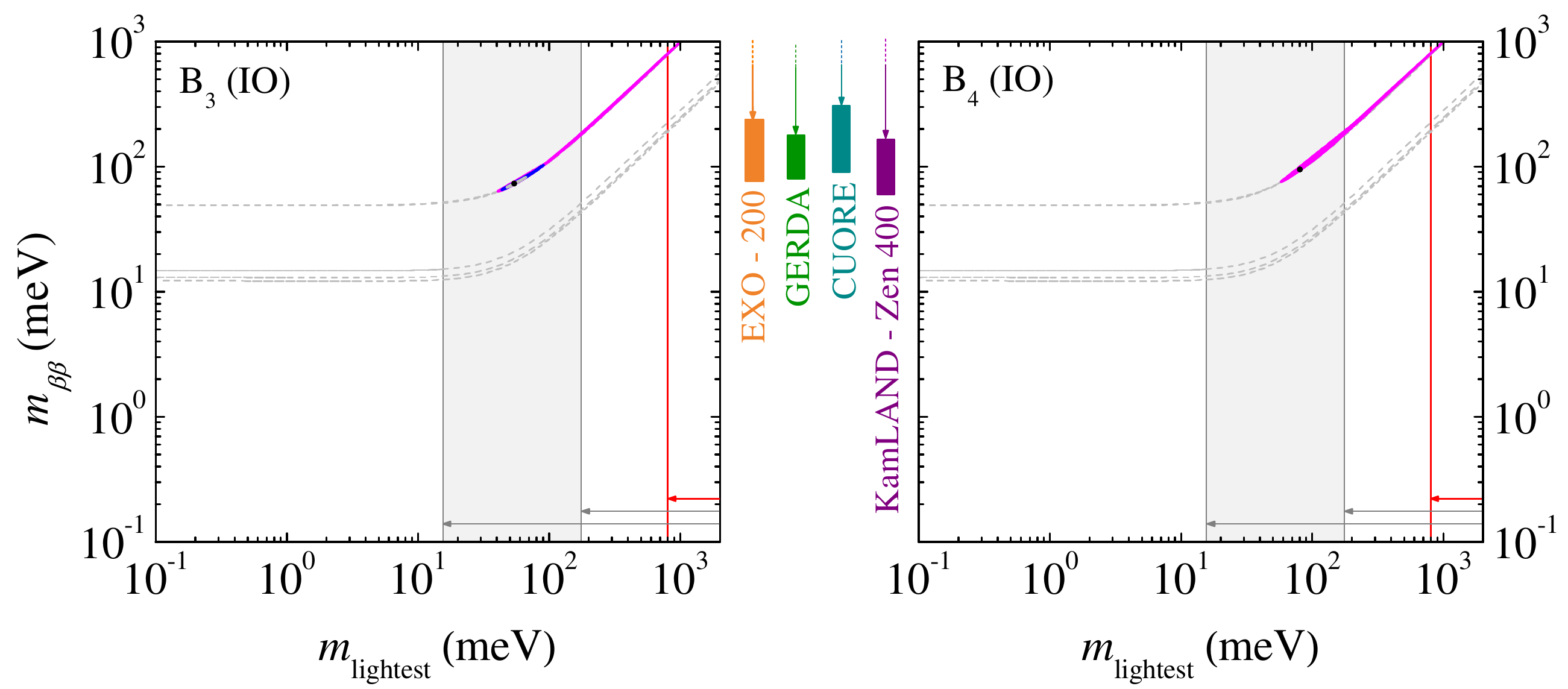}
    \caption{Allowed regions in the ($m_\text{lightest}$,$m_{\beta\beta}$) plane for NO (upper plots) and IO (lower plots). The presented colour scheme is the same as in figs.~\ref{fig:predictionst23delta} and~\ref{fig:predictionsmlightestdelta}. The dashed contours delimit the 1, 2 and 3$\sigma$ regions allowed by neutrino oscillation data only, without considering any extra constraint on $\Mnu$. The coloured vertical bars correspond to the upper-bound ranges on $m_{\beta\beta}$ from EXO-200~\cite{EXO-200:2019rkq}, GERDA~\cite{GERDA:2020xhi}, CUORE~\cite{CUORE:2021mvw} and KamLAND-Zen 400~\cite{KamLAND-Zen:2016pfg}.}
    \label{fig:predictionsmlightestmbb}
\end{figure}
In fig.~\ref{fig:predictionsmlightestmbb}, the ($m_\text{lightest}$,$m_{\beta\beta}$) model regions allowed by neutrino oscillation data at the 1$\sigma$, 2$\sigma$ and 3$\sigma$ level are shown for B$_{3,4}$ (using the same colour code as in fig.~\ref{fig:predictionst23delta} and~\ref{fig:predictionsmlightestdelta}), and for NO (upper plots) and IO (lower plots) neutrino masses.~\footnote{Note that case A$_1$ for NO is not included since it predicts $m_{\beta\beta}=0$ -- see eq.~\eqref{eq:neutrinomasstextures}.} Besides the current limits on $m_\text{lightest}$, presented previously in fig.~\ref{fig:predictionsmlightestdelta}, we show as well the upper-limit ranges imposed on $m_{\beta\beta}$ by the $0\nu\beta\beta$ experiments EXO-200~\cite{EXO-200:2019rkq}, GERDA~\cite{GERDA:2020xhi}, CUORE~\cite{CUORE:2021mvw} and KamLAND-Zen 400~\cite{KamLAND-Zen:2016pfg}. These ranges are indicated by vertical bars which reflect the uncertainty in theoretical calculations of the nuclear matrix elements. From the results of fig.~\ref{fig:predictionsmlightestmbb}, one can see that for B$_3$ and B$_4$ the lower bounds on $m_{\beta\beta}$ are within the sensitivity of current $0\nu\beta\beta$ decay experiments. In fact, the current KamLAND-Zen 400 result strongly disfavours B$_3$ (NO and IO) and B$_4$ (IO). The case B$_4$ (NO) is still compatible with all data and will be tested by future projects such as KamLAND2-Zen~\cite{KamLAND-Zen:2016pfg}, AMORE II~\cite{Lee:2020rjh}, CUPID~\cite{CUPID:2015yfg}, LEGEND~\cite{LEGEND:2017cdu}, nEXO\cite{nEXO:2017nam}, PandaX-III~\cite{Chen:2016qcd} or SNO+ I~\cite{SNO:2015wyx}. It is interesting to note that although our model is constrained by a very simple $\mathcal{Z}_8$ symmetry, it still imposes a lower limit on $m_{\beta\beta}$. 

An appealing feature of our model is that the SCPV phase $\theta$ is determined by low-energy observables and the loop function ratio $\mathcal{R}$, as given in eq.~\eqref{eq:theta} for $\mathcal{Z}_8^{e-\mu}$. In table~\ref{tab:thetaintervals}, we show the allowed intervals for $\theta$ when considering the best-fit values for the low-energy parameters and varying $\mathcal{R}$ in the range $[0,1]$.~\footnote{We limit $\mathcal{R}$ to this set of values since the condition $\mathcal{F}_{12}\geq \mathcal{F}_{ii}$ with $i=1,2$, is always verified in our model (see discussion at the end of appendix~\ref{sec:scalar}).} From that table, one can see that $|\sin\theta| \ll 1$, suggesting that, although the CPV vacuum phase is small, large CP-violating effects in the lepton sector can be observed -- see figs.~\ref{fig:predictionst23delta} and \ref{fig:predictionsmlightestdelta}.
\begin{table}[t!]
    \centering
    \begin{tabular}{|l|c|}
        \hline
        Case&$\theta$ interval\\
        \hline
         A$_1$ (NO)& $[1.942;1.980]\,\pi$\\
         B$_3$ (NO)& $[0.071;0.072]\,\pi$\\
         B$_3$ (IO)& $[1.953;1.954]\,\pi$\\
         B$_4$ (NO)& $[1.917;1.918]\,\pi$\\
         B$_4$ (IO)& $[0.052;0.053]\,\pi$\\
        \hline
    \end{tabular}
    \caption{Intervals for the SCPV phase $\theta$ obtained by fixing $\Mnuh$ to its best-fit value and varying~$\mathcal{R}$ in the range $[0,1]$ using eq.~\eqref{eq:theta}, for cases A$_1$, B$_3$ and B$_4$.}
    \label{tab:thetaintervals}
\end{table}
Also, since $\theta\sim 2 k\pi$, the condition required for SCPV given in eq.~\eqref{eq:SCPVsol} leads to the approximate result 
\begin{align}
    m^{\prime\, 2}_\sigma\simeq \pm\, 2u^2\lambda'_\sigma\,,
\end{align}
meaning that compatibility with neutrino data requires a nonvanishing value for the $\mathcal{Z}_8$ soft-breaking parameter $m^{\prime\,2}_\sigma$, evading the cosmological domain-wall problem.

\section{Charged-lepton flavour violation}
\label{sec:cLFVmain}

In our model, the presence of new (charged and neutral) particles interacting with SM leptons opens up the possibility for cLFV processes like radiative $\ell_{\alpha}^- \rightarrow \ell_{\beta}^- \gamma$, three-body  $\ell_{\alpha}^- \rightarrow \ell_{\beta}^- \ell_{\gamma}^+ \ell_{\lambda}^-$ decays and $\mu - e$ conversion in nuclei. These phenomena have been widely studied in the literature for SM extensions with scalar triplets~\cite{Pich:1984uoh,Ma:2000xh,Akeroyd:2009nu,Chakrabortty:2012vp,Dinh:2013vya,BhupalDev:2013xol,Antusch:2018svb,Primulando:2019evb}, as well as in scotogenic frameworks~\cite{Toma:2013zsa,Vicente:2014wga,Ahriche:2016cio,Hagedorn:2018spx} (the reader is addressed to these references for details on the decay width and conversion-rate calculations). We will not present the full analytical results here since, as will be later detailed, we will use existent packages for our computations. Before presenting numerical results, it is worth stressing some important features of our model regarding cLFV. Namely:
\begin{itemize}

\item the cLFV contributions involving the scalar triplet $\Delta$ always depend on either elements of $\mathbf{Y}_{\Delta} \mathbf{Y}_{\Delta}^\dagger$ or on products of elements of $\mathbf{Y}_{\Delta}$. For instance, in the one-loop $\ell_{\alpha}^- \rightarrow \ell_{\beta}^- \gamma$ process, stemming from photon-diagram dipole contribution, $(\mathbf{Y}_{\Delta} \mathbf{Y}_{\Delta}^\dagger)_{\alpha \beta}$ appears. As can be readily seen from eq.~\eqref{eq:yukawastructures}, $\mathbf{Y}_{\Delta} \mathbf{Y}_{\Delta}^\dagger$ is diagonal for all possible $\mathcal{Z}_8$ charge assignments. Hence, that LFV process receives no triplet contribution. Note that for $\mathcal{Z}_8^{e-\mu}$  and $\mathcal{Z}_8^{e-\tau}$ we have $(\mathbf{Y}_{\Delta})_{e e} \equiv y_{1}$ and $(\mathbf{Y}_{\Delta})_{\mu \tau} \equiv y_{2}$, while $(\mathbf{Y}_{\Delta})_{\mu \mu} \equiv y_{1}$ and $(\mathbf{Y}_{\Delta})_{e \tau} \equiv y_{2}$ for $\mathcal{Z}_8^{\mu-\tau}$. $\mu - e$ conversion in nuclei arises solely from photon-exchange diagrams leading to dipole and monopole contributions proportional to $(\mathbf{Y}_{\Delta} \mathbf{Y}_{\Delta}^\dagger)_{\mu e}$ and $ (\mathbf{Y}^*_{\Delta})_{e \lambda} (\mathbf{Y}_{\Delta})_{\lambda \mu}$, respectively. As mentioned, the dipole term vanishes and the two Yukawa couplings in $(\mathbf{Y}_{\Delta})_{\alpha \beta}$ do not have a common index $\alpha$ or $\beta$ for all symmetry cases, implying that $\mu - e$ conversion via $\Delta$ mediation is not possible. As for the three-body $\ell_{\alpha}^- \rightarrow \ell_{\beta}^- \ell_{\gamma}^+ \ell_{\lambda}^-$ decays, mediated at tree-level by the doubly-charged scalar $\Delta^{++}$, these are proportional to $|(\mathbf{Y}_{\Delta})_{\alpha \gamma}|^2 |(\mathbf{Y}_{\Delta})_{\beta \lambda}|^2$. Thus, contributions from $\Delta$ to $\ell_{\alpha} \rightarrow 3 \ell_{\beta}$, $\tau^- \rightarrow e^- \mu^+ \mu^-$ and $\tau^- \rightarrow \mu^- e^+ e^-$ vanish. In summary, the only allowed $\Delta$-mediated cLFV processes are: $\tau^- \rightarrow e^+ \mu^- \mu^-$ and $\tau^- \rightarrow \mu^+ e^- e^-$.

\item the scotogenic contributions to cLFV decays involving the \emph{dark} fermion $f$ and scalars $\eta_{1,2}$ are sensitive to the Yukawa parameters in $\mathbf{Y}_{f}^1$ and $\mathbf{Y}_{f}^2$ -- see eqs.~\eqref{eq:LYuk} and \eqref{eq:yukawastructures}. For the case $\mathcal{Z}_8^{e-\mu}$ ($\mathcal{Z}_8^{e-\tau}$) [$\mathcal{Z}_8^{\mu-\tau}$], we have $\mathbf{Y}_{f}^1 \equiv y_e$  and $\mathbf{Y}_{f}^2 \equiv y_\mu$ ($\mathbf{Y}_{f}^1 \equiv y_e$ and $\mathbf{Y}_{f}^2 \equiv y_\tau$) [$\mathbf{Y}_{f}^1 \equiv y_\mu$ and $\mathbf{Y}_{f}^2 \equiv y_\tau$]. Furthermore, all processes will be mediated at one-loop level by the odd charged scalars~$S_1^+$ and $S_2^+$. Performing the rotation $\mathbf{R}$ presented in eq.~\eqref{eq:rotcharged} of appendix~\ref{sec:scalarmass}, which relates the weak states $\eta_1^+$ and $\eta_2^+$ to the mass states $S_1^+$ and $S_2^+$, the Yukawa matrices in the mass-eigenstate basis will be given by
\begin{equation}
    \widetilde{\mathbf{Y}}_f^1 = \mathbf{Y}_f^1 \cos \varphi + \mathbf{Y}_f^2 \sin \varphi \; , \; \widetilde{\mathbf{Y}}_f^2 = - \mathbf{Y}_f^1 \sin \varphi + \mathbf{Y}_f^2 \cos \varphi \; .
    \label{eq:Yukmix}
\end{equation}
The cLFV amplitudes for $\mathcal{Z}_8^{\beta-\alpha}$ will depend on products of the Yukawa couplings $y_{\alpha}$ and $y_{\beta}$, the mixing angle $\varphi$ and the masses of the odd charged scalars. Each $\mathcal{Z}_8^{\beta-\alpha}$ case will lead to distinct $\ell_{\alpha}^- \rightarrow \ell_{\beta}^- \gamma$ and $\ell_{\alpha} \rightarrow 3 \ell_{\beta}$ processes. The three-body decays are limited to $\ell_{\alpha} \rightarrow 3 \ell_{\beta}$ since there are only two nonvanishing $f$ Yukawa couplings in $\mathbf{Y}_{f}^{1,2}$. Lastly, $\mu - e$ conversion in nuclei is only possible for $\mathcal{Z}_8^{e-\mu}$.

\end{itemize}
\begin{table}[!t]
\renewcommand{\arraystretch}{1.2}
\centering
\begin{tabular}{|l|l|l|}   
\hline
Cases & Type-II seesaw & Scotogenic \\ \hline
$\mathcal{Z}_8^{e-\mu} \quad (\text{B}_4)$ & $\tau^{-} \rightarrow \mu^{+} e^{-} e^{-}$ &  $\mu \rightarrow e \gamma , \; \mu  \rightarrow 3 e , \; \mu - e$ conversion \\
$\mathcal{Z}_8^{e-\tau} \quad (\text{B}_3)$ & $\tau^{-} \rightarrow \mu^{+} e^{-} e^{-}$ & $\tau \rightarrow e \gamma , \; \tau \rightarrow 3 e$ \\
$\mathcal{Z}_8^{\mu-\tau} \quad (\text{A}_1)$ & $\tau^{-} \rightarrow e^{+} \mu^{-} \mu^{-}$ & $\tau \rightarrow \mu \gamma , \; \tau \rightarrow 3 \mu$ \\
\hline
\end{tabular}
\caption{cLFV contribution from the type-II seesaw and scotogenic part of our model for the different symmetry cases $\mathcal{Z}_8^{e-\mu}$, $\mathcal{Z}_8^{e-\tau}$ and $\mathcal{Z}_8^{\mu-\tau}$ (see table~\ref{tab:part&sym}) with effective neutrino mass textures~$\text{B}_4$, $\text{B}_3$ and $\text{A}_1$ [see eq.~\eqref{eq:neutrinomasstextures}], respectively.}
\label{tab:CasesCLFV}
\end{table}
\begin{table}[!t]
\renewcommand{\arraystretch}{1.2}
\centering
\begin{tabular}{|l|l|l|}   
\hline
cLFV process & Present limit ($90 \%$ CL) & Future sensitivity \\ \hline
$\BR(\mu \rightarrow e \gamma)$ & $ 4.2 \times 10^{-13}$ (MEG \cite{MEG:2016leq}) &  $ 6 \times 10^{-14}$ (MEG II \cite{MEGII:2018kmf}) \\
$\BR(\tau \rightarrow e \gamma)$ & $ 3.3 \times 10^{-8}$ (BaBar \cite{BaBar:2009hkt}) & $3 \times 10^{-9}$ (Belle II \cite{Belle-II:2018jsg})\\
$\BR(\tau \rightarrow \mu \gamma)$ & $ 4.4 \times 10^{-8}$ (BaBar \cite{BaBar:2009hkt}) & $10^{-9}$ (Belle II \cite{Belle-II:2018jsg}) \\
\hline
$\BR(\mu \rightarrow 3 e)$ & $ 1.0 \times 10^{-12}$ (SINDRUM \cite{SINDRUM:1987nra}) & $10^{-16}$ (Mu3e \cite{Blondel:2013ia})\\
$\BR(\tau \rightarrow 3 e)$ & $ 2.7 \times 10^{-8}$  (Belle \cite{Hayasaka:2010np})& $5 \times 10^{-10}$ (Belle II \cite{Belle-II:2018jsg})\\
$\BR(\tau^{-} \rightarrow e^{+} \mu^{-} \mu^{-})$ & $ 1.7 \times 10^{-8}$  (Belle \cite{Hayasaka:2010np})& $3 \times 10^{-10}$ (Belle II \cite{Belle-II:2018jsg})\\
$\BR(\tau^{-} \rightarrow \mu^{+} e^{-} e^{-})$ & $ 1.5 \times 10^{-8}$  (Belle \cite{Hayasaka:2010np})& $3 \times 10^{-10}$ (Belle II \cite{Belle-II:2018jsg})\\
$\BR(\tau \rightarrow 3 \mu)$ & $ 2.1 \times 10^{-8}$  (Belle \cite{Hayasaka:2010np})& $4 \times 10^{-10}$ (Belle II \cite{Belle-II:2018jsg})\\
\hline
$\CR(\mu - e, \text{Al})$ & $-$ & $3 \times 10^{-17}$ (Mu2e \cite{Mu2e:2014fns}) \\
&  & $10^{-15}-10^{-17}$ (COMET~\cite{COMET:2018auw}) \\
$\CR(\mu - e, \text{Ti})$ & $ 4.3 \times 10^{-12}$ (SINDRUM II \cite{SINDRUMII:1993gxf})& $10^{-18}$ (PRISM/PRIME \cite{Alekou:2013eta}) \\
$\CR(\mu - e, \text{Au})$ & $ 7 \times 10^{-13}$ (SINDRUM II \cite{SINDRUMII:2006dvw})& $-$ \\
$\CR(\mu - e, \text{Pb})$ & $ 4.6 \times 10^{-11}$ (SINDRUM II \cite{SINDRUMII:1996fti}) & $-$ \\ 
\hline
\end{tabular}
\caption{Current experimental bounds and future sensitivities for the branching ratios (BRs) and conversion rates (CRs) of cLFV processes.}
\label{tab:boundsCLFV}
\end{table}
In table~\ref{tab:CasesCLFV} we list the cLFV processes to which the triplet and scotogenic degrees of freedom contribute at tree- and one-loop levels, respectively, for the three possible $\mathcal{Z}_8$ charge assignments. Additionally, in table~\ref{tab:boundsCLFV} we present the current experimental bounds and future sensitivities for these cLFV observables.
A couple of interesting conclusions can be drawn from the former table. Namely, one can see that in all cases the triplet and scotogenic contributions never overlap. Thus, each of the cLFV processes identified above is driven by either the $\Delta$ or the $f/\eta_{1,2}$ fields. This means that if, for instance, $\tau^- \rightarrow e^+ \mu^- \mu^-$ and/or $\tau^- \rightarrow \mu^+ e^- e^-$ are observed, then they can be only mediated by the scalar triplet. This reasoning also applies for the remaining processes which occur via scotogenic mediation. Apart from this mechanism-dominance feature, one can also see that the simultaneous observation of certain cLFV processes would automatically exclude our model for all possible $\mathcal{Z}_8$ charge assignments. This would be the case if, for example, $\mu \rightarrow e \gamma$ and $\tau \rightarrow \ell_{\beta} \gamma$, or $\tau^- \rightarrow e^- \mu^+ \mu^-$ and $\tau^- \rightarrow \mu^- e^+ e^-$ were simultaneously observed.

Having described some qualitative aspects of our framework regarding cLFV, we now proceed to a more detailed analysis which will allow us to probe the parameter space of the model. Before presenting the results, we discuss some important aspects that are relevant for the numerical procedure. Namely, 
\begin{itemize}

\item the neutrino low-energy parameters (mass-squared differences and mixing angles) are fixed to their best-fit values for each possible charge assignment and neutrino-mass ordering (see section~\ref{sec:neutrinomass}). This guarantees that the Yukawa couplings $y_{1,2}$, $y_{e,\mu}$ and SCPV phase $\theta$, reconstructed through eqs.~\eqref{eq:yscoto}-\eqref{eq:Rratio}, lead to compatibility with neutrino data for any given set of input masses $m_{S_k}$, $M_f$ and \textit{dark} neutral-scalar mixing~$\mathbf{V}$ [see eq.~\eqref{eq:MixneutralDM} in appendix~\ref{sec:scalarmass}]. In all cases, as a perturbativity requirement, we impose the upper bound $y_{1,2}, y_{\alpha,\beta} < 1$ on the Yukawa couplings. 

\item We will mainly focus on the $\mathcal{Z}_8^{e-\mu}$ (texture $\text{B}_4$) case which leads to the most stringent muon cLFV constraints (we comment on the remaining cases at the end of this section). At the best-fit point (see the top-right and middle-right plots in figs.~\ref{fig:predictionst23delta} and~\ref{fig:predictionsmlightestdelta}, respectively), the magnitude of the $\Mnuh$ matrix elements, is
\begin{align}
    \left|\Mnuh\right|\simeq\begin{pmatrix}
    5.60\times10^{-2}&3.78\times10^{-3}&0\\
    \cdot&1.87\times10^{-2}&6.46\times10^{-2}\\
    \cdot&\cdot&0
    \end{pmatrix} \; \text{eV} \,,
    \label{eq:Mnumod}
\end{align}
for NO neutrino masses. Furthermore, as shown in table~\ref{tab:thetaintervals}, the SCPV phase is approximately $\theta \simeq 1.92\,\pi$, leading to $\delta \sim 3\pi/2$. Note that, we choose to present the cLFV results for NO since these are not significantly altered for the IO case.
\item The VEV hierarchy $w \ll v \lesssim u$ is assumed, i.e. the triplet VEV is very small when compared with the EW scale and the scalar singlet VEV $u$. The reason for this is twofold: first, from electroweak precision measurements on the parameter $\rho \simeq 1.00039 \pm 0.00019$~\cite{ParticleDataGroup:2020ssz}, the upper bound $w \lesssim 8$ GeV can be derived~\cite{Kanemura:2012rj,Arhrib:2011uy,Arhrib:2011vc,Aoki:2012yt,Aoki:2012jj}. Second, since we will consider $M_f$ around or above the EW scale, it is reasonable to take $u \gtrsim v$. This VEV configuration implies that the contribution of $w$ to the scalar masses, namely those of the inert doublets components, can be safely neglected as explained in section~\ref{sec:model} and appendix~\ref{sec:scalarmass}. Hence, the loop functions $\mathcal{F}_{i j} \left(M_f,m_{S_k}\right)$ given in eqs.~\eqref{eq:F11loop}, \eqref{eq:F12loop} and \eqref{eq:F22loop}, will be computed in that limit, i.e. only the diagrams of fig.~\ref{fig:neutrinomassdiagrams} are relevant.  

\item As explained above, the scalar triplet contributes to the processes $\tau^- \rightarrow e^+ \mu^- \mu^-$ and $\tau^- \rightarrow \mu^+ e^- e^-$ at tree level. The branching ratios~(BRs), normalized to their current experimental upper bounds provided by the Belle collaboration~\cite{Hayasaka:2010np} (see table~\ref{tab:boundsCLFV}), are given by~\cite{Leontaris:1985qc,Bernabeu:1985na,Bilenky:1987ty,Akeroyd:2009nu,Dinh:2013vya}
\begin{align}
\frac{\BR(\tau^{-} \rightarrow \mu^{+} e^{-} e^{-})}{1.5 \times 10^{-8}} & \approx 35.28 \left(\frac{10 \; \text{TeV}}{m_{\Delta^{+ +}}} \right)^4 \left|(\mathbf{Y}_{\Delta})_{e  e} \right|^2 \left|(\mathbf{Y}_{\Delta})_{\mu\tau} \right|^2 \; , \nonumber \\
\frac{\BR(\tau^{-} \rightarrow e^{+} \mu^{-} \mu^{-})}{1.7 \times 10^{-8}} & \approx 29.40 \left(\frac{10 \; \text{TeV}}{m_{\Delta^{+ +}}} \right)^4 \left|(\mathbf{Y}_{\Delta})_{\mu  \mu} \right|^2 \left|(\mathbf{Y}_{\Delta})_{e \tau} \right|^2 \;.
\label{eq:BRTII}
\end{align}
The doubly-charged scalar mass $m_{\Delta^{+ +}}$ is given by eq.~\eqref{eq:doublychargeddeltamass} and can be freely adjusted by varying the couplings $\mu_{\Delta}$ and $\lambda_{\Delta 4}$, for any fixed VEV~$w$. Taking $w = 8$~GeV and using as reference value $\left|(\Mnuh)_{23}\right| \approx 6.46 \times 10^{-2}$ eV [see eq.~\eqref{eq:Mnumod}], we have $y_2 \sim 10^{-11}$ [see eq.~\eqref{eq:yDelta} and \eqref{eq:Mnumod}], leading to extremely suppressed BRs.

\item  The precise LEP-I measurements on the $W$ and $Z$-boson decay widths rule out SM-gauge boson decays into \textit{dark} particles~\cite{Cao:2007rm,Gustafsson:2007pc}. This imposes the following lower bounds on the scalar masses of our model:
    \begin{align}
    & m_{S_i} + m_{S^+_j} > m_W  \; , \; m_{h_l} + m_{H^+} > m_W   \; , \; m_{H^+} + m_{\Delta^{++}} > m_W \; , \nonumber \\
    & m_{S_i} + m_{S_k} > m_Z   \; , \; m_{h_l} + m_{h_n} > m_Z \; , \nonumber \\
    & 2 m_{S^+_j} > m_Z \; , \; 2 m_{H^+} > m_Z \; , \; 2 m_{\Delta^{++}} > m_Z \,,
    \label{eq:LEPbound}
    \end{align}
where $m_W$ and $m_Z$, are respectively, the $W$ and $Z$-boson masses. Hence, the above conditions ensure that the decays $W^+ \rightarrow \{ S_i S_j^+ ; h_l H^+ ; H^- \Delta^{++} \}$, $Z \rightarrow \{ S_i S_k ; h_l h_n \}$ and  $Z \rightarrow \{  S_j^+ S_j^- ; H^+ H^-; \Delta^{++} \Delta^{--} \}$, with $i , k = 1, \cdots, 4$; $l , n = 1, \cdots, 5$ and $j=1,2$ (see appendix~\ref{sec:scalarmass} for details), are kinematically forbidden. Reinterpreting the LEP-II results for chargino searches in the context of singly-charged scalar production $e^+ e^- \rightarrow h^+ h^-$, the bounds $\{m_{H^{+}}, m_{S^+_j}\} > 70 - 90$ GeV can be considered~\cite{Pierce:2007ut}. 

\item The LHC ATLAS and CMS collaborations have been searching for doubly-charged scalar decays into same-sign dileptons $\Delta^{++} \rightarrow \ell_{\alpha}^{+} \ell_{\beta}^{+}$, at different centre of mass energies $\sqrt{s} = \{7, 8, 13\}$ TeV~\cite{ATLAS:2012hi,CMS:2011sqa,ATLAS:2014kca,CMS:2016cpz,ATLAS:2017xqs,CMS:2017pet}. The strongest bound comes from ATLAS analyses with $36.1 \; \text{fb}^{-1}$ of data at $\sqrt{s} = 13$~TeV for same-sign $e e$, $\mu \mu$ and $e \mu$ final-state searches. In fact, depending on the value of $\BR(\Delta^{++} \rightarrow \ell_{\alpha}^{+} \ell_{\beta}^{+})$, we obtain the bound $m_{\Delta^{++}} \geq 450\,[770]$ GeV for $\BR(\Delta^{++} \rightarrow \ell_{\alpha}^{+} \ell_{\beta}^{+}) = 0.1\,[1.0]$~\cite{ATLAS:2017xqs}. Note that the $e \mu$ decay channel is absent in our model for all symmetry cases. Additionally, the ATLAS collaboration has been looking for $\Delta^{++} \rightarrow W^{+} W^{+}$~\cite{ATLAS:2018ceg} with $36.1 \; \text{fb}^{-1}$ of data at $\sqrt{s} = 13$ TeV establishing an upper bound of $m_{\Delta^{++}} \geq 220$ GeV, improving the LEP bound of eq.~\eqref{eq:LEPbound}. Besides the mentioned bounds, other constraints stemming from, e.g. boundedness-from-below requirements, oblique parameters and contributions to Higgs invisible decays can be obtained. These studies are beyond the scope of this work (for some phenomenological analyses we refer the reader to refs.~\cite{Huitu:1996su,Akeroyd:2005gt,Kanemura:2006bh,Melfo:2011nx,BhupalDev:2013xol,delAguila:2013mia,Babu:2016rcr,Antusch:2018svb,Primulando:2019evb,Chun:2019hce,Padhan:2019jlc,deMelo:2019asm,Ashanujjaman:2021txz}). 

\end{itemize}
Taking into account the considerations above, it is clear that we are able to satisfy the $\tau$ cLFV decay constraints for smaller $m_{\Delta^{+ +}}$ and higher $w$ VEV values, as compared to the typical bounds obtained in the literature~\cite{BhupalDev:2013xol,Antusch:2018svb,Primulando:2019evb}. Taking the $\mathcal{Z}_{8}^{e-\mu}$ ($\text{B}_4$) NO case, and considering $y_2 \le 1$, implies a lower bound on the triplet VEV $w \geq \left|(\Mnuh)_{23}\right|/\sqrt{2}\simeq 4.6\times 10^{-2}$~eV. In turn, for this value of $w$, we have $y_1 \sim 0.9$. Consequently, in order to satisfy the current Belle constraint on $\BR(\tau^{-} \rightarrow \mu^{+} e^{-} e^{-})$, $m_{\Delta^{+ +}} \gtrsim 22.55$ TeV is required. Obviously, this lower bound on $m_{\Delta^{+ +}}$ decreases for higher $w$ values, and vice-versa. For example, taking the lowest possible value $m_{\Delta^{+ +}} \gtrsim 45.6 \; \text{GeV}$, allowed by the LEP constraints (see discussion above), we obtain $w \lesssim 15.98$ eV.
\begin{table}[!t]
\renewcommand{\arraystretch}{1.2}
\centering
\begin{tabular}{|l|l|}   
\hline
Parameters & Scan range \\ \hline
$M_f$ & $[10 , 1000]$ (GeV) \\
$m_{\eta_1}^2 , m_{\eta_2}^2 $ &  $[10^2 , 1000^2]$ (GeV$^2$)  \\
$|\mu_{1 2}|$ & $[10^{-6} , 10^3]$ (GeV) \\
$|\lambda_{3}| , \; |\lambda_{4}| , \; |\lambda_{3}^\prime| , \; |\lambda_{4}^\prime|$ & $[10^{-5} , 1]$ \\
$|\lambda_{5}|$ & $[10^{-12} , 1]$ \\
\hline
\end{tabular}
\caption{Input parameters of our model and corresponding ranges used in our numerical scan. Scalar couplings involving the Higgs triplet are discussed in the text. The Yukawa couplings and SCPV phase $\theta$ are reconstructed through eqs.~\eqref{eq:yscoto}-\eqref{eq:Rratio} using the best-fit values for low-energy neutrino observables.}
\label{tab:ParameterScan}
\end{table}
For definiteness, we will consider $w = 1$~eV and, for the triplet scalar couplings (see appendix~\ref{sec:scalar}), we set $\mu_{\Delta} = - 2.34 \times 10^{-6}$ GeV, $\lambda_{\Delta 1-4}=1$, $\lambda_{\Delta 5-7}=\lambda_{\Delta 5-7}^\prime =0$ and $\mu_{\Delta}^\prime=0$. This parameter choice leads to a quasi-degenerate spectrum for the triplet scalars, namely the doubly and singly charged, as well as, the neutral scalars, have masses around $10$ TeV. These values respect all the electroweak precision data and collider constraints mentioned above and the current cLFV bounds for the $\tau$ three-body decays mediated at tree level by $\Delta^{++}$.

Given the fact that the triplet-scalar is not relevant for muon cLFV, we will now concentrate our discussion on the scotogenic contributions to the $\mu \rightarrow e\gamma$, $\mu \rightarrow 3e$ and $\mu-e$ conversion in nuclei. Our input parameters, namely the \textit{dark} fermion mass and scalar potential couplings (see Appendix~\ref{sec:scalar}), vary as shown in table~\ref{tab:ParameterScan}. Throughout this work we assume that the SM Higgs doublet, as well as the inert doublets, do not mix with the singlet $\sigma$, i.e. we are working in the alignment limit. This is reflected by taking $\lambda_{\Phi \sigma}=\lambda_{\eta \sigma}=\lambda_{\eta \sigma}^\prime= 0$. Later on we will comment on the impact of relaxing this limit in our DM analysis of section~\ref{sec:DM}. Additionally, we set the self-interacting \textit{dark} scalar couplings to $\lambda_{2}=\lambda_{2}^\prime=\lambda_{2}^{\prime \prime}= 0.5$. Note that, these parameters have no implications for the analysis performed in this work since they do not affect the \textit{dark} scalar masses used in our cLFV and WIMP-DM studies (see section~\ref{sec:DM}). Moreover, since $h_1$ is the lightest \textit{non-dark} scalar, we identify it as the CP-even Higgs boson discovered by the ATLAS and CMS collaborations with mass $m_{h_1} = 125.25$ GeV~\cite{ParticleDataGroup:2020ssz}. For the remaining \textit{non-dark} scalar masses we take $m_{h_2} = 1.5$ TeV and $m_{h_3} = 6$ TeV, the singlet VEV is set to $u = 5.5$ TeV. One can justify the benchmark values for $m_{h_3}$ and $u$ by looking at the corresponding neutral scalar mass matrix. For the case with decoupled Higgs triplet states and assuming the alignment limit, we have [see eq.~\eqref{eq:ENSMassMatrixlim}]
\begin{align}
    \mathcal{M}^2_{\phi\sigma}=\begin{pmatrix}
    v^2\,\lambda_1&0&0\\
    \cdot&u^2\,[\lambda_\sigma+4\lambda_\sigma'\cos(4\theta)]&4u^2\,\lambda_\sigma'\sin(4\theta)\\
    \cdot&\cdot&8u^2\,\lambda_\sigma'\sin^2(2\theta)
    \end{pmatrix}\, ,
    \label{eq:ENSMassMatrixNoMixLimit}
\end{align}
leading to the $h_2$ and $h_3$ masses
\begin{align}
    m_{h_{2,3}}^2=\dfrac{u^2}{2}\left[\lambda_\sigma+4\lambda'_\sigma\pm\sqrt{\lambda_\sigma^2-8\lambda_\sigma\lambda'_\sigma+80\lambda^{'2}_\sigma+16(\lambda_\sigma-4\lambda_\sigma')\lambda_\sigma'\cos(4\theta)}\right],
\end{align}
with $m_{h_{2}}<m_{h_{3}}$. To guarantee that for given $m_{h_{2,3}}$ the couplings $\lambda_\sigma$ and $\lambda^\prime_\sigma$ remain real, the condition
\begin{align}
    \dfrac{m_{h_{3}}^2}{m_{h_{2}}^2} \geq \text{cotan}^2 \theta + \tan^2\theta \; ,
    \label{eq:conditionrealcouplings}
\end{align}
must be satisfied. Recall that the best-fit values of the SCPV phase given in table~\ref{tab:thetaintervals} are around $\theta\sim 2 k\pi$ ($k=0,1$). In the particular case of interest $\mathcal{Z}_8^{e-\mu}$ (B$_4$) with NO, the best-fit point corresponds to $\theta\simeq1.92\pi$ and the relation in eq.~\eqref{eq:conditionrealcouplings} leads to $m_{h_3}\gtrsim 3.90\, m_{h_2}$. For $m_{h_3}\simeq3.90\, m_{h_2}$, one has
\begin{align}
    \lambda_{\sigma}\simeq \left(\dfrac{0.89\, m_{h_3}}{u}\right)^2\; \text{ and }\; \lambda'_{\sigma}\simeq \left(\dfrac{0.26\, m_{h_3}}{u}\right)^2,
    \label{eq:ujustification}
\end{align}
and, hence, $u\gsim 0.89\, m_{h_3}$ has to be verified to ensure $\lambda_{\sigma}$ perturbativity. Thus, by taking as benchmark $m_{h_2} = 1.5$~TeV, $u$ and $m_{h_3}$ may be respectively at, $5.5$ and $6$~TeV.

For model implementation we use the Mathematica package \texttt{SARAH 4.14.5}~\cite{Staub:2013tta,Staub:2015kfa}. Particle mass and mixing spectra, BRs and conversion rates (CRs) are computed with \texttt{SPheno 4.0.4}~\cite{Porod:2003um,Porod:2011nf} and \texttt{FlavourKit}~\cite{Porod:2014xia}. We start the analysis looking at muon cLFV, namely $\mu\rightarrow e\gamma$, $\mu\rightarrow 3e$ and $\mu-e$ conversion in nuclei, for which the current bounds on the BRs and CRs and corresponding projected sensitivities are shown in table~\ref{tab:boundsCLFV}. Our results for these quantities are presented in figs.~\ref{fig:FlavourVS} and~\ref{fig:FlavourMassYuk}, from which the following conclusions can be drawn:
\begin{figure}[!ht]
   \centering
   \includegraphics[scale=0.321]{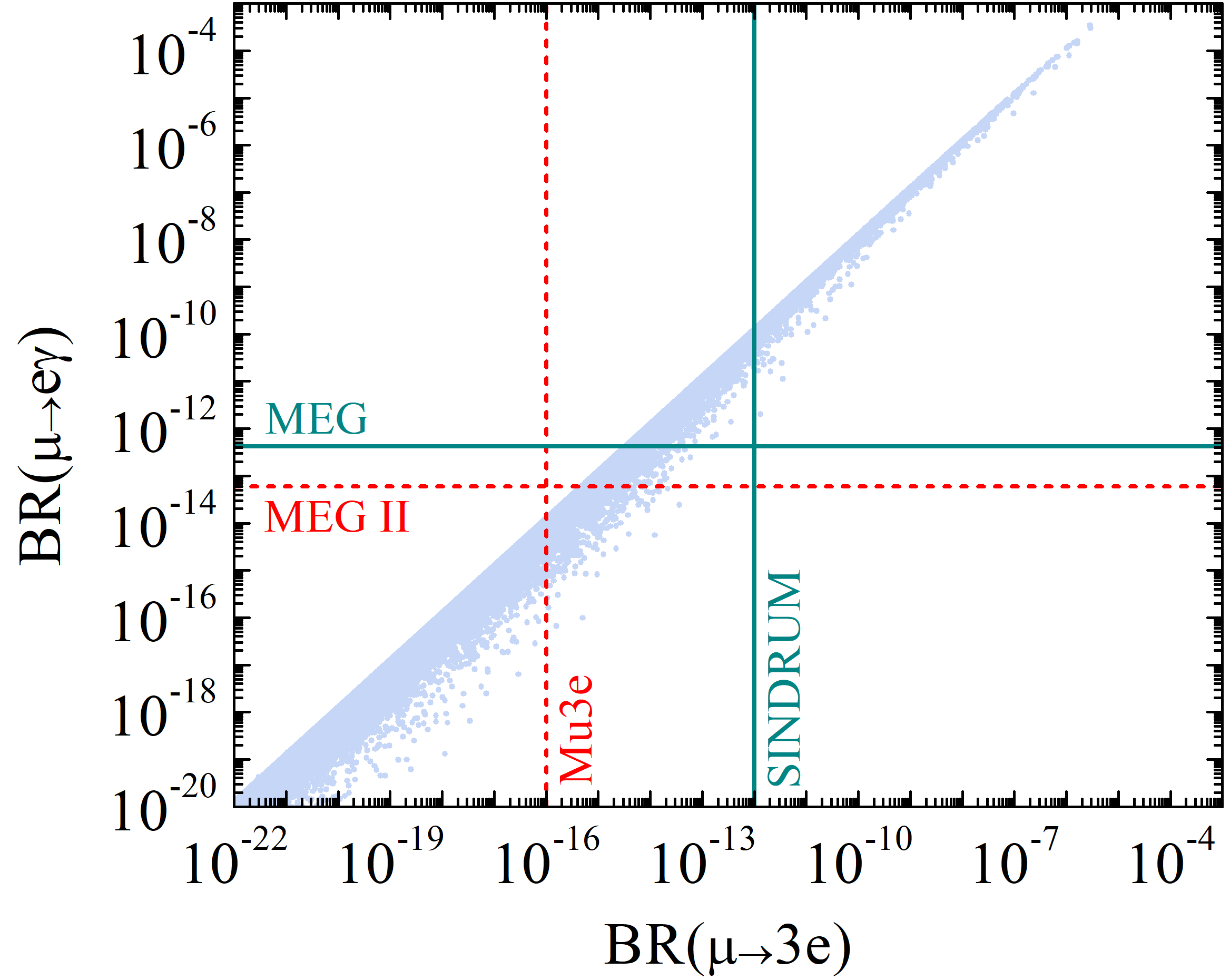} \\
   \vspace{+0.3cm}
   \includegraphics[scale=0.321]{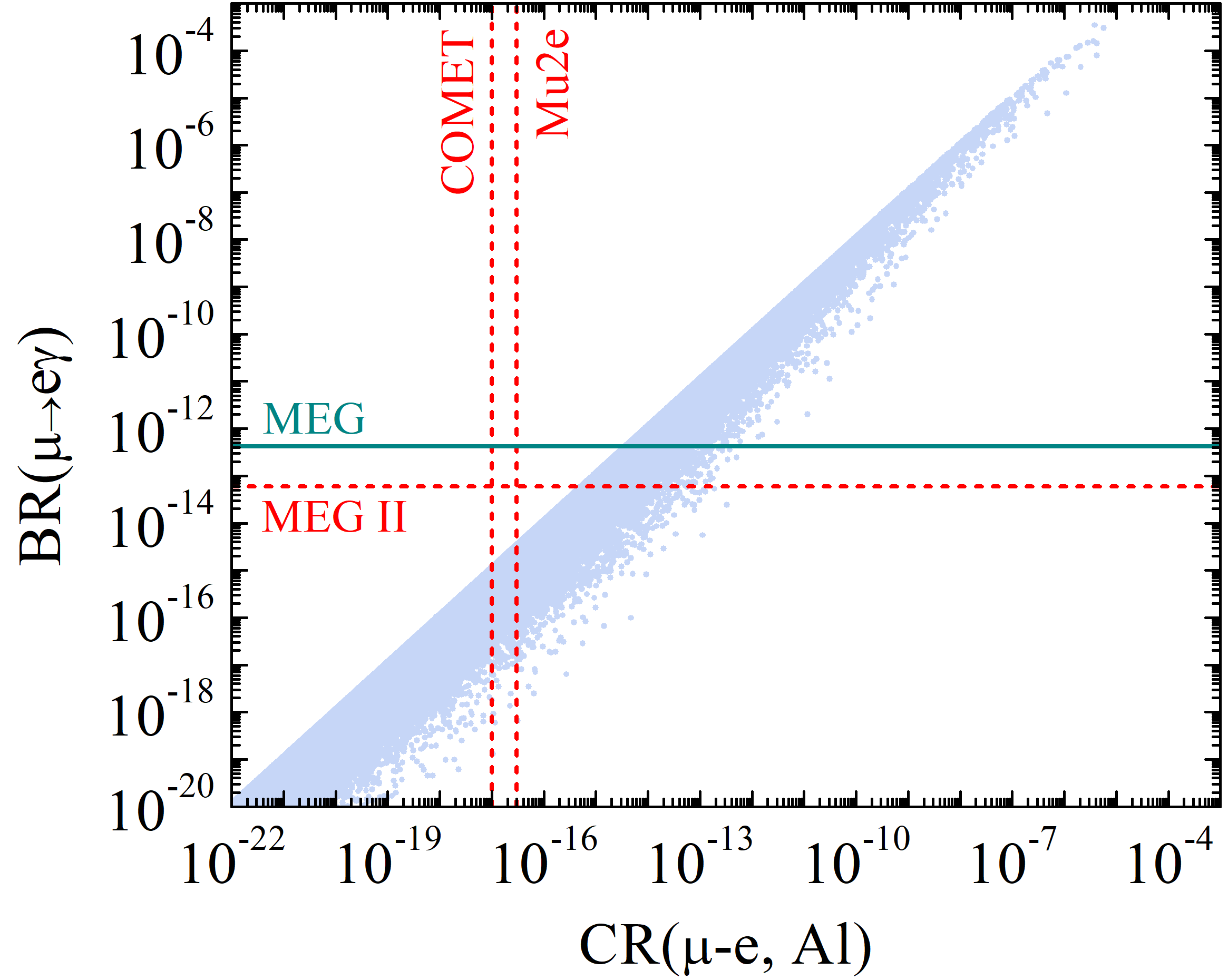}
   \includegraphics[scale=0.321]{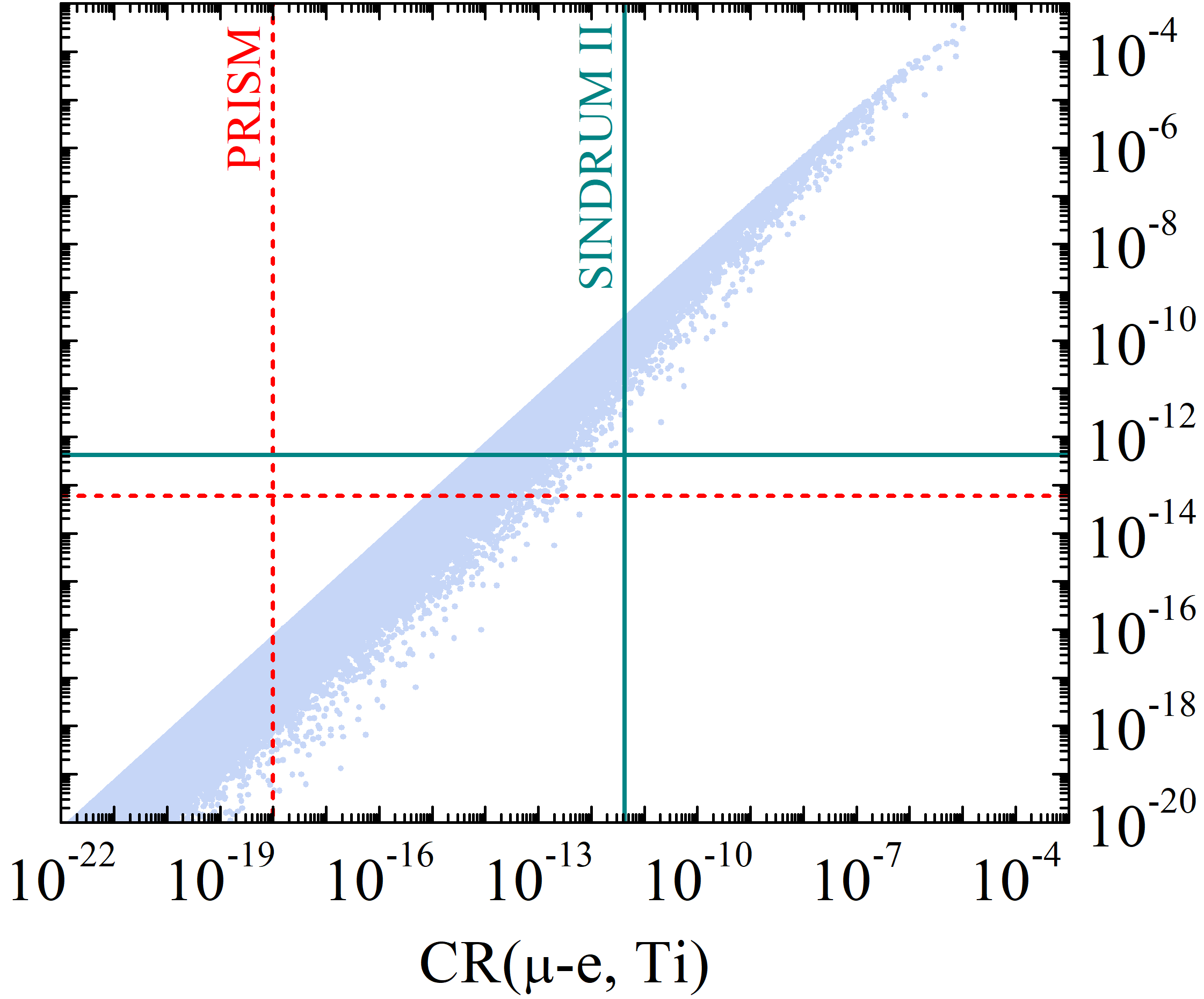} \\
   \vspace{+0.3cm}
   \includegraphics[scale=0.321]{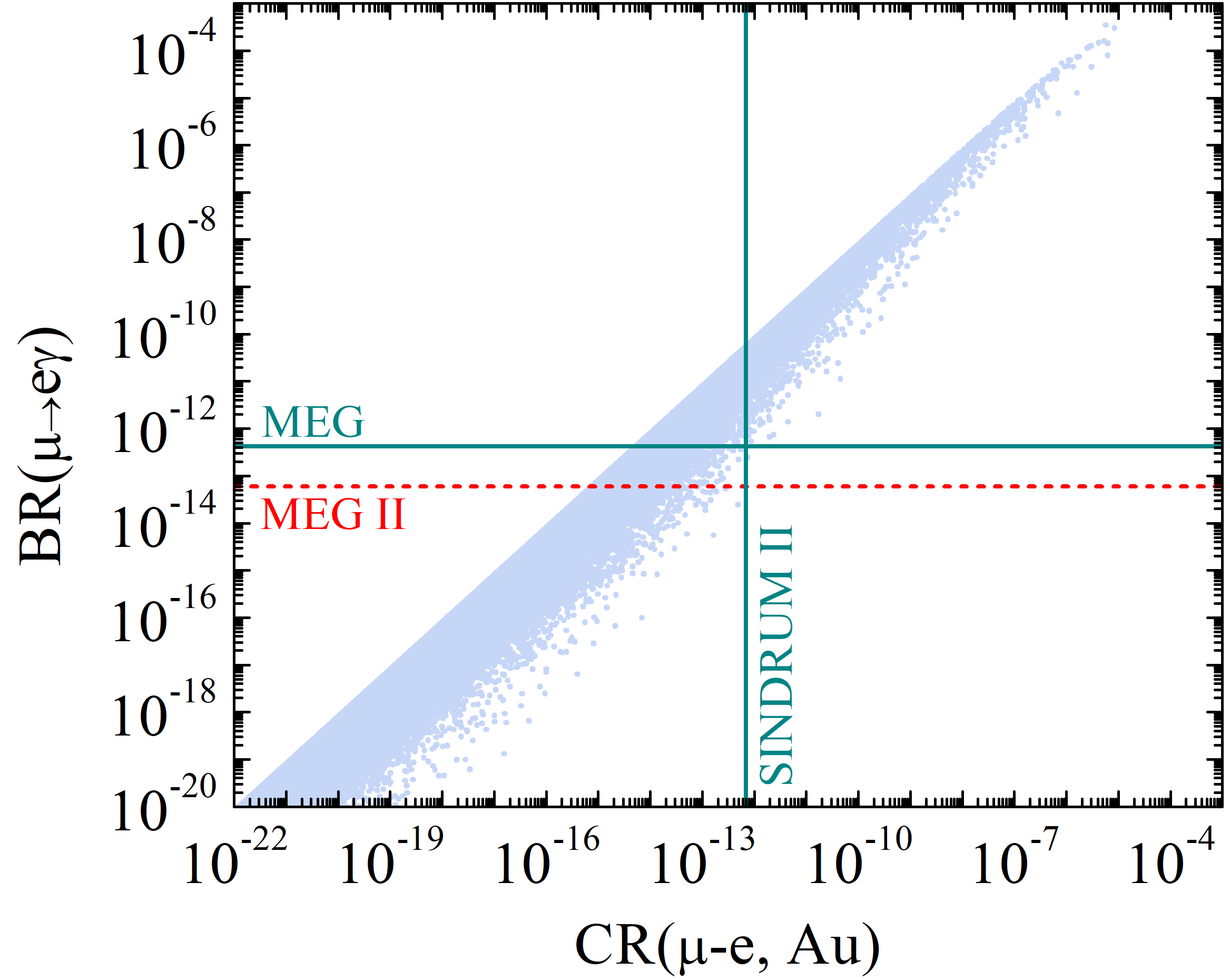}
   \includegraphics[scale=0.321]{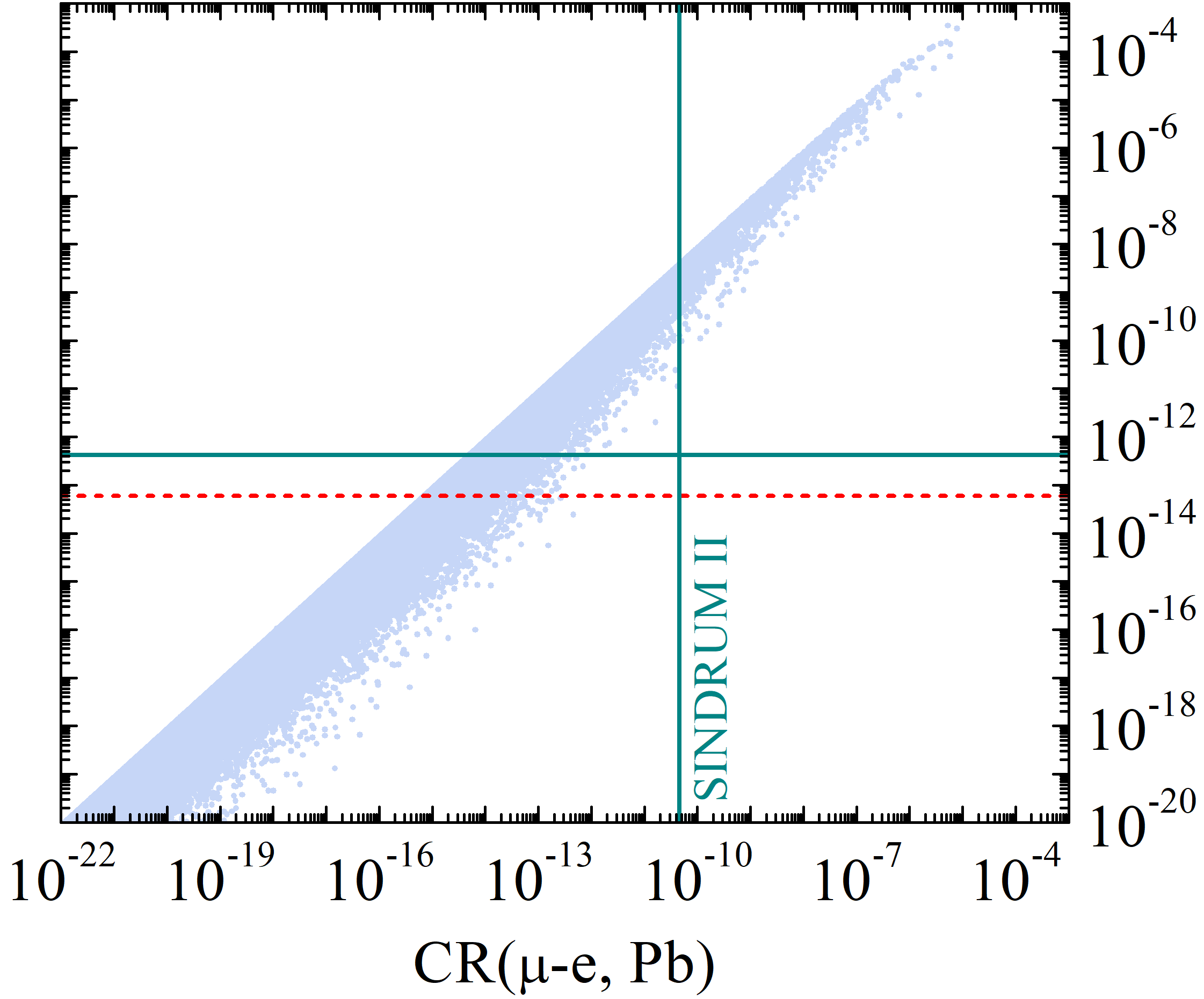}
  \caption{Comparison between muon cLFV processes that occur for the  $\mathcal{Z}_8^{e-\mu}$ ($\text{B}_4$) case. Namely, we plot $\BR(\mu \rightarrow e \gamma)$ in terms of $\BR(\mu \rightarrow 3 e)$ and the conversion rates for the Al, Ti, Au and Pb nuclei. The current bounds and future sensitivities for these observables, presented in table~\ref{tab:boundsCLFV}, are respectively indicated by a solid green and dashed red line.}
  \label{fig:FlavourVS}
\end{figure}
\begin{figure}[!ht]
   \centering
   \includegraphics[scale=0.32]{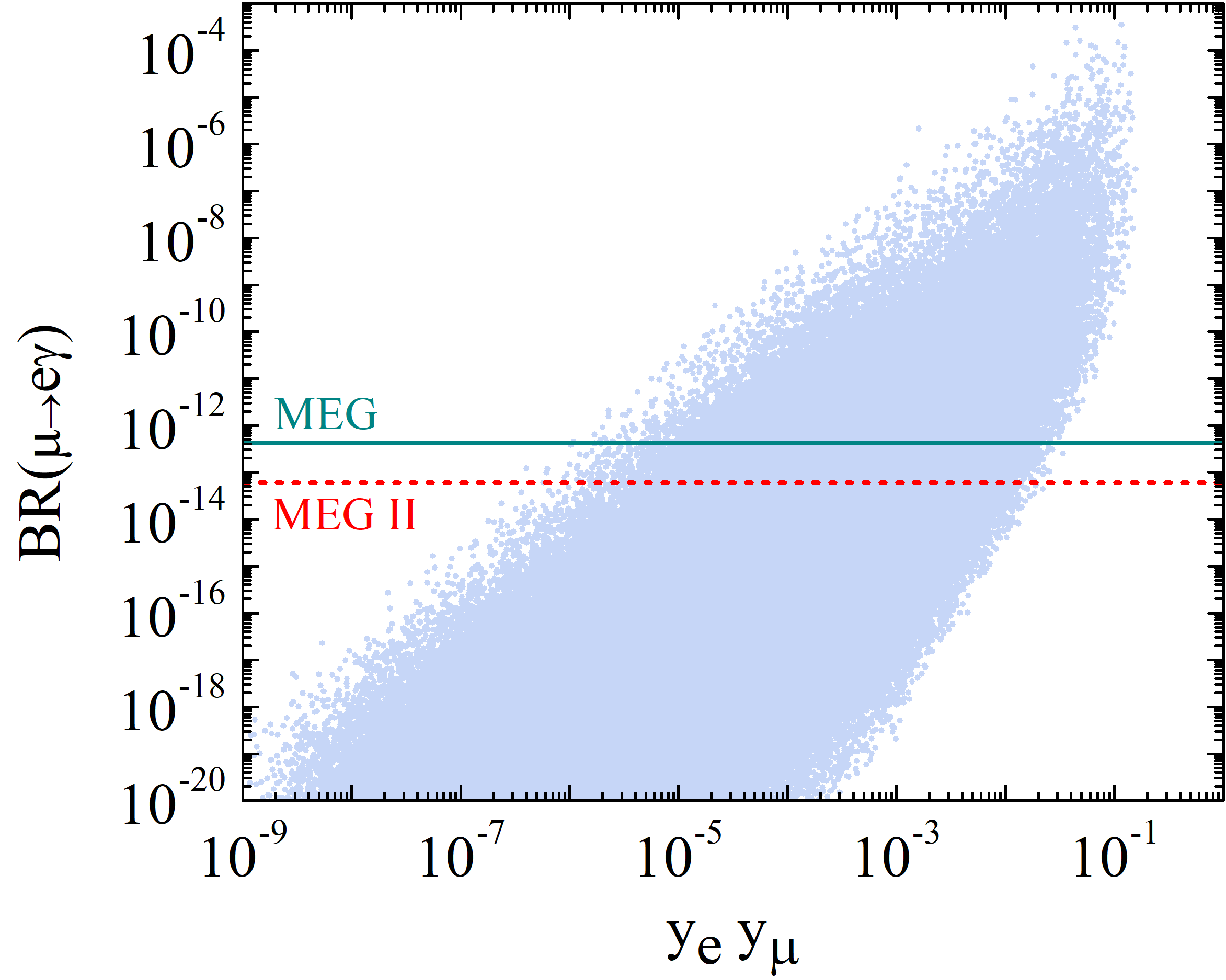}
   \includegraphics[scale=0.32]{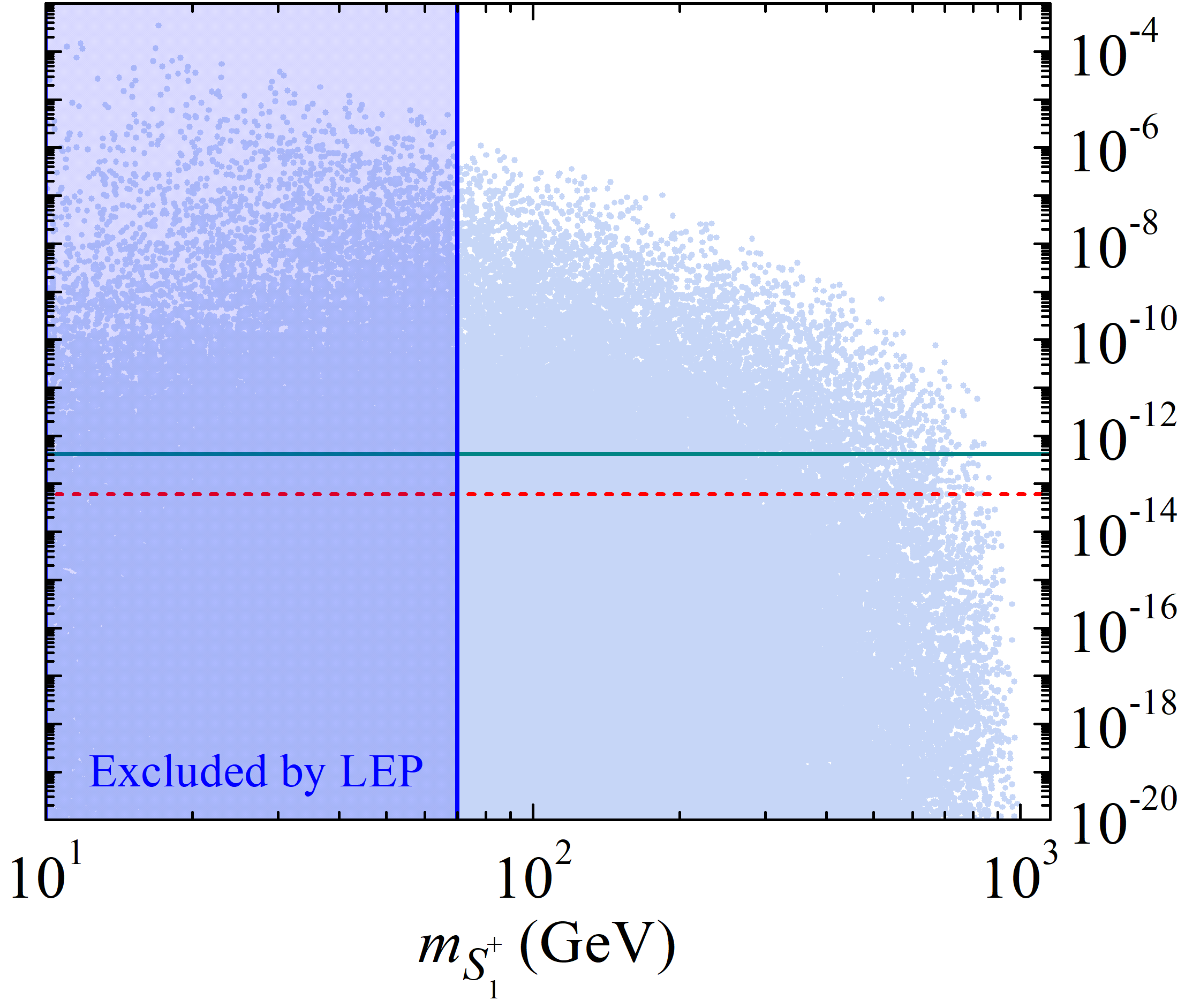}
  \caption{$\BR(\mu \rightarrow e \gamma)$ as a function of the Yukawa couplings product $y_e y_{\mu}$, on the left, and, in terms of the lightest \textit{dark} charged scalar mass $m_{S^+_1}$, on the right. We use the same colour code as in fig.~\ref{fig:FlavourVS} to indicate the current bound and projected sensitivity for $\BR(\mu \rightarrow e \gamma)$ showed in table~\ref{tab:boundsCLFV}.}
  \label{fig:FlavourMassYuk}
\end{figure}
\begin{itemize}

    \item The relation between $\BR(\mu \rightarrow e \gamma)$ and $\BR(\mu \rightarrow 3 e)$ or $\CR(\mu-e, \text{N})$ for the nuclei $\text{N} = \{\text{Al}, \text{Ti}, \text{Au}, \text{Pb}\}$ is nearly linear, showing that the dipole contribution coming from the photon-penguin diagram is dominant. This linear relation is more evident when the dipole approximation holds, i.e. when the mass~$M_f$ and~$m_{S_1^+}$ are approximately equal. The observed spreading of the scatter points is due to the monopole contribution coming from the photon and the difference between the~$S_1^+$ and~$S_2^+$ masses. The box and $Z$-penguin diagrams have negligible contributions to the processes. This is due to the fact that their amplitudes depend, respectively, on the fourth power of Yukawa couplings $y_{\alpha}^2 y_{\beta}^2$ and on the charged-lepton masses, suppressed by the large odd scalar masses. The current experimental bounds are indicated by the solid green lines and the projected sensitivities by the red dashed ones. We also refer from which experiments these bounds on the BRs and CRs come from, being their value reported in table~\ref{tab:boundsCLFV}. We notice that these bounds are easily surpassed for  a large number of points.
    
\item In fig.~\ref{fig:FlavourMassYuk}, we show the variation of the BR for $\mu \rightarrow e \gamma$ in terms of the product of the Yukawa couplings $y_ey_\mu$ (left plot) and on the odd-charged scalar mass $m_{S_1}^+$ (right plot). By looking at the left plot we notice that there is an upper bound $y_e y_{\mu} \leq 0.2$. This can be understood by considering eq.~\eqref{eq:yscoto} which, for a maximum value $y_{\mu}=1$, implies $M_f \mathcal{F}_{2 2} = \left|(\Mnuh)_{22}\right| = 1.87 \times 10^{-2}$~eV. Given that $\mathcal{F}_{1 2} \geq \mathcal{F}_{2 2}$ (see section~\ref{sec:scalarmass}) or, equivalently, $M_f \mathcal{F}_{1 2} \geq \left|(\Mnuh)_{22}\right|$, one has $y_e \leq 0.2$ and thus $y_e y_{\mu} \leq 0.2$. Furthermore, the quadratic dependence on $y_e y_{\mu}$ is evident and the BR value increases for larger $y_e y_{\mu}$ and lower $m_{S^+_1}$.

    \item  Using eq.~\eqref{eq:Yukmix}, we can derive the following analytical expression~\cite{Hagedorn:2018spx}:
    \begin{align}
    \frac{\BR(\mu \rightarrow e \gamma)}{4.2 \times 10^{-13}} & \approx 1.98 \times 10^{10} \left(\frac{70 \; \text{GeV}}{m_{S_1^+}}\right)^4 \sin^2(2 \varphi) y_e^2 y_{\mu}^2 \left|g\left(\frac{M_f^2}{m_{S_1^+}^2}\right) - \frac{{m_{S_1^+}^2}}{m_{S_2^+}^2} g\left(\frac{M_f^2}{m_{S_2^+}^2}\right)\right|^2, \nonumber \\
    g(x) & = \frac{1-6 x + 3 x^2 + 2 x^3 - 6 x^2 \log x}{6 (1-x)^4} \; .
    \label{eq:BRmueg}
    \end{align}
    Note that we have taken the reference value of $70$~GeV for $m_{S_1^+}$ since it corresponds to the lower bound imposed by LEP-II on that mass (this excludes the blue-shaded region on the right plot of fig.~\ref{fig:FlavourMassYuk}). From the above result, it is also clear that charged odd-scalar mixing is crucial. In fact, for vanishing $\mu_{12}$ in the scalar potential there is no $\eta_1^+-\eta_2^+$ mixing and, consequently, $\varphi=0$ -- see eq.~\eqref{eq:beta}. This would imply vanishing contributions to cLFV from the scotogenic sector. 
    In order to understand the impact of Yukawa couplings and charged-scalar masses/mixing on cLFV rates, let us take the approximate result of eq.~\eqref{eq:BRmueg} with maximal $\eta_1^+-\eta_2^+$ mixing ($\varphi = \pi/4$) and $M_f=m_{S_1^+}$. Defining a mass-degeneracy parameter $\epsilon = m_{S_2^+}/ m_{S_1^+} - 1$, we get
    \begin{align}
    \frac{\BR(\mu \rightarrow e \gamma)}{4.2 \times 10^{-13}} & \approx 1.4 \times 10^{8} \left(\frac{70 \; \text{GeV}}{m_{S_1^+}}\right)^4 y_e^2 y_{\mu}^2 \left|1 - \frac{1}{\left(1+ \frac{\epsilon}{2}\right)^3}\right|^2 \,,
    \end{align}
    which, for $\epsilon \ll 1$, can be approximated by
    \begin{align}
    \frac{\BR(\mu \rightarrow e \gamma)}{4.2 \times 10^{-13}} \approx 3.1 \times 10^{8} \left(\frac{70 \; \text{GeV}}{m_{S_1^+}}\right)^4 y_e^2 y_{\mu}^2 \, \epsilon^2\,.
    \label{eq:BRmueglimit}
    \end{align}
  It is clear that, as expected, the $\BR$ vanishes when the charged scalars are degenerate ($\epsilon \rightarrow 0$). Furthermore, taking  $y_e y_\mu = 0.2$ (upper bound on $y_e y_\mu$), $m_{S_1^+}=70$~GeV (LEP lower bound on the charged scalar mass) and the limit $m_{S_1^+} \ll m_{S_2^+}$, i.e. $\epsilon \gg 1$, we have $\BR(\mu \rightarrow e \gamma) \approx 2.4 \times 10^{-6}$. This value is consistent with the results shown in the right plot of fig.~\ref{fig:FlavourMassYuk}. For $m_{S_1^+}=70$~GeV, $y_e y_\mu$ must be lowered to $8.5 \times 10^{-5}$ or $\epsilon \leq 2.8 \times 10^{-4}$, in order to satisfy the current MEG bound. For the maximum value $y_e y_\mu = 0.2$, and for non-degenerate masses, the MEG bound implies $m_{S_1^+} \gtrsim 3.4$ TeV.  This interplay among the several parameters leads to a wide parameter-space region compatible with current experimental bounds.
\end{itemize}

\begin{figure}[!t]
   \centering
   \includegraphics[scale=0.31]{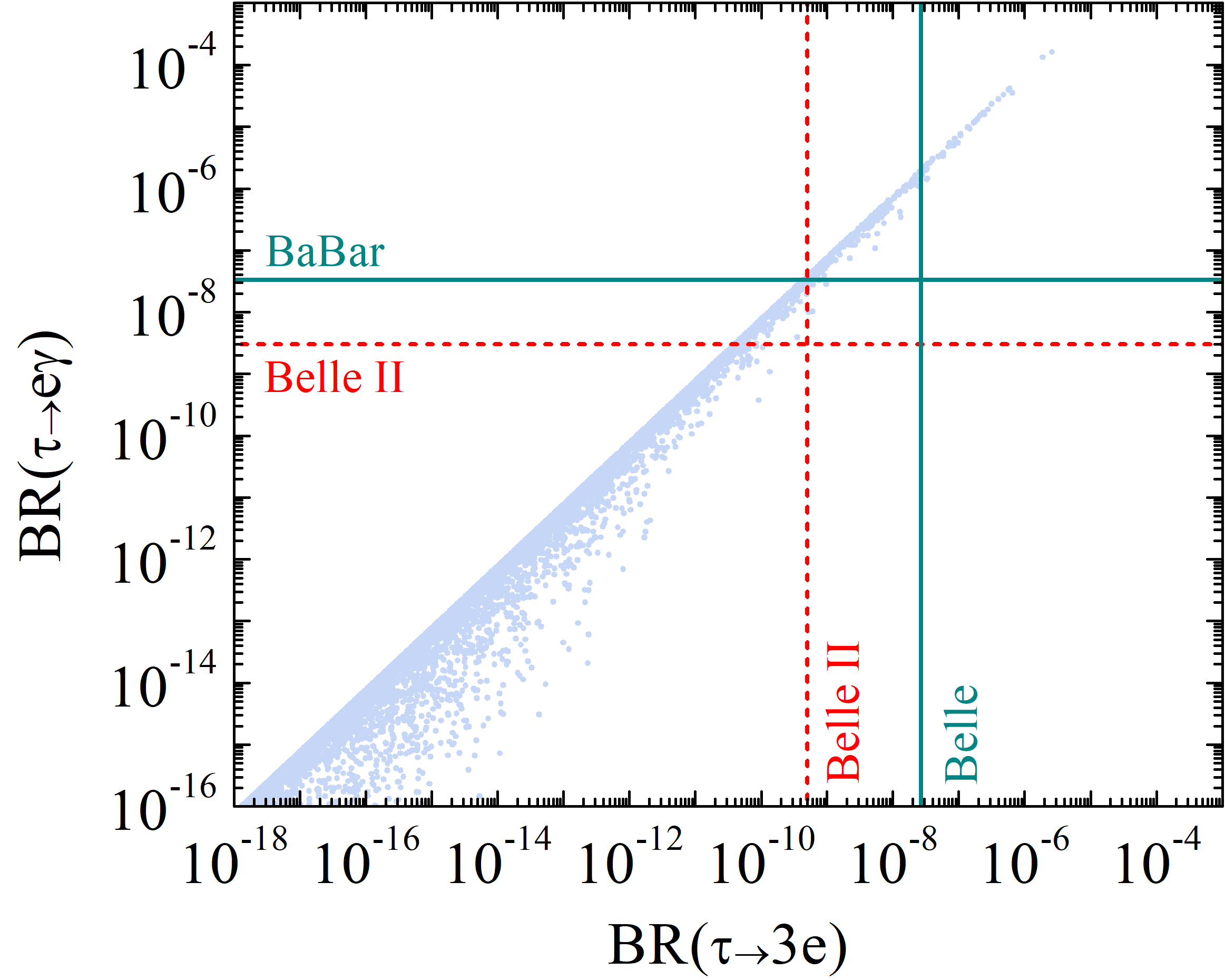}
   \includegraphics[scale=0.31]{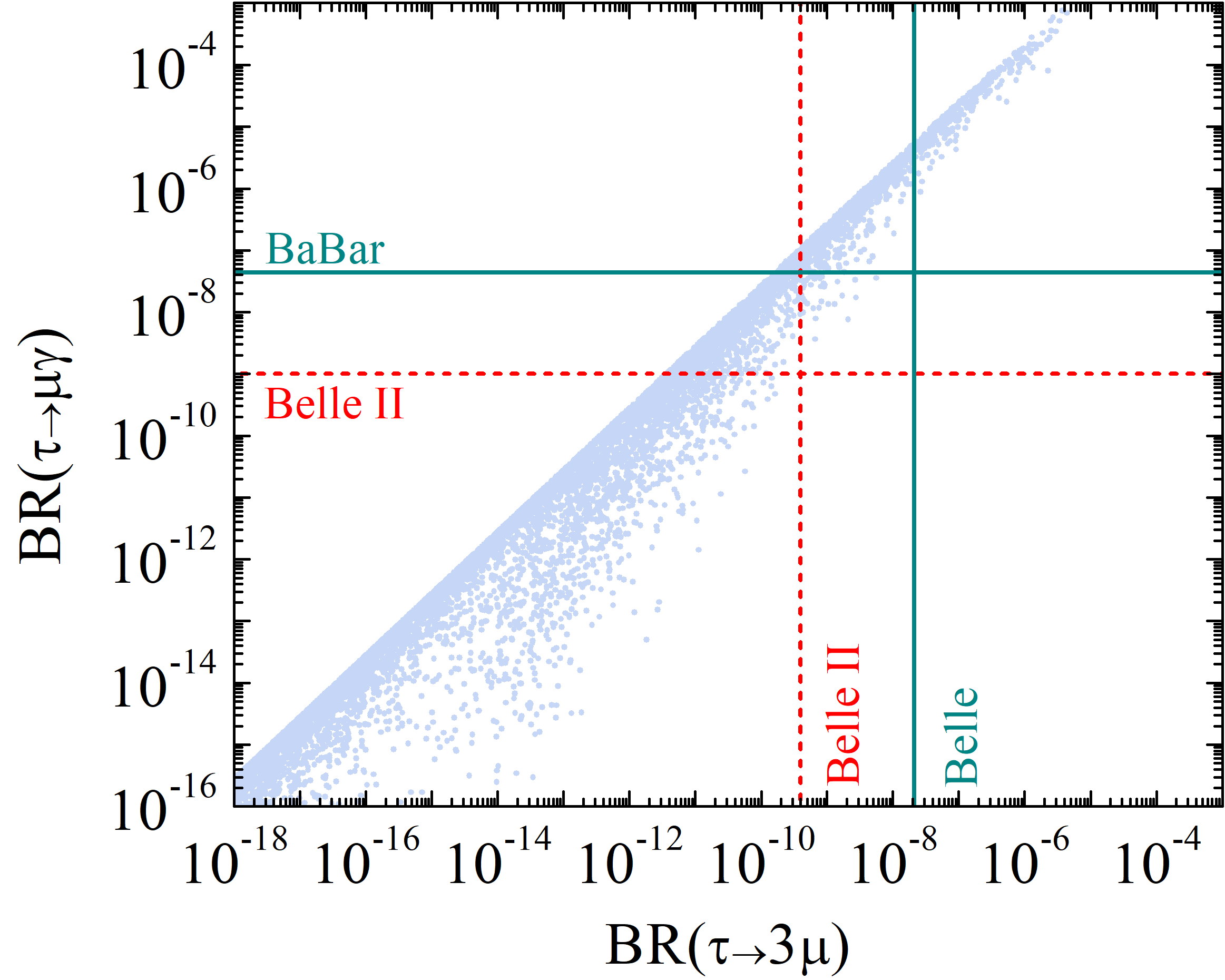}
   \caption{On the left [right] we present the $\BR(\tau \rightarrow e \gamma)$ [$\BR(\tau \rightarrow \mu \gamma)$] in terms of $\BR(\tau \rightarrow 3 e)$ [$\BR(\tau \rightarrow 3 \mu)$]. These cLFV observables are allowed for the $\mathcal{Z}_8^{e-\tau}$ ($\text{B}_3$) [$\mathcal{Z}_8^{\mu-\tau}$ ($\text{A}_1$)] case. We apply, the same colour code used in fig.~\ref{fig:FlavourVS} and~\ref{fig:FlavourMassYuk} to indicate the current bounds and projected sensitivities for these processes, which are reported in table~\ref{tab:boundsCLFV}.}
  \label{fig:Flavourtau}
\end{figure}
Although we have focused our discussion to the $\mathcal{Z}_8^{e-\mu}$ case, it is worth commenting on the remaining $\mathcal{Z}_8^{e-\tau}$ ($\text{B}_3$) and $\mathcal{Z}_8^{\mu-\tau}$ ($\text{A}_1$) scenarios. The cLFV observables allowed by the latter cases are shown in fig.~\ref{fig:Flavourtau} where we compare on the left $\BR(\tau \rightarrow e \gamma)$ to $\BR(\tau \rightarrow 3 e)$, and on the right $\BR(\tau \rightarrow \mu \gamma)$ to  $\BR(\tau \rightarrow 3 \mu)$. Note that we reconstructed the Yukawa parameters for the best-fit NO values of low-energy observables, as done for $\mathcal{Z}_8^{e-\mu}$ (see previous discussion). We remark that, even though the current bounds and future sensitivities (see table~\ref{tab:boundsCLFV}) of these $\tau$ cLFV observables are orders of magnitude above the muon cLFV ones, there is significant parameter space where the constraints are saturated. As detailed above, this is due to the large range of Yukawa coupling values and \textit{dark} charged scalar mass splittings, provided by the parameter variation displayed in table~\ref{tab:ParameterScan}. The conclusions for these cases are similar to the ones reached by our analysis of the most stringent muon cLFV processes. Having investigated the cLFV implications of our hybrid seesaw/scotogenic framework, we now study the DM phenomenology of our model.

\section{Dark matter}
\label{sec:DM}

The \textit{dark}-sector particles of our model are potential candidates for DM thanks to the $\mathcal{Z}_8$ charge assignments given in table~\ref{tab:part&sym}. After SSB, a $\mathcal{Z}_2$ symmetry, under which the \textit{dark} sector is odd, remains unbroken. The WIMP DM candidate will obviously be the lightest \textit{dark} particle~(LDP), for which there are  two possibilities in our model: the (lightest) neutral scalar~$S_1$ and the fermion~$f$. We will contemplate both scenarios in what follows. The computation of the DM relic density~$\Omega h^2$ and DM-nucleon spin-independent cross section at tree level, $\sigma^{\text{SI}}$, is performed with the \texttt{MicrOmegas 5.2.7.a} package~\cite{Belanger:2014vza}. We use \texttt{SSP 1.2.5}~\cite{Staub:2011dp} to establish the link with \texttt{SARAH}, \texttt{SPheno} and \texttt{FlavourKit} -- used in the previous section to obtain the particle mass/mixing spectrum, decays and cLFV observables. We will restrict our analysis to the NO $\mathcal{Z}_8^{e-\mu}$ case, for which cLFV has been studied in section~\ref{sec:cLFVmain}. Note that DM phenomenology is not significantly altered for $\mathcal{Z}_8^{e-\tau}$ and $\mathcal{Z}_8^{\mu-\tau}$. Compatibility with neutrino oscillation data is always ensured by extracting the Yukawa parameters and SCPV phase $\theta$ via eqs.~\eqref{eq:yscoto}-\eqref{eq:Rratio}. From now on, we will take best-fit values for the low-energy neutrino observables preferred by the model. Additionally, all generated points satisfy the current cLFV bounds. The remaining parameters vary in the ranges shown in table~\ref{tab:ParameterScan}. Note that, for the scalar (fermion) DM case, we impose $m_{S_1} < M_f$ ($M_f < m_{S_1}$), making $S_1$ ($f$) the LDP.

In appendix~\ref{sec:DMdiag}, we show the annihilation and coannihilation diagrams that contribute to the thermally average cross section $ \left<\sigma(\text{DM} \ \text{DM} \rightarrow \text{nonDM} \ \text{nonDM}) v\right>$, where "nonDM" refers to SM and non-SM even particles, i.e. all particles that are not \textit{dark}. These contribute to the DM relic density as $\Omega h^2 \propto \left<\sigma v\right>^{-1}$. We will use the $3 \sigma$ range for the cold dark matter~(CDM) relic density obtained by the Planck satellite data~\cite{Planck:2018vyg},
\begin{equation}
   0.1126 \leq \Omega_{\text{CDM}} h^2 \leq 0.1246 \; .
   \label{eq:Oh2Planck}
\end{equation}

\subsection{Scalar dark matter}
\label{sec:DMscalar}

For the scalar DM candidate $S_1$ the annihilation diagrams are shown in fig.~\ref{fig:DiagAnnScalarDM}, while the coannihilation ones with the odd-charged scalars $S^+$ and fermion $f$, are presented in figs.~\ref{fig:DiagCoAnn1} and~\ref{fig:DiagCoAnn3}, respectively.
\begin{figure}[t!]
    \centering
    \includegraphics[scale=0.45]{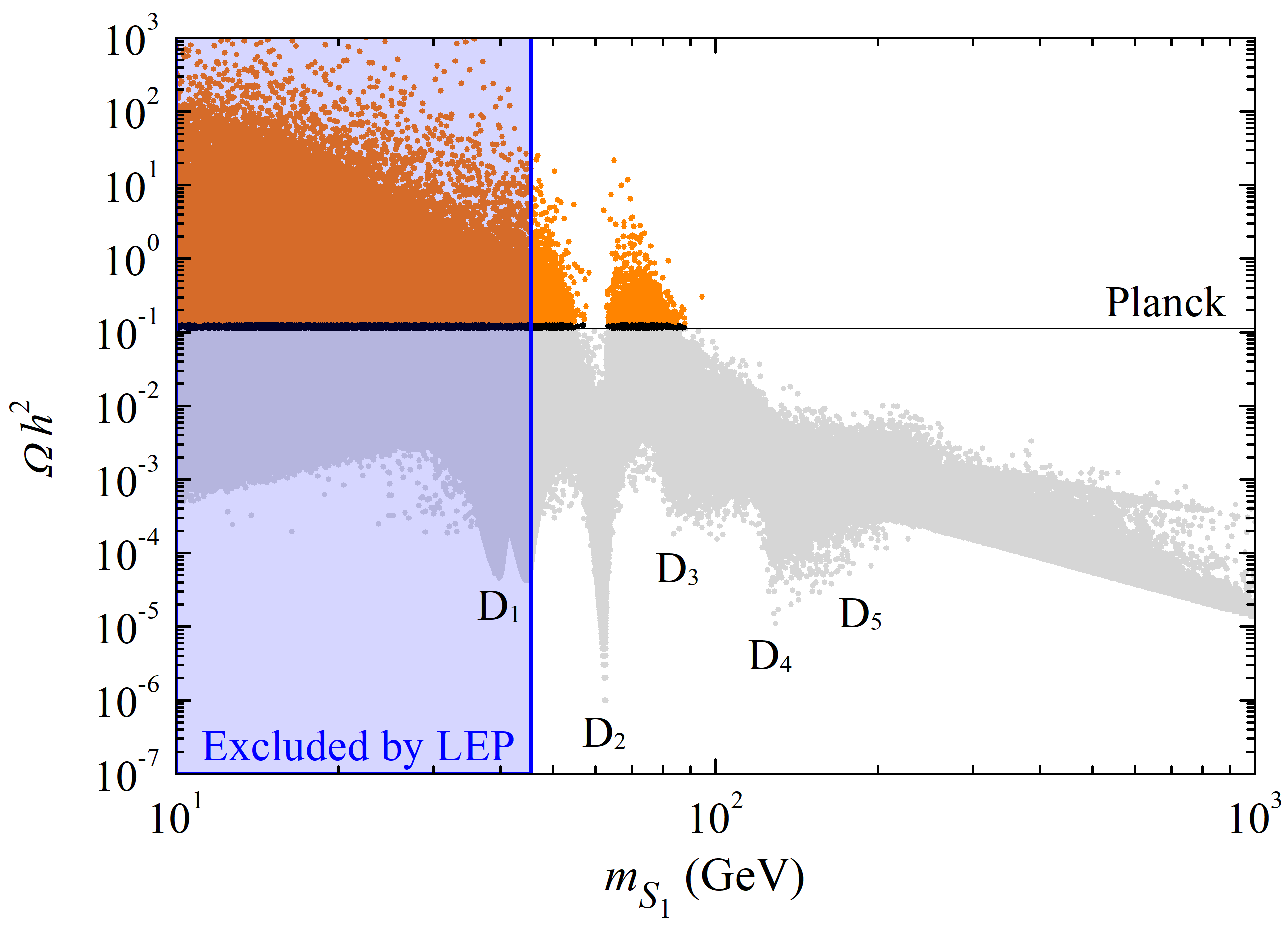}
    \caption{Relic density $\Omega h^2$ as a function of the scalar DM mass $m_{S_1}$. The black points lie within the $3 \sigma$ range of the measured CDM relic abundance obtained by Planck~\cite{Planck:2018vyg} [see eq.~\eqref{eq:Oh2Planck}]. The orange (grey) points give an over (under) abundance of DM. The blue-shaded region is excluded by the LEP data that forbids the $Z$ and $W$-bosons to decay into $S_1$. The labels (D$_1$)-(D$_5$) refer to the different dips in relic density observed in the plot -- see the main text for more details.}
    \label{fig:ScalarDM_Relic}
\end{figure}
In fig.~\ref{fig:ScalarDM_Relic} we show $\Omega h^2$ versus the LDP or DM mass $m_{S_1}$. The black points are within the Planck $3 \sigma$ range for $\Omega h^2$ given in eq.~\eqref{eq:Oh2Planck}. Instead, the orange and grey points correspond to an over and under DM abundance, respectively. The vertical blue-shaded band is excluded by collider data, namely by the LEP precision measurements on the $W$ and $Z$-boson decay widths. This forbids the decay~$Z \rightarrow S_1 S_1$, excluding the low-mass region $m_{S_1} > m_Z/2$ [see eq.~\eqref{eq:LEPbound}]. The key features of our results concerning the DM abundance for scalar DM are:
\begin{itemize}

    \item There are only two viable mass regions that lead to the correct relic density, namely $m_{S_1} \lesssim 60$ GeV and $68 \ \text{GeV} \lesssim m_{S_1} \lesssim 90 \ \text{GeV}$. As already mentioned, the former is almost completely excluded by LEP. 
    
    \item The dips in the relic density, labeled as (D$_1$)-(D$_5$) in the figure, can be understood by looking at the~$S_1$ (co)annihilation diagrams shown in appendix~\ref{sec:DMdiag}. Dips D$_{1}$ occur at $m_{S_1} \sim  m_W/2, m_Z/2$. They are due to the coannihliation $s$-channel mediated by the $W$-boson $S_1 S^+ \rightarrow W^* \rightarrow \text{nonDM} \ \text{nonDM}$, and the (co)annihilation through $s$-channel $Z$-boson mediation $S_1 S \rightarrow Z^* \rightarrow \text{nonDM} \ \text{nonDM}$, respectively. D$_2$ at $m_{S_1} \sim m_{h_1}/2$ stems from annihilation through $s$-channel mediated by the SM Higgs-boson, i.e. $S_1 S_1 \rightarrow h_1^* \rightarrow \text{nonDM} \ \text{nonDM}$. Notice that D$_2$ is deeper than D$_{1}$ since the $W$ and $Z$-boson contributions are momentum suppressed. Consequently, within the mass interval $60 \ \text{GeV} \lesssim m_{S_1} \lesssim 68 \ \text{GeV}$, there are no viable relic density points. For $m_{S_1} \gtrsim 80 - 90$ GeV, multiple decay channels open up, namely, the annihilation into two massive gauge bosons $S_1 S_1 \rightarrow W^+ W^- , Z Z$ through quartic-coupling interactions, leading to a decrease of~$\Omega h^2$ (D$_3$). When $m_{S_1} \gtrsim 125$~GeV, the decay $S_1 S_1 \rightarrow h_1 h_1$ becomes possible -- D$_4$ and for $m_{S_1} \gtrsim 180$~GeV the decay into a pair of top quarks opens up -- D$_5$.
 
    \item For $m_{S_1} \gtrsim 90 \ \text{GeV}$, our model is not compatible with the Planck $\Omega h^2$ $3 \sigma$ interval. This is a distinct feature with respect to the usual inert-doublet/scotogenic models~\cite{Diaz:2015pyv,Belyaev:2016lok,Rojas:2018wym}. In these frameworks, for small charged and neutral odd-scalar mass splittings of $\mathcal{O}(1)$ GeV, a viable relic-density value is reached for scalar DM masses above $\sim 500$ GeV. This happens because the thermally-averaged cross section scales as $\sim 1/m_{S_1}^2$, leading to increasing~$\Omega h^2$ with increasing DM mass. One may consider scenarios with two inert scalar doublets, as in the case of the 3HDM supplemented with symmetries in the scalar potential studied in refs.~\cite{Fortes:2014dca,Keus:2015xya,Cordero-Cid:2016krd,Hernandez-Sanchez:2020aop,Aranda:2019vda,Khater:2021wcx}. It was noted that a low/intermediate and high-mass regions, with correct $\Omega h^2$, do not coexist. In fact, a high-mass region was shown to be viable requiring a strong degeneracy among scalar masses, stemming from fine-tuning of scalar potential parameters or from restrictive symmetries. Multiple coannihilations among scalars lead to cancellations between pure (quartic) gauge and Higgs mediated channels, lowering the thermally-averaged cross section. As a result, a correct value for the relic density is achieved, being this effect more pronounced for masses above 500~GeV. In turn, the absence of an intermediate DM mass region is due to multiple coannihilations mediated by the $W$, $Z$ and Higgs bosons. In our two inert-doublet case, due to the link between \textit{dark} scalar masses and scotogenic neutrino mass generation (see section~\ref{sec:neutrinomass}), a highly-degenerate mass spectrum is not natural since it would lead to non-perturbative Yukawa couplings.
    
\end{itemize}

\begin{figure}[t!]
    \centering
    \includegraphics[scale=1,trim={3cm 22.5cm 2cm 1.0cm},clip]{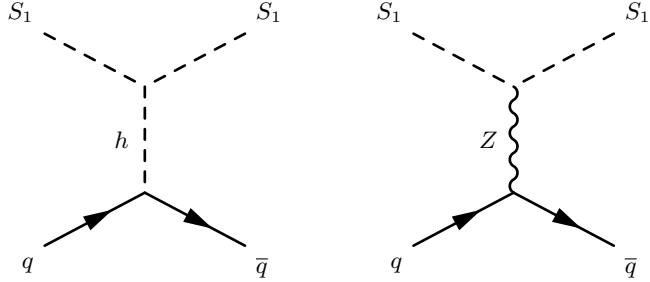}
    \caption{Tree-level diagrams contributing to WIMP-nucleon spin-independent elastic cross-section for scalar DM.}
    \label{fig:diagDDScalarDM}
\end{figure}
We now turn our attention to the DD implications for scalar DM. In fig.~\ref{fig:diagDDScalarDM}, we present the tree-level diagrams that contribute to the spin-independent elastic-scattering cross section $\sigma^{\text{SI}}$ via \textit{non-dark} neutral scalar $h_i$ (left diagram) and $Z$-boson (right diagram) exchange. The dominant contribution to $\sigma^{\text{SI}}$ arises from the SM Higgs-mediated diagram, which in the alignment limit ($\lambda_{\Phi\sigma}=0$) is given by,
\begin{align}
\sigma^\text{SI}= \dfrac{v^2}{4 \pi m_{h_1}^4} [ (\lambda_3 + \lambda_4) c_+^2 + (\lambda_3^\prime + \lambda_4^\prime) s_+^2 - \lambda_5 c_+ s_+ ]^2 \frac{m_N^2}{(m_{S_1}+m_N)^2} f_N^2 \; ,
\label{eq:sigmaSIScalar}
\end{align}
where $c_+ \equiv \cos \varphi_+$ and $s_+ \equiv \sin \varphi_+$, being $\varphi_+$ the mixing angle given in eqs.~\eqref{eq:MixneutralDM} and \eqref{eq:phimix} (see appendix~\ref{sec:scalarmass}). In the above equation, $m_N$ is the average nucleon mass, and $f_N$ is the nucleon form factor
\begin{align}
f_N=\dfrac{Z f_p+(A-Z) f_n}{A},
\label{eq:fN}
\end{align}
being $Z$ and $A$ the atomic and nucleon number of the atoms in the detector, respectively, and 
\begin{align}
    f_{p,n}=\dfrac{m_{p,n}}{v}\left[\sum_{q=u,d,s} f_q^{p,n}+\dfrac{2}{9}\left(1-\sum_{q=u,d,s} f_q^{p,n}\right)\right],
\label{eq:fpn}
\end{align}
where $f_q^{p,n}$ are the individual quark form factors in protons and neutrons. The \texttt{MicrOmegas} package default values for $f_q^{p,n}$~\cite{Belanger:2008sj} yield $f_N\simeq 1.8\times10^{-3}$. An upper limit for the cross section in eq.~\eqref{eq:sigmaSIScalar} can be obtained considering $\lambda_{3}^{(')}=\lambda_{4}^{(')}=1$, $\lambda_5=0$ and $\varphi_+=\pi/4$, as
\begin{align}
\sigma^\text{SI}\lesssim 3.5\times 10^{-43}\left(\dfrac{500\, \text{GeV}}{m_{S_1}}\right)^2~\text{cm}^2\; .
\label{eq:sigmaSIScalarupperbound}
\end{align}
This bound is well above the current limits from DD experiments on $\sigma^{\text{SI}}$ such as LUX~\cite{LUX:2016ggv}, PandaX-II~\cite{PandaX-II:2016vec} and XENON1T~\cite{XENON:2018voc}. 

\begin{figure}[t!]
    \centering
    \includegraphics[scale=0.45]{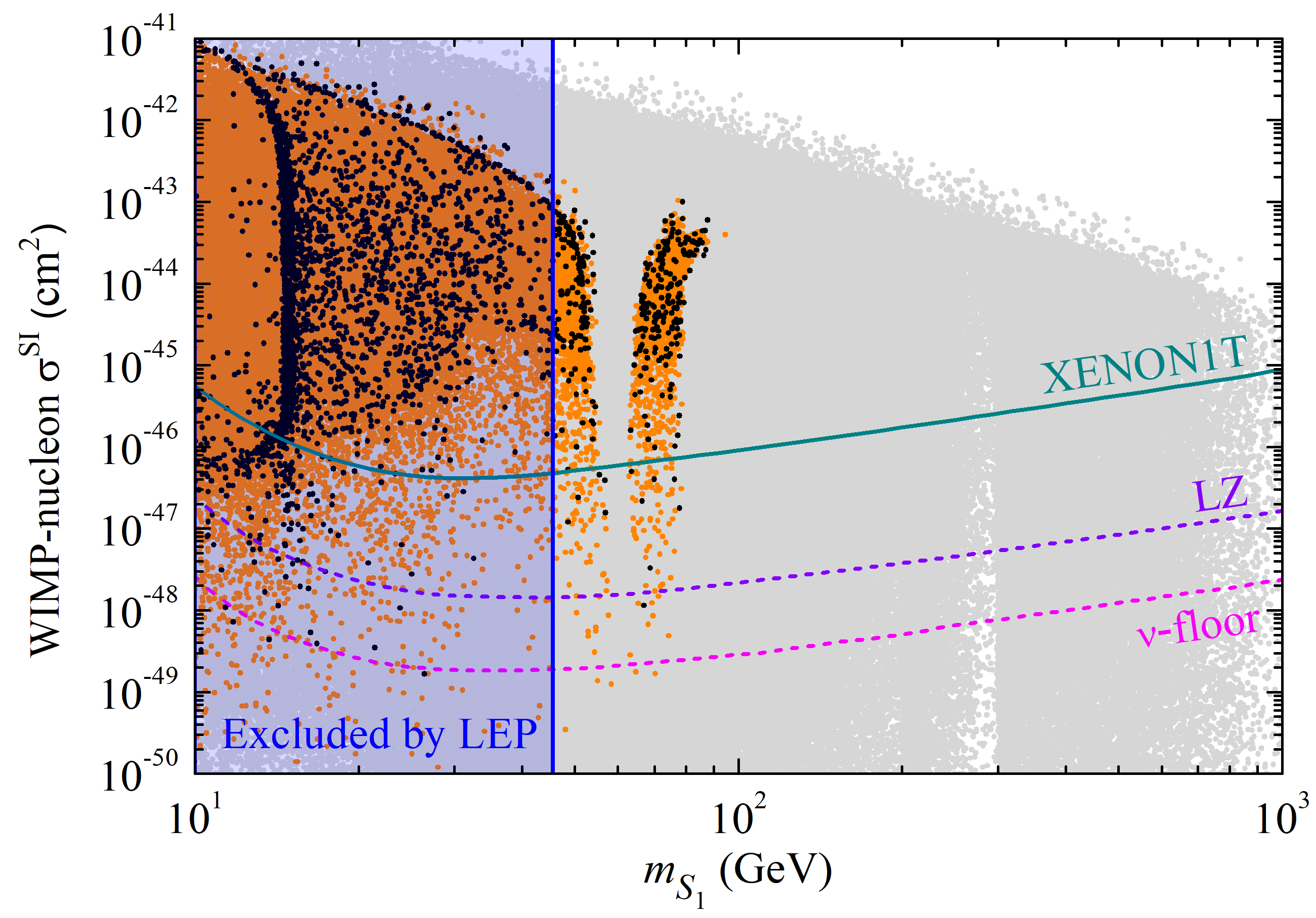}
    \caption{WIMP-nucleon spin-independent elastic scattering cross-section $\sigma^{\text{SI}}$ as a function of the scalar DM mass $m_{S_1}$. We use the same colour code as in fig.~\ref{fig:ScalarDM_Relic} to differentiate points in terms of their relic abundance and indicate the LEP excluded region. The solid turquoise line indicates the current bound from the XENON1T experiment~\cite{XENON:2018voc}, the dashed violet line represents the future sensitivity of the LUX-ZEPLIN~(LZ) proposal~\cite{LUX-ZEPLIN:2018poe} and the dashed magenta line shows the "neutrino floor" coming from coherent elastic neutrino scattering~\cite{Billard:2013qya}.}
    \label{fig:ScalarDM_DD}
\end{figure}
Scanning the quartic couplings $\lambda_{3}, \lambda_{4}, \lambda_{3}^\prime, \lambda_{4}^\prime, \lambda_{5}$, and the parameters $\mu_{12}$, $m_{\eta_1}^2$, $m_{\eta_2}^2 $ in the ranges of table~\ref{tab:ParameterScan}, we obtain the results for $\sigma^\text{SI}$ as function of the scalar DM mass~$m_{S_1}$ shown in fig.~\ref{fig:ScalarDM_DD}. The same color code as in fig.~\ref{fig:ScalarDM_Relic} is used, where the orange (grey) points yield an over(under)-abundance of DM relic density, while for the black dots a relic abundance within the observed 3$\sigma$ range is obtained. For comparison, we also show the most stringent current constraint on $\sigma^{\text{SI}}$ provided by the XENON1T experiment~\cite{XENON:2018voc} (solid turquoise line). It can be seen that most of the viable relic density points are excluded by current DD bounds. Those surviving the XENON1T constraint will be probed by future experimental searches, as can be seen by looking at the dashed violet line which indicates the projected sensitivity of the LUX-ZEPLIN (LZ)~\cite{LUX-ZEPLIN:2018poe} experiment. Other forthcoming DD experiments will be able to further probe the parameter space of our model, such as, DARWIN~\cite{DARWIN:2016hyl}, PandaX~\cite{PandaX:2018wtu} and  XENONnT~\cite{XENON:2020kmp} (for recent review on DD experimental prospects see ref.~\cite{Billard:2021uyg}). The dashed magenta line shows the "neutrino floor" coming from coherent elastic neutrino scattering~\cite{Billard:2013qya}. Notice that in the upper right corner there is a region with no points since there the perturbativity condition imposed on the quartic couplings $\lambda_{3}, \lambda_{4}, \lambda_{3}^\prime, \lambda_{4}^\prime$ and $\lambda_{5}$ -- that contribute to $\sigma^{\text{SI}}$ as in eq.~\eqref{eq:sigmaSIScalar} -- is violated.

As already mentioned, for scalar DM, most of our model's parameter space is excluded by the LEP bound, relic density Planck interval and DD current constraints. Only a few points within the mass interval $ 45.6 \; \text{GeV} \lesssim m_{S_1} \lesssim 90 \; \text{GeV}$ survive. There exist further collider constraints to take into account in this mass region when considering LHC Higgs boson data~\cite{ATLAS:2016neq,ATLAS:2019nkf,CMS:2018yfx,ATLAS:2018hxb}. Namely:
\begin{itemize}

    \item \textit{Higgs invisible decay}: The additional odd-neutral scalars $S_i$ ($i= 1, \cdots, 4$) contribute to the Higgs invisible decay width through the channel $h_1 \rightarrow S_i S_j$ ($i,j= 1, \cdots, 4$) which opens up for masses $m_{S_i} + m_{S_j} < m_{h_1}$ when $i \neq j$, or $m_{S_i} < m_{h_1}/2 \approx 62.6$~GeV for $i = j$. The decay width and BR are,
    \begin{align}
    &\Gamma(h_1 \rightarrow \text{inv}) = \frac{1}{2} \sum_{i, j = 1}^4 \Gamma(h_1 \rightarrow S_i S_j) \; , \nonumber \\  
    & \BR(h_1 \rightarrow \text{inv}) = \frac{\Gamma(h_1 \rightarrow \text{inv})}{\Gamma_{h_1}^{\text{tot}}}  = \frac{\Gamma(h_1 \rightarrow \text{inv})}{\Gamma_{h_1}^{\text{SM}} + \Gamma(h_1 \rightarrow \text{inv})} \; ,
    \label{eq:BoundBRh1inv_1}
    \end{align}
    where $\Gamma_{h_1}^{\text{tot}} = 3.2^{+2.8}_{-2.2} \; \text{MeV}$ is the SM Higgs decay width~\cite{ParticleDataGroup:2020ssz}. The BR above, is constrained by the LHC Higgs data, being the current bound~\cite{ParticleDataGroup:2020ssz},
    \begin{equation}
         \BR(h_1 \rightarrow \text{inv}) \leq 0.19 \; .
         \label{eq:BoundBRh1inv}
    \end{equation}
    
    \item \textit{Higgs to photon-photon}: Note that, there is no phase space for the decay $h_1 \rightarrow S_i^+ S_i^-$  ($i= 1, 2$) due to the LEP-II bound on $m_{S^+_i}$ [see eqs.~\eqref{eq:LEPbound} and discussion therein]. However, our odd-charged scalars $S_i^+$ will contribute at one-loop to the diphoton Higgs decay $h_1 \rightarrow \gamma \gamma$.  We define the $h_1 \rightarrow \gamma \gamma$ signal strength as,
    \begin{equation}
    R_{\gamma \gamma} = \frac{\BR(h_1 \rightarrow \gamma \gamma)}{\BR_{\text{SM}}(h_1 \rightarrow \gamma \gamma)} \; ,
    \end{equation}
    where we use the following SM value for $\BR(h_1 \rightarrow \gamma \gamma)$ and constraint on $R_{\gamma \gamma}$~\cite{ParticleDataGroup:2020ssz},
    \begin{equation}
        \BR_{\text{SM}}(h_1 \rightarrow \gamma \gamma) = 2.27 \times 10^{-3} \; , \; R_{\gamma \gamma} = 1.11^{+0.10}_{-0.09} \; .
    \label{eq:BoundRgg}
    \end{equation}
    
\end{itemize}

\begin{figure}[!t]
   \centering
   \includegraphics[scale=0.320]{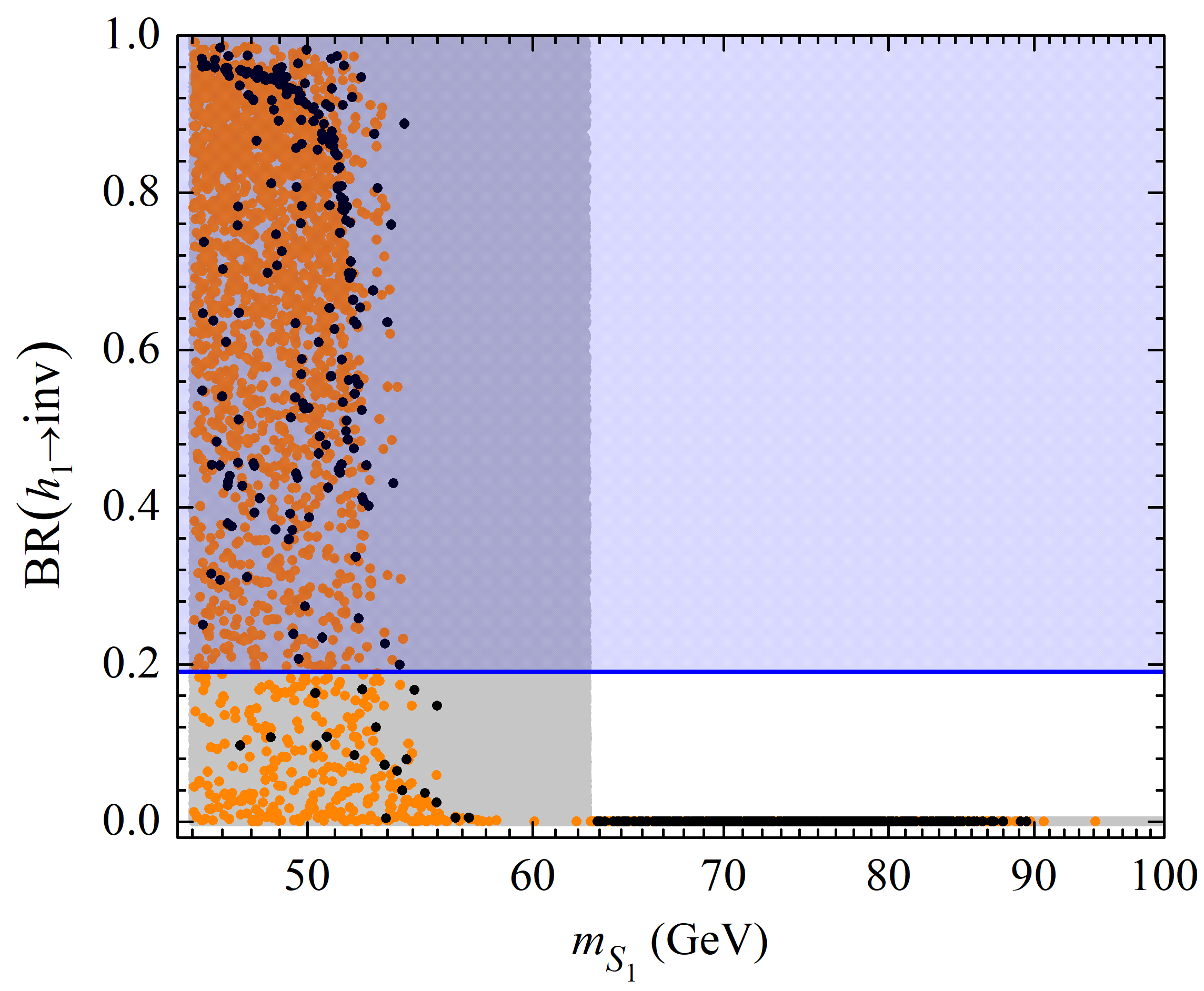} \hspace{-0.2cm}
   \vspace{-0.2cm}\includegraphics[scale=0.320]{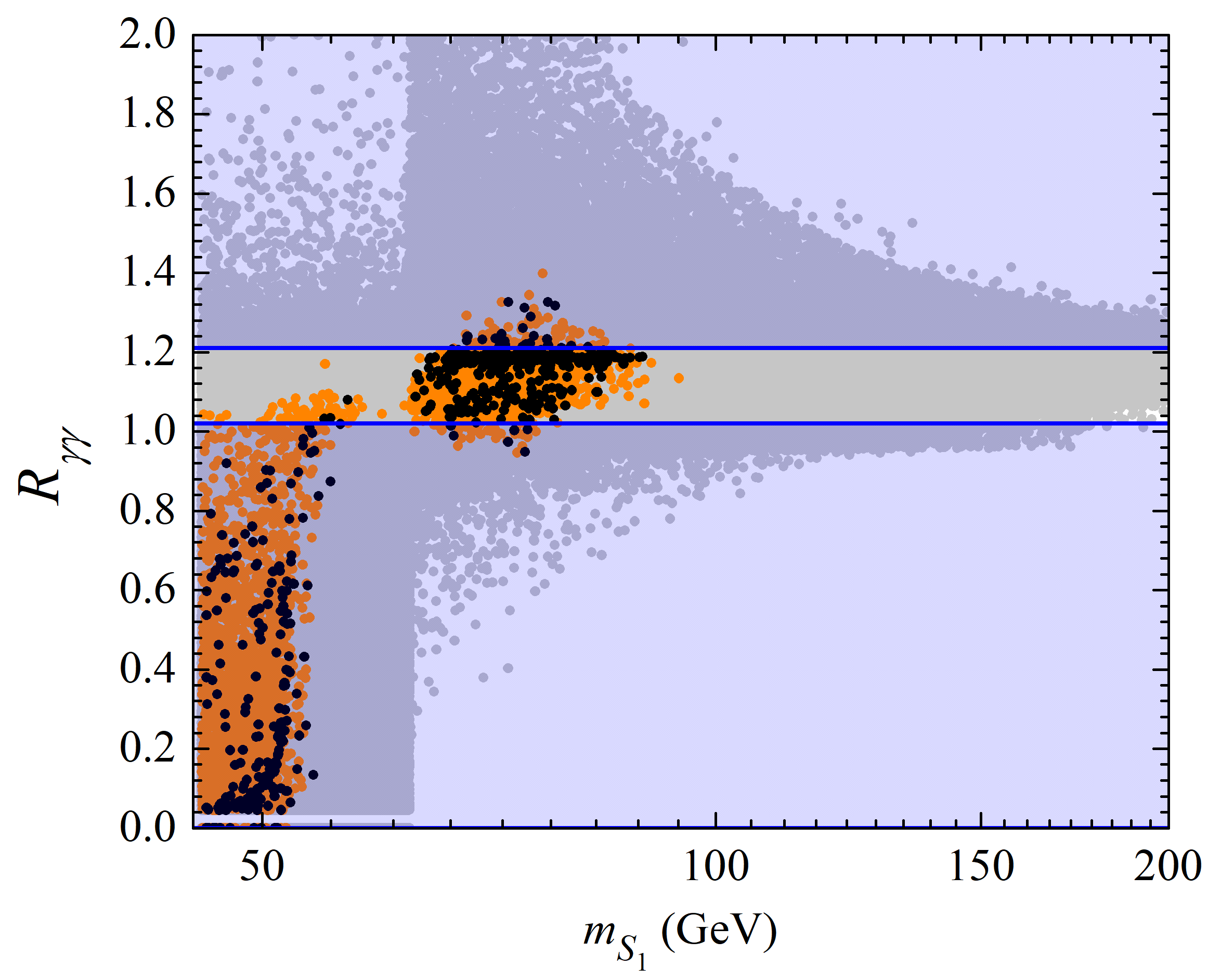}
  \caption{On the left [right], the $\BR(h_1 \rightarrow \text{inv})$ [$R_{\gamma \gamma}$] in terms of the scalar DM mass $m_{S_1}$, with the blue-shaded region[s] being excluded by the constraint in eq.~\eqref{eq:BoundBRh1inv} [eq.~\eqref{eq:BoundRgg}] (see text for details). To distinguish points, in terms of relic density, we use the colour code employed in fig.~\ref{fig:ScalarDM_Relic}.}
  \label{fig:ScalarDM_Higgs}
\end{figure}
The impact of the above two constraints can be appreciated in fig.~\ref{fig:ScalarDM_Higgs}. On the left and right panels are shown, respectively, the $\BR(h_1 \rightarrow \text{inv})$ and $R_{\gamma \gamma}$, in terms of the scalar DM mass $m_{S_1}$. By looking at the results, we see that the low-mass region $45.6 \; \text{GeV} \lesssim m_{S_1} \lesssim 60 \; \text{GeV}$ is simultaneously ruled out by the bound on the Higgs invisible decay and the $R_{\gamma \gamma}$ constraint of eqs.~\eqref{eq:BoundBRh1inv}~and~\eqref{eq:BoundRgg} (the exclusion regions are indicated in blue). 
However, the intermediate mass region $68 \ \text{GeV} \lesssim m_{S_1} \lesssim 90 \ \text{GeV}$ is compatible with those constraints. Thus, we conclude that this mass region for scalar DM is viable since it leads to a correct relic density while evading current collider and DD experimental constraints. 

Before proceeding to the study of fermionic DM we briefly comment on the impact of deviations from the alignment limit, by allowing $\{\lambda_{\Phi\sigma}, \lambda_{\eta \sigma}, \lambda_{\eta \sigma}^\prime \} \neq 0$. In this scenario, additional decay channels are possible as $S_1 S_i \rightarrow h_j \rightarrow h_k h_l$ with $i=1, \cdots, 4$ and $j,k,l=1,2,3$ (see appendix~\ref{sec:DMdiag}). This will have the overall effect of lowering the relic-density value. However, for a wide range of variation of $\lambda_{\Phi\sigma}, \lambda_{\eta \sigma}, \lambda_{\eta \sigma}^\prime$, as we did for the remaining couplings, the overall presented picture will not be significantly altered. Regarding the spin-independent scattering cross section, one can show that considering $\{\lambda_{\Phi\sigma}, \lambda_{\eta \sigma}, \lambda_{\eta \sigma}^\prime \} \neq 0$ does not lead to significant changes on the results of fig.~\ref{fig:ScalarDM_DD}. This can be seen by including in  eq.~\eqref{eq:sigmaSIScalar} the corrections from $\eta-\sigma$ interactions and mixing among \textit{non-dark} scalars. The complete~$\sigma^\text{SI}$ expression reads,
\begin{align}
\sigma^\text{SI}= \dfrac{\kappa^2}{4 \pi} \frac{m_N^2}{(m_{S_1}+m_N)^2} f_N^2 \; ,
\label{eq:sigmaSIScalarnondec}
\end{align}
where
\begin{align}
\kappa=\sum_{i=1,3}\left(\dfrac{ v[ (\lambda_3 + \lambda_4) c_+^2 + (\lambda_3^\prime + \lambda_4^\prime) s_+^2 - \lambda_5 c_+ s_+ ]\,\mathbf{K}_{1i}^2}{m_{h_i}^2}\right.\nonumber\\
\left.+\dfrac{ [u(c_+^2\lambda_{\eta\sigma}+s_+^2\lambda'_{\eta\sigma})-\sqrt{2}\,\mu_{12}\,c_+ s_+]\,\mathbf{K}_{2i}\mathbf{K}_{1i}}{m_{h_i}^2}\right)\,.
\label{eq:kappafactorgenscalar}
\end{align}
Note that since we consider $m_{h_1} < m_{h_2} < m_{h_3}$, the $h_2$ and $h_3$ contributions to $\sigma^\text{SI}$ are subdominant with respect to that of $h_1$. In the above expression, the $\mathbf{K}$ matrix defined in eq.~\eqref{eq:mixingnondarklim}, encodes the \textit{non-dark} neutral scalar mixing. In general, $\mathbf{K}$ is parametrised by three mixing angles, $\alpha_{12}$, $\alpha_{13}$ and  $\alpha_{23}$, quantifying the admixture between $\phi_\text{R}^0$ and $\sigma_\text{R}$, $\phi_\text{R}^0$ and $\sigma_\text{I}$, and $\sigma_\text{R}$ and $\sigma_\text{I}$, respectively. However, assuming $h_3$ to be very heavy ($m_{h_3}\gg m_{h_{1,2}}$), it can be shown that $\alpha_{12}\gg\alpha_{13},\alpha_{23}$ and, thus, the contribution of $\alpha_{13}$ and $\alpha_{23}$ to $\kappa$ can be safely neglected. In this limit, eq.~\eqref{eq:kappafactorgenscalar}, reduces to
\begin{align}
\kappa\simeq\; &v[ (\lambda_3 + \lambda_4) c_+^2 + (\lambda_3^\prime + \lambda_4^\prime) s_+^2 - \lambda_5 c_+ s_+ ]\left(\dfrac{c_\alpha^2}{m_{h_1}^2}+\dfrac{s_\alpha^2}{m_{h_2}^2}\right)\nonumber\\
&-[u(c_+^2\lambda_{\eta\sigma}+s_+^2\lambda'_{\eta\sigma})-\sqrt{2}\,\mu_{12}\,c_+ s_+]c_\alpha s_\alpha\left(\dfrac{1}{m_{h_1}^2}-\dfrac{1}{m_{h_2}^2}\right),
\label{eq:kappafactorappscalar}
\end{align}
with $\alpha\equiv\alpha_{12}$. From eqs.~\eqref{eq:ENSMassMatrixlim} and~\eqref{eq:mixingnondarklim}, one can show that $\alpha$ is given by
\begin{align}
\tan(2\alpha)\simeq\dfrac{2uv\lambda_{\Phi\sigma}}{\sqrt{(m_{h_1}^2-m_{h_2}^2)^2-4u^2v^2\lambda_{\Phi\sigma}^2}}.
\label{eq:mixingh1h2}
\end{align}
As expected, the mixing between the two lightest \textit{non-dark} neutral scalar states is controlled by their mass hierarchy and the quartic coupling $\lambda_{\Phi \sigma}$. By setting $\alpha=0$ (alignment limit), we recover the $\sigma^\text{SI}$ expression in eq.~\eqref{eq:sigmaSIScalar}. For $m_{h_3}\gg m_{h_{1,2}}$ we obtain
\begin{align}
\sigma^\text{SI}\lesssim 5.5\times 10^{-43}\left(\dfrac{500\, \text{GeV}}{m_{S_1}}\right)^2~\text{cm}^2\; ,
\label{eq:sigmaSIScalarupperboundnonalign}
\end{align}
with $\lambda_{3}^{(')}=\lambda_{4}^{(')}=1$, $\lambda_5=0$, $\varphi_+=\pi/4$, $\mu_{12}=10^3$, $\lambda_{\eta \sigma}=\lambda'_{\eta \sigma}=0$ and $|\sin\alpha|\simeq0.235$ for $m_{h_2}=450$~GeV. Note that, we take $\alpha$ to be at its maximum allowed value from Higgs signal rate observables at the LHC~\cite{Ilnicka:2018def,Robens:2019ynf}. The obtained upper limit is comparable to the one of eq.~\eqref{eq:sigmaSIScalarupperbound} and by varying all couplings in their perturbativity ranges, the results would be similar to those in fig.~\ref{fig:ScalarDM_DD}.

\subsection{Fermionic dark matter}
\label{sec:DMfermionic}

The annihilation and coannihilation diagrams for the fermion DM candidate $f$ are shown in fig.~\ref{fig:DiagAnnFermDM}, and figs.~\ref{fig:DiagCoAnn2} and~\ref{fig:DiagCoAnn3}, respectively.
\begin{figure}[t!]
   \centering
   \includegraphics[scale=0.45]{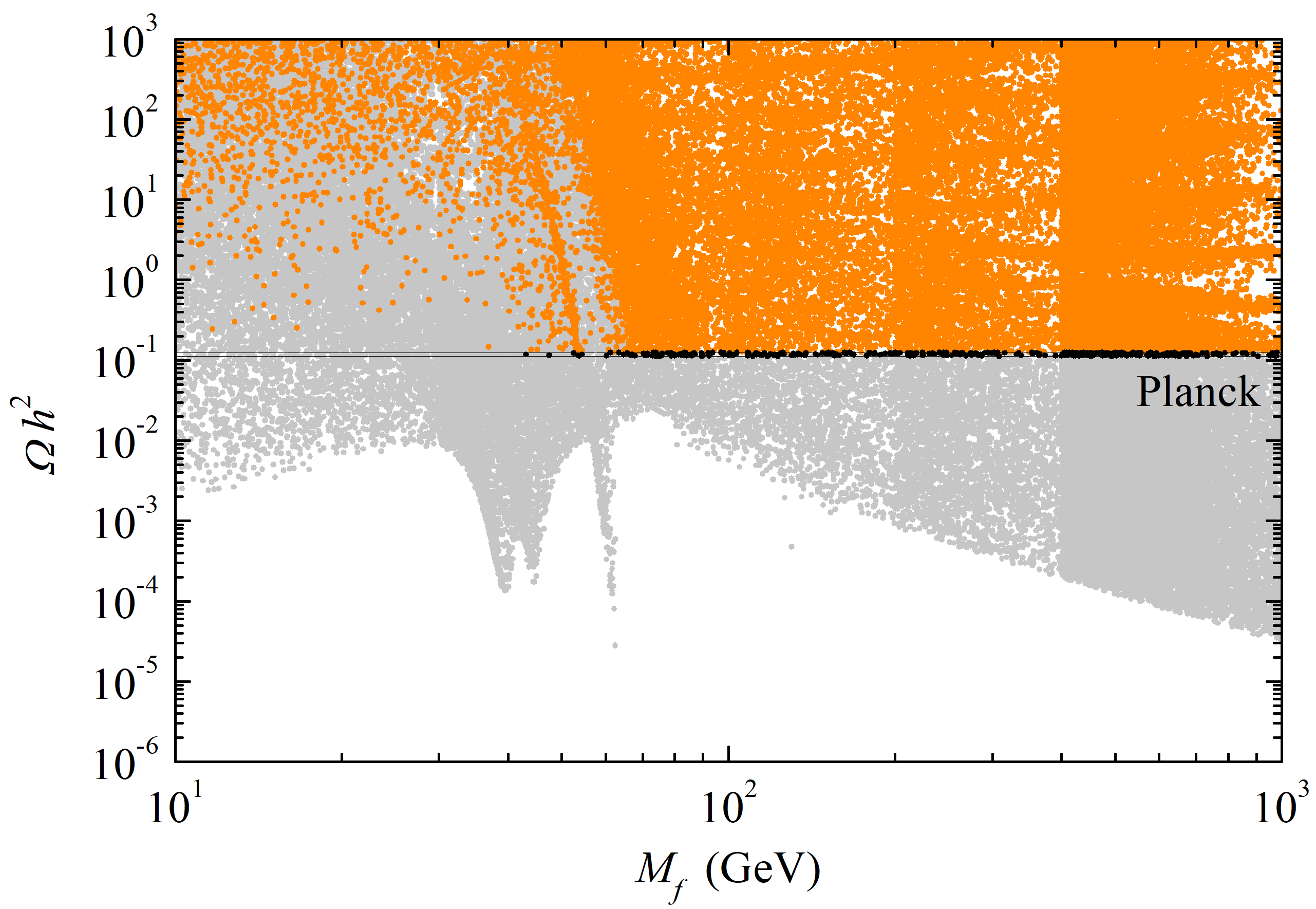}
   \caption{Relic density $\Omega h^2$ as a function of the fermion DM mass $M_f$. We use the same colour code as in fig.~\ref{fig:ScalarDM_Relic}. However, the points in grey indicate an under abundance of relic density and/or are excluded by the LEP bounds on scalar masses [see eqs.~\eqref{eq:LEPbound} and discussion therein].}
 \label{fig:FermionicDM_RelicI}
\end{figure}
\begin{figure}[t!]
  \centering
  \hspace{-0.5cm}\includegraphics[scale=0.31]{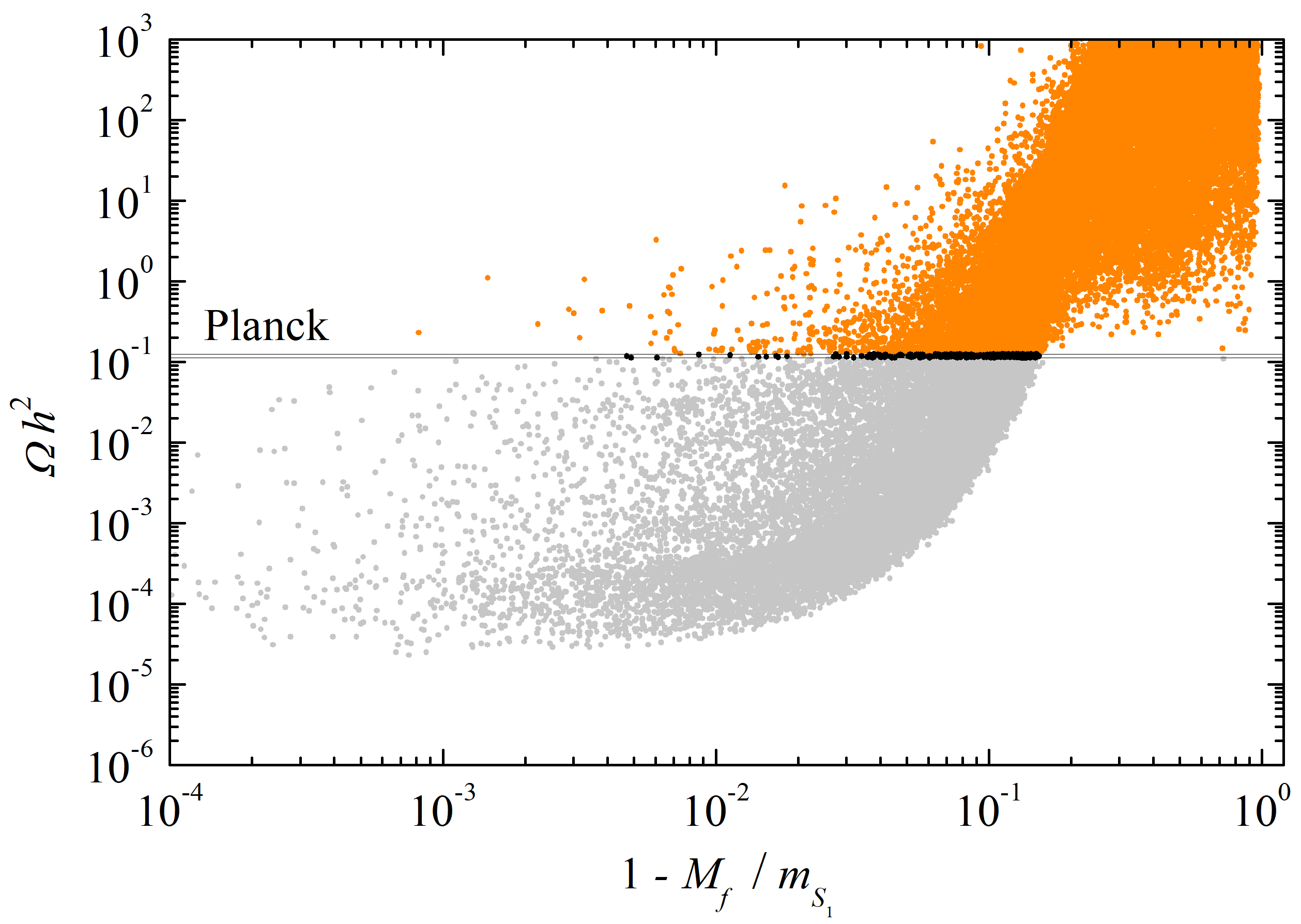}  \hspace{-0.2cm} \includegraphics[scale=0.31]{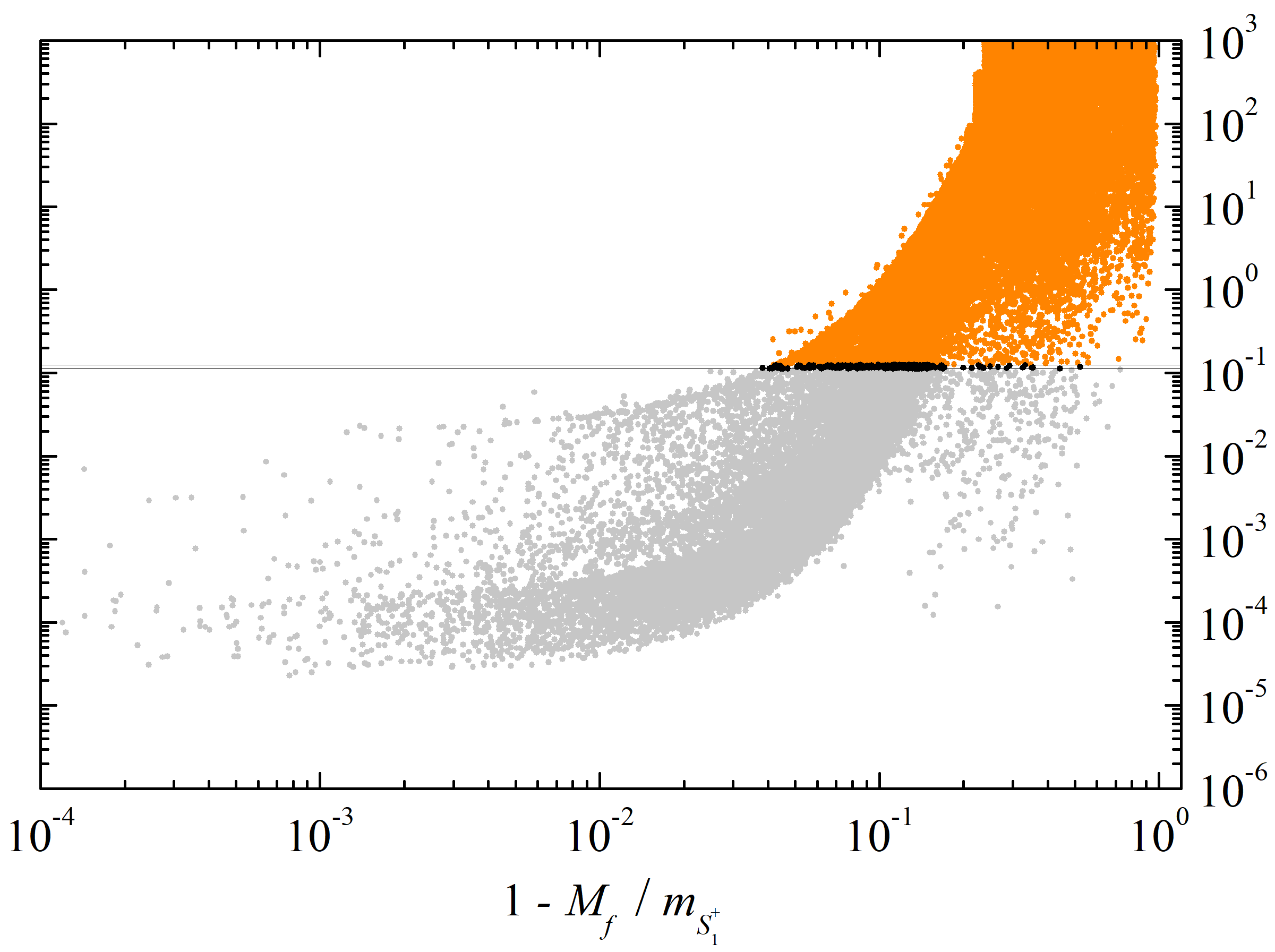}
  \caption{Relic density $\Omega h^2$ as a function of the relative difference between the DM mass $M_f$ and the lightest odd neutral (charged) scalar mass $m_{S_1}$ ($m_{S_1^+}$) is shown on the left (right) figure. The same colour code as in fig.~\ref{fig:ScalarDM_Relic} is being employed. Note that, all points in the plots above satisfy the LEP bounds on scalar masses [see eqs.~\eqref{eq:LEPbound} and discussion therein].}
  \label{fig:FermionicDM_RelicII}
\end{figure}
In fig.~\ref{fig:FermionicDM_RelicI}, we present the relic density $\Omega h^2$ as a function of the fermionic DM mass~$M_{f}$. We employ the same colour code of figs.~\ref{fig:ScalarDM_Relic},~\ref{fig:ScalarDM_DD} and~\ref{fig:ScalarDM_Higgs}, with the addition that LEP excluded points [see eqs.~\eqref{eq:LEPbound} and discussion therein] are also indicated in grey. By looking at the black points, one concludes that correct relic density can be achieved for any value of $M_f$ lying above $\sim45$~GeV. Furthermore, a similar relic density "dip pattern" as the one of fig.~\ref{fig:ScalarDM_Relic} for scalar DM, is visible. This indicates that coannihilation decays between~$f$ and the odd-scalars were included, being these crucial for achieving viable $\Omega h^2$. In fact, only considering $f$ annihilation channels leads to a small thermally average cross section resulting in relic density over abundance. In fig.~\ref{fig:FermionicDM_RelicII}, we present $\Omega h^2$ as a function of the relative difference between $M_f$ and the mass of the lightest \textit{dark} neutral (charged) scalar, $m_{S_1}$ ($m_{S_1^+}$), on the left (right) plot. Globally, it can be inferred that for a relative difference inferior to $\sim 10 \%$, viable relic density is achieved.

\begin{figure}[t!]
    \centering
    \includegraphics[scale=1,trim={3cm 22.5cm 2cm 1.0cm},clip]{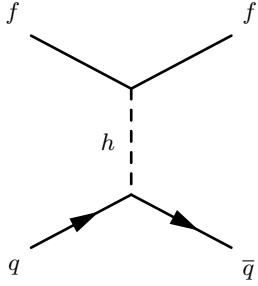}
    \caption{Tree-level diagram contributing to WIMP-nucleon spin independent elastic scattering cross-section for fermionic DM.}
    \label{fig:diagDDfermDM}
\end{figure}

Let us finally discuss the DD prospects for the fermionic DM. Since we are working in the alignment limit, i.e. $\lambda_{\Phi \sigma} = 0$, the Higgs doublet neutral scalars do not mix with $\sigma$. This implies the absence of tree-level contributions to the spin-independent cross section $\sigma^{\text{SI}}$ coming from the scattering of $f$. Additionally, since~$f$ is a Majorana fermion, the contributions to the electric and magnetic dipole moment identically vanish at one-loop level~\cite{Herrero-Garcia:2018koq}. Thus, in the alignment limit, there are no DD prospects for fermion DM. Once we allow $\lambda_{\Phi \sigma} \neq 0$, the fermion $f$ contributes at tree level to $\sigma^{\text{SI}}$ as show in fig.~\ref{fig:diagDDfermDM}, thanks to the mixing between \textit{non-dark} neutral scalars. The spin-independent nucleon-DM scattering cross-section in this case reads
\begin{align}
    \sigma^\text{SI}=\dfrac{4y_f^2\kappa^2}{\pi}\dfrac{M_f^2 m_N^2}{(M_f+m_N)^2}f_N^2,
    \label{eq:sigmaSImixinggen}
\end{align}
where $f_N$ is defined in eq.~\eqref{eq:fN}, and
\begin{align}
\kappa=\left|\sum_{j=1,3}\dfrac{\mathbf{K}_{1j}(\mathbf{K}_{2j}+i\,\mathbf{K}_{3j})}{m_{h_j}^2}\right|.
    \label{eq:kappafactorgen}
\end{align}
In the limit $m_{h_3}\gg m_{h_{1,2}}$, the above expression reduces to
\begin{align}
\kappa\simeq\dfrac{\sin(2\alpha)}{2}\left(\dfrac{1}{m_{h_1}^2}-\dfrac{1}{m_{h_2}^2}\right),
    \label{eq:kappafactorsimp}
\end{align}
being $\alpha$ defined in eq.~\eqref{eq:mixingh1h2}. Thus, for fermion DM, $\sigma^\text{SI}$ can be simplified to
\begin{align}
    \sigma^\text{SI}\simeq\dfrac{y_f^2\sin^2(2\alpha)}{\pi}\dfrac{M_f^2 m_N^2}{(M_f+m_N)^2}\left(\dfrac{1}{m_{h_1}^2}-\dfrac{1}{m_{h_2}^2}\right)^2f_N^2.
    \label{eq:sigmaSImixing}
\end{align}
Notice that since $M_f\gg m_N$, the dependence of $\sigma^{\text{SI}}$ on $M_f$ comes almost exclusively from $y_f=\sqrt{2}M_f/u$. Also, for $m_{h_2}\gg m_{h_1}$ the contribution from the heaviest scalar to $\sigma^{\text{SI}}$ is subdominant. In such case, $u$ is required to be large enough to guarantee $\lambda_{\sigma}$ perturbativity for $\theta\simeq 1.92\pi$ [see eq.~\eqref{eq:ujustification} and discussion therein], leading to a further suppression of $\sigma^{\text{SI}}$ through the Yukawa coupling $y_f$.

Maximizing the expression in eq.~\eqref{eq:sigmaSImixing} with $y_f=1$, $\alpha=\pi/4$ and $m_{h_2}\gg m_{h_1}$, leads to the bound $\sigma^\text{SI} \lsim 1.4\times10^{-42}$~cm$^2$. For a more realistic case where $\alpha$ is at its maximally-allowed value from Higgs signal rate observables at the LHC ($|\sin\alpha|\simeq0.235$ for $m_{h_2}=450$~GeV~\cite{Ilnicka:2018def,Robens:2019ynf}) and $y_f=\sqrt{2}M_f/u$, with $u=1565$~GeV, we obtain
\begin{align}
    \sigma^\text{SI}\lsim 5.2\times 10^{-44}\left(\dfrac{M_f}{500\, \text{GeV}}\right)^2~\text{cm}^2.
\end{align}
In order to guarantee $y_f\leq1$, $M_f$ can be at most 1.1~TeV, leading to $\sigma^\text{SI}\lsim 2.5\times 10^{-43}$~cm$^2$, which stands well above the current XENON1T limits.

\begin{figure}[t!]
   \centering
   \includegraphics[scale=0.45]{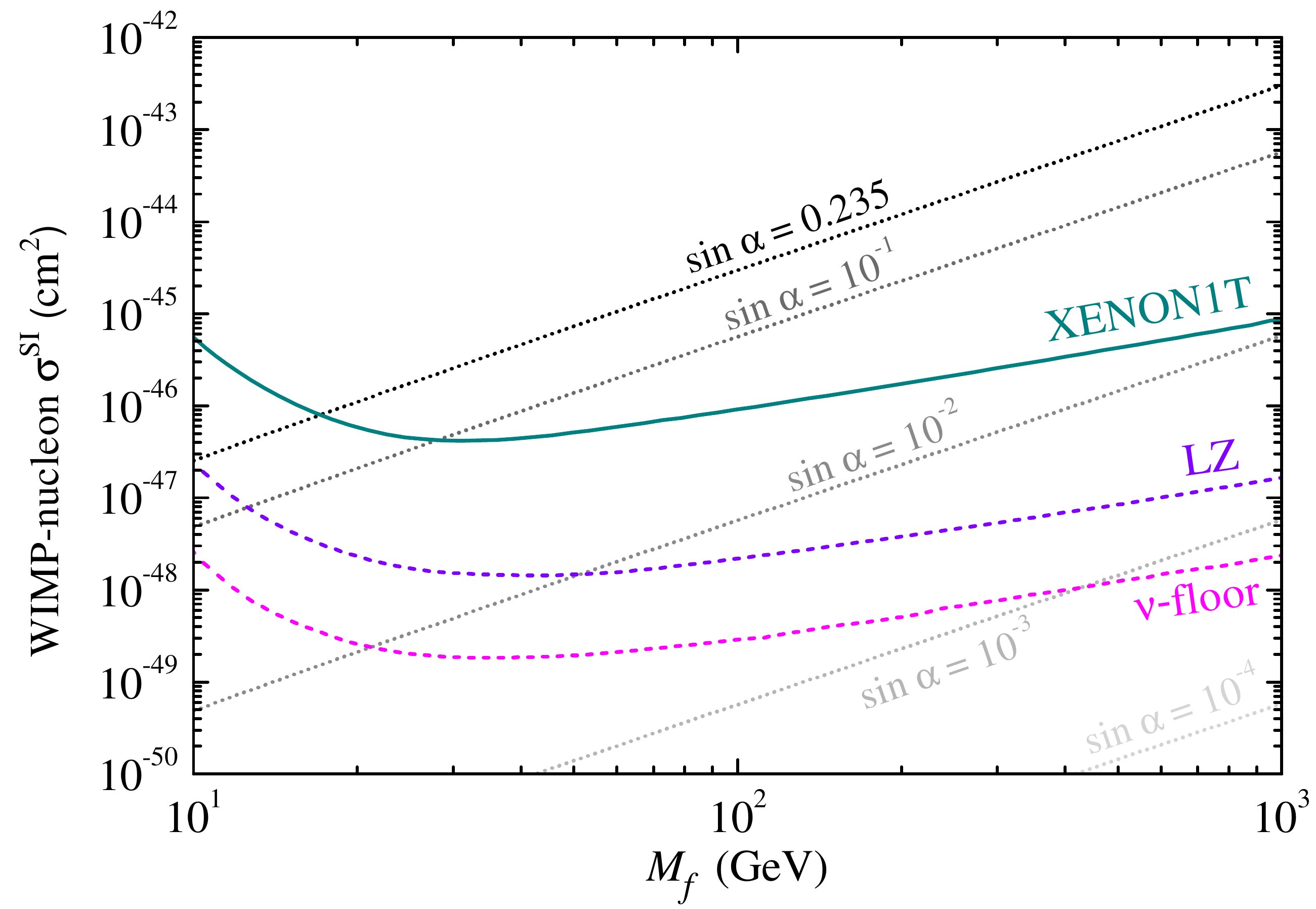}
   \caption{Spin-independent WIMP-nucleon cross section $\sigma^\text{SI}$ as a function of the fermion DM mass $M_f$, for different values of $\sin\alpha$. The maximum value $|\sin\alpha|\simeq0.235$ corresponds to the experimental limit for $m_{h_2}=450$~GeV~\cite{Ilnicka:2018def,Robens:2019ynf}. The presented DD experimental bounds are the same as those in fig.~\ref{fig:ScalarDM_DD}.}
 \label{fig:FermionicDM_DD}
\end{figure}
In fig.~\ref{fig:FermionicDM_DD}, we show the $(M_f,\sigma^\text{SI})$ contours for different values of $\sin\alpha$ up to the experimental limit $\sin\alpha=0.235$~\cite{Ilnicka:2018def,Robens:2019ynf}. We conclude that, for $M_f\gsim 100$~GeV, contours with $\sin\alpha\lsim 0.05$ are below the XENON1T bound. Such mixing can be achieved for $\lambda_{\Phi\sigma}\lsim 0.02$.

\section{Concluding remarks}
\label{sec:concl}

In this work, we have considered a SM extension where neutrino masses are generated via an interplay between the tree-level type-II seesaw and the radiative scotogenic mechanisms. The model features a $\mathcal{Z}_8$ flavour symmetry which leads to specific Yukawa patterns with texture zeros. Imposing CP conservation at the Lagrangian level, we show that SCPV leads to observable CP-violating effects in the lepton sector. The two texture-zero patterns for the effective neutrino mass matrix are tested in light of the most recent neutrino oscillation data. The main results and findings of our work can be summarised as follows:

\begin{itemize}
    
    \item The imposed $\mathcal{Z}_8$ symmetry leads to an effective neutrino mass matrix $\Mnu$ with two texture zeros. Depending on the $\mathcal{Z}_8$ charges of the lepton fields, there are three different cases, namely $\mathcal{Z}_8^{e-\mu}$, $\mathcal{Z}_8^{e-\tau}$ and $\mathcal{Z}_8^{\mu-\tau}$, which  lead to $\Mnu$ of type $\text{B}_4$, $\text{B}_3$ and $\text{A}_1$, respectively. For A$_1$ we have $(\Mnu)_{11}=0$, which is not compatible with IO. In this case, for NO, $\delta$ is constrained to be in the range $[0.8,1.6]\pi$ (3$\sigma$), while $m_\text{lightest}$ is required to be in the interval $[4,9]$~meV (3$\sigma$) -- well below the current limits from Planck cosmology and KATRIN. Significant constraints on $\theta_{23}$ and $\delta$ are obtained for B$_3$ and B$_4$, which select one of the $\theta_{23}$ octants and sharply predict $\delta\sim 3\pi/2$. These two cases also lead to a lower limit on $m_\text{lightest}$ in the range probed by cosmology and to a neutrinoless double beta decay mass parameter $m_{\beta\beta}$ within the sensitivity of current $0\nu\beta\beta$ decay experiments. In particular, we have shown that the KamLAND-Zen 400 result strongly disfavours B$_3$ (NO and IO) and B$_4$ (IO).
    
    \item The three body decays $\tau^- \rightarrow e^+ \mu^- \mu^-$ and $\tau^- \rightarrow \mu^+ e^- e^-$ are induced by the scalars stemming from the $\Delta$ triplet responsible for the type-II seesaw contributions to neutrino masses. Instead, the processes $\ell_{\alpha}^- \rightarrow \ell_{\beta}^- \gamma$ and $\ell_{\alpha} \rightarrow 3 \ell_{\beta}$ are mediated at loop level by the {\em dark}-sector degrees of freedom. While these decays are possible for all $\mathcal{Z}_8^{\beta-\alpha}$ cases,  $\mu - e$ conversion occurs only for $\mathcal{Z}_8^{e-\mu}$. We have shown that a large fraction of the model's parameter space is excluded by current cLFV bounds, leaving some regions which will be probed by future experiments.

    \item We have studied the DM phenomenology for both the scalar and fermion cases. In the former, 
    two viable mass regions for the lightest stable scalar $S_1$ are compatible with the experimental interval for the DM relic density, namely $m_{S_1} \lesssim 60$ GeV and $68 \ \text{GeV} \lesssim m_{S_1} \lesssim 90 \ \text{GeV}$. The lower mass region is excluded by current DM DD experiments and by collider constraints from LEP on scalar masses and LHC Higgs data (invisible and diphoton decay).  The intermediate mass interval is viable and will be probed by future DD searches. On the other hand, for $m_{S_1} \gsim 90$ GeV (high-mass region), our model leads to under abundant DM. This happens due to the link between \textit{dark} scalar masses and scotogenic neutrino mass generation in our model (see section~\ref{sec:neutrinomass}), which makes a highly-degenerate mass spectrum (required for obtaining correct $\Omega h^2$ for high DM masses) not natural since it would lead to non-perturbative Yukawa couplings. When the {\em dark} fermion $f$ is the lightest stable particle, a viable relic-density value is always obtained for $M_f \gsim 45$ GeV. This happens for \textit{dark} fermion-scalar mass spittings below $10 \%$. In the alignment limit, there is no DD contribution (neither tree-level nor loop since $f$ is Majorana). Moving away from the alignment limit, \textit{non-dark} neutral scalars mix, leading to tree-level contribution to the spin-independent cross section.
    
\end{itemize}

In conclusion, we have studied the phenomenology of a hybrid scoto-seesaw model which, albeit being based on a very simple flavour symmetry, leads to significant constraints in view of present and future data coming from neutrino, cLFV and DM experiments. We remark that the model presented here is much more restrictive and testable than that of ref.~\cite{Barreiros:2020gxu}. This is mainly due to the fact that, in the present case, the seesaw is of type II instead of type I as in that work.

\acknowledgments
This work is supported by Fundação para a Ciência e a Tecnologia (FCT, Portugal) through the projects CFTP-FCT Unit UIDB/00777/2020 and UIDP/00777/2020, CERN/FIS-PAR/0004/2019 and PTDC/FIS-PAR/29436/2017, which are partially funded through POCTI (FEDER), COMPETE, QREN and EU. The work of D.B. and H.C. is supported by the PhD FCT grants SFRH/BD/137127/2018 and 2021.06340.BD, respectively.

\appendix

\section{The scalar sector}
\label{sec:scalar}

The scalar sector of our model presented in section~\ref{sec:model} contains, besides of the usual Higgs doublet~$\Phi$, two doublets $\eta_{1,2}$, a scalar triplet $\Delta$ and a complex scalar singlet~$\sigma$, which we define as
\begin{align}
&\Phi =\begin{pmatrix}
\phi^{+} \\
\phi^0
\end{pmatrix}= \frac{1}{\sqrt{2}}  \begin{pmatrix}
 \sqrt{2} \phi^{+} \\
 v + \phi^0_{\text{R}} + i \phi^0_{\text{I}}
\end{pmatrix}; \, \eta_a =\begin{pmatrix}
\eta_a^{+} \\
\eta_a^0
\end{pmatrix}= \frac{1}{\sqrt{2}}  \begin{pmatrix}
 \sqrt{2} \eta_a^{+} \\
\eta^0_{\text{R} a } + i \eta^0_{\text{I} a}
\end{pmatrix} , \; a=1,2 \; ; \nonumber \\
& \Delta = \begin{pmatrix}
\Delta^{+}/\sqrt{2} &  \Delta^{++} \\
 \Delta^0 & - \Delta^{+}/\sqrt{2}
\end{pmatrix} , \; \Delta^0 = \frac{1}{\sqrt{2}} \left(w+\Delta^0_{\text{R}} + i \Delta^0_{\text{I}}\right) ; \;  \sigma = \frac{1}{\sqrt{2}}\left( u\, e^{i \theta} + \sigma_{\text{R}} + i \sigma_{\text{I}}\right) ;
\label{eq:scalarparam}
\end{align}
with the following vacuum configuration,
\begin{align}
  \left< \phi^0 \right> = \frac{v}{\sqrt{2}} \; , \; \left< \eta_{1,2}^0 \right> = 0 \; , \; \left< \Delta^0 \right> = \frac{w}{\sqrt{2}} \; , \left< \sigma \right> = \frac{u\, e^{i \theta}}{\sqrt{2}} \; .
  \label{eq:vaccum}
\end{align}
Note that the \textit{dark} scalar doublets $\eta_a$ are inert, i.e. they are VEV-less. This condition is needed to guarantee DM stability. The full scalar potential allowed by the $\mathcal{Z}_{8}$ flavour symmetry of our model (see table~\ref{tab:part&sym}) is given by
\begin{equation}
V = V_{\Phi} + V_\Delta + V_{\sigma} + V_{\eta}  + V_{\Phi \Delta} + V_{\Phi \sigma} + V_{\Delta \sigma} + V_{\eta \Phi} + V_{\eta \Delta} + V_{\eta \sigma} + V_{\eta \Delta \sigma}  \; ,
\label{eq:Vpotgeneric}
\end{equation}
with
\begin{align}
V_{\Phi} =& \ m_{\Phi}^2 \Phi^{\dagger} \Phi + \frac{\lambda_1}{2} \left(\Phi^{\dagger} \Phi\right)^2 \; , \label{eq:VPhi} \\
V_\Delta =& \ m_{\Delta}^2 \text{Tr}\left[ \Delta^{\dagger} \Delta \right] + \frac{\lambda_{\Delta 1}}{2} \text{Tr}\left[ \left(\Delta^{\dagger} \Delta\right)^2 \right] + \frac{\lambda_{\Delta 2}}{2} \text{Tr}\left[ \Delta^{\dagger} \Delta\right]^2  \; , \label{eq:VDelta} \\
V_{\sigma} =& \ m_{\sigma}^2 \left|\sigma\right|^2 + \frac{\lambda_{\sigma}}{2} \left|\sigma\right|^4 + m^{\prime\,2}_\sigma \left(\sigma^2 + \sigma^{* 2}\right) + \frac{\lambda_{\sigma}^{\prime}}{2} \left(\sigma^4 + \sigma^{* 4}\right) \; , \label{eq:Vsig} \\
V_{\eta} =& \ m_{\eta_1}^2 \eta_1^{\dagger} \eta_1 + m_{\eta_2}^2 \eta_2^{\dagger} \eta_2 + \frac{\lambda_2}{2} \left(\eta_1^{\dagger} \eta_1\right)^2 + \frac{\lambda_2^{\prime}}{2} \left(\eta_2^{\dagger} \eta_2\right)^2 + \lambda_2^{\prime \prime} \left(\eta_1^{\dagger} \eta_1\right) \left(\eta_2^{\dagger} \eta_2\right)   \; , \label{eq:Veta} \\
V_{\Phi \Delta} =& \ \mu_{\Delta} \left( \Phi^{\dagger} \Delta i \tau_2 \Phi^{*} + \text{H.c.}\right) + \lambda_{\Delta 3} \left(\Phi^{\dagger} \Phi\right) \text{Tr}\left[ \Delta^{\dagger} \Delta \right] + \lambda_{\Delta 4} \left(\Phi^{\dagger} \Delta^{\dagger} \Delta \Phi \right)\; , \label{eq:VPhiDelta} \\
V_{\Phi \sigma} =& \ \lambda_{\Phi \sigma} \left(\Phi^{\dagger} \Phi\right)  \left|\sigma\right|^2 \; , \; 
V_{\Delta \sigma} = \ \lambda_{\Delta \sigma} \text{Tr}\left[ \Delta^{\dagger} \Delta \right]  \left|\sigma\right|^2  \; , \label{eq:VDeltasig} \\
V_{\eta \Phi} =& \ \lambda_3 \left(\Phi^{\dagger} \Phi\right) \left(\eta_1^{\dagger} \eta_1 \right) +  \lambda_3^{\prime} \left(\Phi^{\dagger} \Phi\right) \left(\eta_2^{\dagger} \eta_2 \right) + \lambda_4 \left(\Phi^{\dagger} \eta_1\right) \left(\eta_1^{\dagger} \Phi \right) \nonumber\\
& +  \lambda_4^{\prime} \left(\Phi^{\dagger} \eta_2\right) \left(\eta_2^{\dagger} \Phi \right) + \frac{\lambda_{5}}{2} \left[ \left(\eta_1^{\dagger} \Phi\right) \left(\eta_2^{\dagger} \Phi \right) + \text{H.c.} \right]  \; , \label{eq:VetaPhi} \\
V_{\eta \Delta} =& \ \lambda_{\Delta 5} \left(\eta_1^{\dagger} \eta_1\right) \text{Tr}\left[ \Delta^{\dagger} \Delta \right] + \lambda_{\Delta 5}^{\prime} \left(\eta_2^{\dagger} \eta_2\right) \text{Tr}\left[ \Delta^{\dagger} \Delta \right] + \lambda_{\Delta 6} \left(\eta_1^{\dagger} \Delta^{\dagger} \Delta \eta_1 \right) \nonumber\\
&+ \lambda_{\Delta 6}^{\prime} \left(\eta_2^{\dagger} \Delta^{\dagger} \Delta \eta_2 \right) + \mu_{\Delta}^{\prime} \left(\eta_1^{\dagger} \Delta i \tau_2 \eta_2^{*} + \text{H.c.}\right) \; , \label{eq:VetaDelta} \\
V_{\eta \sigma} =& \ \lambda_{\eta \sigma}  \left(\eta_1^{\dagger} \eta_1 \right) \left|\sigma\right|^2+ \lambda_{\eta \sigma}^{\prime} \left(\eta_2^{\dagger} \eta_2 \right) \left|\sigma\right|^2+ \mu_{1 2} \left[ \left(\eta_1^{\dagger} \eta_2 \right) \sigma^{*}+ \text{H.c.} \right]   \; , \label{eq:Vetasig} \\
V_{\eta \Delta \sigma} =& \ \lambda_{\Delta 7}  \eta_1^{\dagger} \Delta i \tau_2 \eta_1^{*} \sigma^{*} + \lambda_{\Delta 7}^{\prime}  \eta_2^{\dagger} \Delta i \tau_2 \eta_2^{*} \sigma + \text{H.c.} , \label{eq:VetaDeltasig}
\end{align}
where all parameters are real. Note that we added a single $\mathcal{Z}_{8}$ soft-breaking term in eq.~\eqref{eq:Vsig}, namely $m^{\prime\,2}_\sigma\, (\sigma^2 + \sigma^{* 2})$ which, as will be seen later, is crucial for compatibility with neutrino data and also helps avoiding the cosmological domain wall problem.

Minimisation of the scalar potential leads to four non-trivial conditions,
\begin{align}
& m_{\Phi}^2 + \frac{\lambda_{1}}{2} v^2 + \frac{\lambda_{\Phi \sigma}}{2} u^2 + \frac{w}{2} \left(2 \sqrt{2} \mu_{\Delta} + \lambda_{\Delta 3} w \right) = 0 \; ,\\  
& m_{\Delta}^2 + \frac{\lambda_{\Delta 3}}{2} v^2 + \frac{\lambda_{\Delta \sigma}}{2} u^2 + \frac{w^2}{2} \left( \lambda_{\Delta 1} + \lambda_{\Delta 2} \right) + \frac{\mu_{\Delta}}{\sqrt{2} w} v^2 = 0 \; , \\  
& \left( 2 m_{\sigma}^2 + 4 m^{\prime\,2}_\sigma + u^2 \lambda_{\sigma} + v^2 \lambda_{\Phi \sigma} + w^2 \lambda_{\Delta \sigma} \right) \cos \theta + 4 u^2 \lambda_{\sigma}^{\prime} \cos (3\theta)= 0 \; , \\ 
& \left( 2 m_{\sigma}^2 - 4 m^{\prime\,2}_\sigma + u^2 \lambda_{\sigma} + v^2 \lambda_{\Phi \sigma} + w^2 \lambda_{\Delta \sigma}\right) \sin \theta - 4 u^2 \lambda_{\sigma}^{\prime} \sin (3\theta)= 0 \; .
\end{align}
These equations can be solved for $m_{\Phi}^2$, $m_{\Delta}^2$, $m_{\sigma}^2$ and $\theta$ as a function of the VEVs $v$, $w$ and~$u$, as well as the remaining parameters in $V$. We obtain three distinct solutions,
\begin{align}
& \; m_{\Phi}^2 = - \frac{1}{2} \left[ v^2 \lambda_{1} + w \left( 2 \sqrt{2} \mu_{\Delta} + w \lambda_{\Delta 3} \right) + u^2 \lambda_{\Phi \sigma} \right] \; , \label{eq:mincondmphi}\\
& \; m_{\Delta}^2 = - \frac{1}{2 w} \left\{ v^2 \sqrt{2} \mu_{\Delta}  + w \left[u^2 \lambda_{\Delta \sigma} + v^2 \lambda_{\Delta 3} + w^2 \left( \lambda_{\Delta 1} + \lambda_{\Delta 2} \right)\right]\right\}\; , \label{eq:mincondmdelta}\\
\text{(i) :} & \;  m_{\sigma}^2 = - \frac{1}{2} \left[v^2 \lambda_{\Phi \sigma} + u^2 \left(\lambda_{\sigma} + 4 \lambda_{\sigma}^{\prime}\right) + 4 m_{\sigma}^{\prime} + w^2 \lambda_{\Delta \sigma} \right]\; , \; \theta = k \pi \; , k \in \mathbb{Z} \; ; \\
\text{(ii) :} & \; m_{\sigma}^2 = - \frac{1}{2} \left[v^2 \lambda_{\Phi \sigma} + u^2 \left(\lambda_{\sigma} + 4 \lambda_{\sigma}^{\prime}\right) - 4 m_{\sigma}^{\prime} + w^2 \lambda_{\Delta \sigma} \right]\; , \; \theta = \frac{\pi}{2} + k \pi \; , k \in \mathbb{Z} \; ; \\
\text{(iii) :} & \; m_{\sigma}^2 = - \frac{1}{2} \left[ v^2 \lambda_{\Phi \sigma} + u^2 \left(\lambda_{\sigma} - 4 \lambda_{\sigma}^{\prime} \right) + w^2 \lambda_{\Delta \sigma} \right] \; , \; \cos (2 \theta) = - \frac{m^{\prime\,2}_\sigma}{2 u^2 \lambda_{\sigma}^{\prime}} \; .
\label{eq:SCPVsol}
\end{align}
In the above, the expressions for $m_{\Phi}^2$ and $m_{\Delta}^2$ are common to all solutions. Note that (i) and (ii) lead to trivial $\theta$ phases and, therefore, the only viable solution to implement SCPV is (iii). Furthermore, if the soft-breaking parameter $m^{\prime\,2}_\sigma$ vanishes we will have $\theta= \pi/4$ which is still a CPV solution. However, in this model, such value of $\theta$ is only allowed by compatibility with neutrino data  for case A$_{1}$ at the 3$\sigma$ level. In turn, as seen in section~\ref{sec:neutrinomass}, for cases B$_3$ and B$_4$, $\theta$ is fixed to values close to $2k\pi$ ($k\in \mathbb{Z}$), when considering best-fit values for neutrino observables to data. Therefore, a non-vanishing $m^{\prime\,2}_\sigma$ is required, avoiding also the domain wall problem from $\mathcal{Z}_8$ spontaneous breaking. The SCPV solution (\ref{eq:SCPVsol}) corresponds to the global minimum of the potential if $ \left(m_{\sigma}^{\prime \; 4} - 4 u^4 \lambda_{\sigma}^{\prime \; 2} \right) / (4 \lambda_{\sigma}^{\prime}) > 0$.

From eq.~\eqref{eq:mincondmdelta}, the VEV of $\Delta^0$ may be written as
\begin{equation}
w \simeq - \frac{ \sqrt{2} \mu_{\Delta} v^2}{v^2 \lambda_{\Delta 3} + u^2 \lambda_{\Delta \sigma} + 2 m_{\Delta}^2} \; ,
\label{eq:VEVDelta}
\end{equation}
where the hierarchy $w \ll v \lesssim u$ was considered. Note that, as discussed in section~\ref{sec:model}, those VEV relations are natural in the t'Hooft sense~\cite{tHooft:1979rat}. Also, $w$ is further suppressed for very heavy scalar triplet $m_{\Delta} \gg v$.

\subsection{Scalar mass spectrum and scotogenic loop functions}
\label{sec:scalarmass}

The scalar content of our model can be divided in two classes, depending on the charge carried by the fields under the unbroken $\mathcal{Z}_2$ symmetry, i.e. we can have even scalars or odd \textit{dark} scalars. Besides the usual pseudo-Goldstone bosons $G^0$ and $G^{\pm}$, the even scalars are a doubly charged scalar $\Delta^{++}$, a charged Higgs $H^\pm$, and five neutral scalars. The mass of~$\Delta^{++}$ is given by
\begin{align}
    m_{\Delta^{++}}^2=\dfrac{v^2\left(w\lambda_{\Delta4} - \sqrt{2}\mu_\Delta\right)-w^3\lambda_{\Delta1}}{2w},
    \label{eq:doublychargeddeltamass}
\end{align}
while the $H^\pm$ mass reads,
\begin{align}
    m_{H^\pm}^2=\dfrac{\left(v^2+2w^2\right)\left(w\lambda_{\Delta4}-2\sqrt{2}\mu_\Delta\right)}{4w}.
\end{align}
In turn, the mass matrix of the \textit{non-dark} neutral scalar states can be written as,
\begin{align}
\resizebox{\textwidth}{!}{
    $\mathcal{M}^2_{\phi\Delta\sigma}=\begin{pmatrix}
    v^2\,\lambda_1&v\left(w\lambda_{\Delta3}+\sqrt{2}\mu_\Delta\right)&uv\,\lambda_{\Phi \sigma}&0&0&0\\
    \cdot&w^2\left(\lambda_{\Delta1}+\lambda_{\Delta 2}\right)-\dfrac{v^2\mu_\Delta}{\sqrt{2}w}&uw\lambda_{\Delta\sigma}&0&0&0\\
    \cdot&\cdot&u^2\,\left[\lambda_\sigma+4\lambda_\sigma^\prime\cos(4\theta)\right]&0&0&4u^2\,\lambda_\sigma^\prime\sin(4\theta)\\
    \cdot&\cdot&\cdot&-2\sqrt{2}w\mu_\Delta&\sqrt{2}v\mu_\Delta&0\\
    \cdot&\cdot&\cdot&\cdot&-\dfrac{v^2\mu_\Delta}{\sqrt{2}w}&0\\
    \cdot&\cdot&\cdot&\cdot&\cdot&8u^2\,\lambda_\sigma^\prime\sin^2(2\theta)
    \end{pmatrix},$}
    \label{eq:ENSMassMatrixgen}
\end{align}
in the $\left(\phi^0_\text{R}, \Delta^0_\text{R}, \sigma_\text{R}, \phi^0_\text{I}, \Delta^0_\text{I}, \sigma_\text{I} \right)$ basis. In order to obtain the above matrix form, we rephased the fields $\sigma\rightarrow e^{-i\theta}\sigma$, $\eta_1\rightarrow e^{i\theta/2} \eta_1$ and $\eta_2\rightarrow e^{-i\theta/2}\eta_2$ and used eqs.~\eqref{eq:mincondmphi}, \eqref{eq:mincondmdelta} and~\eqref{eq:SCPVsol} to write $m_\Phi^2$, $m_\Delta^2$, $m_\sigma^2$ and $m_\sigma^{'2}$ in terms of the remaining potential parameters. The same method is used to find mass and mixing matrices for the other scalars of the model. In the limit $w \ll v$, achieved for $m_\Delta\gg v$ and small $\mu_\Delta$ [see eq.~\eqref{eq:VEVDelta}], $\Delta_{\rm R,I}^0$ decouple, as can be seen from eq.~\eqref{eq:ENSMassMatrixgen}. The masses of the remaining neutral scalars can then be obtained by computing the eigenvalues of
\begin{align}
    \mathcal{M}^2_{\phi\sigma}=\begin{pmatrix}
    v^2\,\lambda_1&u\,v\,\lambda_{\Phi \sigma}&0\\
   \cdot&u^2\,\left[\lambda_\sigma+4\lambda_\sigma'\cos(4\theta)\right]&4u^2\,\lambda_\sigma'\sin(4\theta)\\
    \cdot&\cdot&8u^2\,\lambda_\sigma'\sin^2(2\theta)
    \end{pmatrix} \; ,
    \label{eq:ENSMassMatrixlim}
\end{align}
written in the $\left(\phi^0_\text{R}, \sigma_\text{R}, \sigma_\text{I} \right)$ basis. The corresponding mixing reads
\begin{equation}
\begin{pmatrix}
\phi^0_\text{R}\\
\sigma_\text{R}\\
\sigma_\text{I}
\end{pmatrix}= \mathbf{K} \begin{pmatrix}
h_1\\
h_2\\
h_3
\end{pmatrix},
\label{eq:mixingnondarklim}
\end{equation}
being $\mathbf{K}$ a $3\times3$ orthogonal matrix parametrised by three mixing angles and $h_i$ ($i=1,2,3$) the neutral scalar mass eigenstates.

The \textit{dark} scalar sector contains two charged $S^{\pm}_{i}$ ($i=1, 2$) and four neutral $S_{j}$ ($j=1, \dots, 4$) scalars. Note that the mass-eigenstates $S^{\pm}_{i}$ with masses $m_{S^{\pm}_i}$ are those relevant for cLFV (see section~\ref{sec:cLFVmain}). The corresponding mass matrix reads,
\begin{align}
    \mathcal{M}^2_{\eta_1^\pm\eta_2^\pm}=\begin{pmatrix}
    m_{\eta_1}^2+\dfrac{v^2\lambda_3+u^2\lambda_{\eta\sigma}+w^2\lambda_{\Delta56}}{2}&\dfrac{u\,\mu_{12}}{\sqrt{2}}\\
    \cdot& m_{\eta_2}^2+\dfrac{v^2\lambda^\prime_3+u^2\lambda^\prime_{\eta\sigma}+w^2\lambda^\prime_{\Delta56}}{2}
    \end{pmatrix},
\end{align}
in the $(\eta_1^\pm,\eta_2^\pm)$ basis, and the respective eigenvalues are
\begin{align}
    m_{S_{1,2}^\pm}^2=&\,\,\dfrac{1}{4}\bigg\{2\,m_\eta^2+v^2(\lambda_3+\lambda_3^\prime)+u^2\lambda^{\prime \prime}_{\eta \sigma}+w^2(\lambda_{\Delta56}+\lambda_{\Delta56}^\prime) \label{eq:massdarkchargedscalars}\\
    &\pm\sqrt{[2\,(m^2_{\eta_1}-m^2_{\eta_2})+v^2(\lambda_3-\lambda_3^\prime)+u^2(\lambda_{\eta \sigma}-\lambda_{\eta \sigma}^\prime)+w^2(\lambda_{\Delta56}-\lambda_{\Delta56}^\prime)]^2+8u^2\mu_{12}^2}\bigg\}, \nonumber
\end{align}
with $m_{S_{1}^\pm}<m_{S_{2}^\pm}$, $m_\eta^2\equiv m_{\eta_1}^2+m_{\eta_2}^2$, $\lambda^{\prime \prime}_{\eta \sigma}\equiv\lambda_{\eta \sigma}+\lambda_{\eta \sigma}^\prime$ and $\lambda_{\Delta56}^{(\prime)}=\lambda_{\Delta5}^{(\prime)}+\lambda_{\Delta6}^{(\prime)}$. The charged components of the inert doublets are related to $S^{\pm}_{j}$ through an orthogonal matrix $\mathbf{R}$ as 
\begin{equation}
\begin{pmatrix}
\eta^+_{1}\\
\eta^+_{2}
\end{pmatrix}  = \mathbf{R} \begin{pmatrix}
S^+_1 \\
S^+_2 \\
\end{pmatrix}, \; \mathbf{R} = \begin{pmatrix}
\cos\varphi &  - \sin\varphi \\
\sin\varphi & \cos\varphi
\end{pmatrix} \;, \label{eq:rotcharged}
\end{equation}
being the mixing angle $\varphi$ given by,
\begin{align}
    \tan(2\varphi)=-\dfrac{2\sqrt{2}\, u\, \mu_{12}}{2 (m^2_{\eta_1}-m^2_{\eta_2})+v^2(\lambda_3-\lambda^\prime_3)+u^2(\lambda_{\eta \sigma}-\lambda_{\eta \sigma}^\prime)+w^2(\lambda_{\Delta56}-\lambda_{\Delta56}^\prime)}.
    \label{eq:beta}
\end{align}

The mass-eigenstates $S_{i}$ ($i=1, \dots, 4$) with masses $m_{{S}_i}$ are the ones relevant for one-loop neutrino mass generation (see sections~\ref{sec:model} and~\ref{sec:neutrinomass}). The corresponding mass matrix in the $(\eta^0_{\text{R} 1},\eta^0_{\text{R} 2},\eta^0_{\text{I} 1},\eta^0_{\text{I} 2})$ basis reads,
\begin{align}
    \mathcal{M}^2_{\eta_1^0\eta_2^0}=\begin{pmatrix}
    \mathcal{M}^2_{\eta_\text{R1}^0\eta_\text{R2}^0}&\mathbb{0}\\
    \mathbb{0}& \mathcal{M}^2_{\eta_\text{I1}^0\eta_\text{I2}^0}
    \end{pmatrix},
    \label{eq:eta1eta2MassMatrix}
\end{align}
with,
\begin{align}
\resizebox{\textwidth}{!}{$
    \mathcal{M}^2_{\eta_\text{R1}^0\eta_\text{R2}^0}=\begin{pmatrix}
    m_{\eta_1}^2+\dfrac{v^2\lambda_{34}+u^2\lambda_{\eta\sigma}+w^2\lambda_{\Delta5}+2uw\lambda_{\Delta 7}}{2}&\dfrac{v^2\lambda_5}{4}+\dfrac{u\,\mu_{12}+w\mu_{\Delta}^\prime}{\sqrt{2}}\\
    \cdot&m_{\eta_2}^2+\dfrac{v^2\lambda^\prime_{34}+u^2\lambda^\prime_{\eta\sigma}+w^2\lambda^\prime_{\Delta5}+2uw\lambda^\prime_{\Delta 7}}{2}
    \end{pmatrix},$}\label{eq:etaR1etaR2MassMatrix}\\
    \resizebox{\textwidth}{!}{$\mathcal{M}^2_{\eta_\text{I1}^0\eta_\text{I2}^0}=\begin{pmatrix}
    m_{\eta_1}^2+\dfrac{v^2\lambda_{34}+u^2\lambda_{\eta\sigma}+w^2\lambda_{\Delta5}-2uw\lambda_{\Delta 7}}{2}&-\dfrac{v^2\lambda_5}{4}+\dfrac{u\,\mu_{12}-w\mu_{\Delta}^\prime}{\sqrt{2}}\\
    \cdot&m_{\eta_2}^2+\dfrac{v^2\lambda^\prime_{34}+u^2\lambda^\prime_{\eta\sigma}+w^2\lambda^\prime_{\Delta5}-2uw\lambda^\prime_{\Delta 7}}{2}
    \end{pmatrix},$}\label{eq:etaI1etaI2MassMatrix}
\end{align}
being $\lambda^{(\prime)}_{34}\equiv\lambda^{(\prime)}_{3}+\lambda^{(\prime)}_{4}$. The $S_{1,2}$ and $S_{3,4}$ masses may then be obtained by computing the eigenvalues of the matrices in eqs.~\eqref{eq:etaR1etaR2MassMatrix} and~\eqref{eq:etaI1etaI2MassMatrix}, respectively. In the limit $w\ll v$, we get
\begin{align}
    m_{S_{1,2}}^2=\dfrac{1}{4}\bigg\{2m_\eta^2+v^2(\lambda_{34}+\lambda_{34}^\prime)+u^2\lambda^{\prime \prime}_{\eta \sigma}\pm\Big[\Lambda-4\sqrt{2}uv^2\lambda_5\mu_{12}\Big]^{1/2}\bigg\},
    \label{eq:S12mass}
\end{align}
with $m_{S_{1}}<m_{S_{2}}$ and
\begin{align}
    m_{S_{3,4}}^2=\dfrac{1}{4}\bigg\{2m_\eta^2+v^2(\lambda_{34}+\lambda_{34}^\prime)+u^2\lambda^{\prime \prime}_{\eta \sigma}\pm\Big[\Lambda + 4\sqrt{2}uv^2\lambda_5\mu_{12}\Big]^{1/2}\bigg\},
    \label{eq:S34mass}
\end{align}
with $m_{S_{3}}<m_{S_{4}}$. In the above expressions,
\begin{align}
    \Lambda & = (2m_\eta^2+v^2(\lambda_{34}+\lambda_{34}^\prime)+u^2\lambda^{\prime \prime}_{\eta \sigma})^2+v^4(4\lambda_{34}\lambda_{34}^\prime-\lambda_5^2)+4u^2v^2(\lambda_{34}\lambda^\prime_{\eta\sigma}+\lambda^\prime_{34}\lambda_{\eta\sigma})\nonumber\\
    &+8v^2(\lambda_{34}m_{\eta_2}^2+\lambda^\prime_{34}m_{\eta_1}^2) +8u^2(\lambda_{\eta\sigma}m_{\eta_2}^2+\lambda^\prime_{\eta\sigma}m_{\eta_1}^2-\mu_{12}^2)+4u^4\lambda_{\eta\sigma}\lambda^\prime_{\eta\sigma}+16m_{\eta_1}^2m_{\eta_2}^2 \; .
    \label{eq:Lambda}
\end{align}
The neutral components of the inert doublets are related to $S_{i}$ through the orthogonal matrix $\mathbf{V}$:
\begin{equation}
\begin{pmatrix}
\eta^0_{\text{R} 1} \\
\eta^0_{\text{R} 2} \\
\eta^0_{\text{I} 1} \\
\eta^0_{\text{I} 2} \\
\end{pmatrix}  = \mathbf{V} \begin{pmatrix}
S_1 \\
S_2 \\
S_3 \\
S_4
\end{pmatrix},\; \mathbf{V}=\begin{pmatrix}
\cos \varphi_{+}&\sin\varphi_{+}&0&0\\
-\sin\varphi_{+}&\cos{\varphi_{+}}&0&0\\
0&0&\cos\varphi_{-}&\sin\varphi_{-}\\
0&0&-\sin\varphi_{-}&\cos\varphi_{-}
\end{pmatrix},
\label{eq:MixneutralDM}
\end{equation}
where the two mixing angles $\varphi_{+}$ and $\varphi_{-}$ are given by,
\begin{equation}
\resizebox{\textwidth}{!}{
    $\tan(2\varphi_{\pm})=-\dfrac{2\sqrt{2}(u\,\mu_{12}\pm w\mu_\Delta^\prime) \pm v^2\lambda_5}{2(m_{\eta_1}^2-m_{\eta_2}^2)+v^2(\lambda_{34}-\lambda^\prime_{34})+u^2(\lambda_{\eta\sigma}-\lambda^\prime_{\eta\sigma})+w^2(\lambda_{\Delta5}-\lambda_{\Delta5}^\prime)\pm2uw(\lambda_{\Delta7}-\lambda_{\Delta7}^\prime)}.$}
    \label{eq:phimix}
\end{equation}

From eq.~\eqref{eq:MixneutralDM}, the weak-basis fields $\eta_1^0$ and $\eta_2^0$ can be written as,
\begin{equation}
\eta_1^0 = \sum_{k=1}^{4} \left(\mathbf{V}_{1 k} + i \mathbf{V}_{3 k} \right) S_k \; , \; \eta_2^0 = \sum_{k=1}^{4} \left(\mathbf{V}_{2 k} + i \mathbf{V}_{3 k} \right) S_k \;,
\end{equation}
and, thus, the scotogenic loop functions relevant for neutrino mass generation, in the mass eigenstate-basis, read:
\begin{align}
\mathcal{F}_{1 1} \left(M_f,m_{S_k}\right) = & \; \frac{1}{32 \pi^2} \sum_{k=1}^{4} \left(\mathbf{V}_{1 k} - i \mathbf{V}_{3 k} \right)^2 \frac{m_{S_k}^2}{M_f^2 - m_{S_k}^2} \ln \left( \frac{M_f^2}{m_{S_k}^2} \right) \; , \label{eq:F11loop} \\
\mathcal{F}_{1 2} \left(M_f,m_{S_k}\right) = & \; \frac{1}{32 \pi^2} \sum_{k=1}^{4} \left(\mathbf{V}_{1 k} - i \mathbf{V}_{3 k} \right) \left(\mathbf{V}_{2 k} - i \mathbf{V}_{4 k} \right) \frac{m_{S_k}^2}{M_f^2 - m_{S_k}^2} \ln \left( \frac{M_f^2}{m_{S_k}^2} \right) \; , \label{eq:F12loop} \\
\mathcal{F}_{2 2} \left(M_f,m_{S_k}\right) = & \; \frac{1}{32 \pi^2} \sum_{k=1}^{4} \left(\mathbf{V}_{2 k} - i \mathbf{V}_{4 k} \right)^2 \frac{m_{S_k}^2}{M_f^2 - m_{S_k}^2} \ln \left( \frac{M_f^2}{m_{S_k}^2} \right) \; . \label{eq:F22loop}
\end{align}
Using the explicit expression for $\mathbf{V}$ in terms of the mixing angles $\varphi^{\pm}$ [see eq.~\eqref{eq:MixneutralDM}]:
\begin{align}
\mathcal{F}_{1 1} \left(M_f,m_{S_k}\right) & = \frac{1}{32 \pi^2} \left[\frac{m_{S_1}^2\cos^2\varphi_+}{M_f^2 - m_{S_1}^2} \ln \left( \frac{M_f^2}{m_{S_1}^2} \right)+\frac{m_{S_2}^2\sin^2\varphi_+}{M_f^2 - m_{S_2}^2} \ln \left( \frac{M_f^2}{m_{S_2}^2} \right)\right.\nonumber\\
& \left.-\frac{m_{S_3}^2\cos^2\varphi_-}{M_f^2 - m_{S_3}^2} \ln \left( \frac{M_f^2}{m_{S_3}^2} \right)-\frac{m_{S_4}^2\sin^2\varphi_-}{M_f^2 - m_{S_4}^2} \ln \left( \frac{M_f^2}{m_{S_4}^2} \right)\right] \; , \label{eq:F11loop_2} \\
\mathcal{F}_{1 2} \left(M_f,m_{S_k}\right) & = \frac{1}{64 \pi^2} \left\{\left[-\frac{m_{S_1}^2}{M_f^2 - m_{S_1}^2} \ln \left( \frac{M_f^2}{m_{S_1}^2} \right)+\frac{m_{S_2}^2}{M_f^2 - m_{S_2}^2} \ln \left( \frac{M_f^2}{m_{S_2}^2} \right)\right]\sin(2\varphi_+)\right.\nonumber\\
& \left.+\left[\frac{m_{S_3}^2}{M_f^2 - m_{S_3}^2} \ln \left( \frac{M_f^2}{m_{S_3}^2} \right)-\frac{m_{S_4}^2}{M_f^2 - m_{S_4}^2} \ln \left( \frac{M_f^2}{m_{S_4}^2} \right)\right]\sin(2\varphi_-)\right\} \; , \label{eq:F12loop_2}
\end{align}
\begin{align}
\hspace{-1cm}\mathcal{F}_{2 2} \left(M_f,m_{S_k}\right) & = \frac{1}{32 \pi^2} \left[\frac{m_{S_1}^2\sin^2\varphi_+}{M_f^2 - m_{S_1}^2} \ln \left( \frac{M_f^2}{m_{S_1}^2} \right)+\frac{m_{S_2}^2\cos^2\varphi_+}{M_f^2 - m_{S_2}^2} \ln \left( \frac{M_f^2}{m_{S_2}^2} \right)\right.\nonumber\\
& \left.-\frac{m_{S_3}^2\sin^2\varphi_-}{M_f^2 - m_{S_3}^2} \ln \left( \frac{M_f^2}{m_{S_3}^2} \right)-\frac{m_{S_4}^2\cos^2\varphi_-}{M_f^2 - m_{S_4}^2} \ln \left( \frac{M_f^2}{m_{S_4}^2} \right)\right].\label{eq:F22loop_2}
\end{align}
Some comments regarding the above loop-functions are in order. Notice that they depend on the $S$ masses shown in eqs.~\eqref{eq:S12mass} and \eqref{eq:S34mass}, as well as mixing among \textit{dark} neutral states defined by the angles $\varphi_{\pm}$ in eq.~\eqref{eq:phimix}. In the limit $w \ll v$ ($\Delta$ decoupled from the remaining scalars), the mass differences $\Delta_{ij}= m_{S_{i}}^2 - m_{S_{j}}^2$ are given by:
\begin{align} 
    \Delta_{1 3} = - \dfrac{1}{4} \left[ \left(\Lambda + 4\sqrt{2}uv^2\lambda_5\mu_{12}\right)^{1/2} - \left(\Lambda - 4\sqrt{2}uv^2\lambda_5\mu_{12}\right)^{1/2}  \right] = - \Delta_{2 4} \; ,
    \label{eq:S13diff}
\end{align}
\begin{align}
    \Delta_{1 4} = \dfrac{1}{4} \left[ \left(\Lambda + 4\sqrt{2}uv^2\lambda_5\mu_{12}\right)^{1/2} + \left(\Lambda - 4\sqrt{2}uv^2\lambda_5\mu_{12}\right)^{1/2}  \right] = - \Delta_{2 3} \; .
    \label{eq:S14diff}
\end{align}
As seen from the above equations, the degeneracy is controlled by $4\sqrt{2}uv^2\lambda_5\mu_{12}$. If $\lambda_5$ and/or $\mu_{1 2}$ are small, which, as argued, is natural in the t'Hooft sense, we have $4\sqrt{2}uv^2\lambda_5\mu_{12} \ll \Lambda$, implying $\Delta_{1 4}>\Delta_{1 3}$.  For the purposes of our analysis, we take some limiting cases of the above loop-functions. If the term proportional to~$\lambda_5$ is absent from the scalar potential [see eq.~\eqref{eq:VetaPhi}], then $\Delta_{1 3,24} = 0$ while $\Delta_{1 4,23} \neq 0$ and $\varphi_- = \varphi_+$. This implies $\mathcal{F}_{1 1}= \mathcal{F}_{1 2}= \mathcal{F}_{2 2}= 0$ and, thus, vanishing scotogenic contributions to neutrino masses (see fig.~\ref{fig:neutrinomassdiagrams} and discussion therein). Still, cLFV mediated by \textit{dark} scalars is possible since $S_1^+$ and $S_2^+$ mix. If, additionally $\mu_{1 2} = 0$ [see eq.~\eqref{eq:Vetasig}], there is no scoto-induced cLFV [see discussion in section~\ref{sec:cLFVmain} and eq.~\eqref{eq:BRmueg}]. If only $\mu_{1 2} = 0$, we obtain $\varphi=0$, $m_{S_1^+} = m_{S_2^+}$ [see eqs.~\eqref{eq:beta} and \eqref{eq:massdarkchargedscalars}], $\Delta_{1 3} = \Delta_{2 4} = 0$ while, $\Delta_{1 4,23} \neq 0$ and $\varphi_- = -\varphi_+$. Thus, $\mathcal{F}_{1 1}= \mathcal{F}_{2 2}= 0$ but $\mathcal{F}_{1 2} \neq 0$ and there is no scotogenic neutrino mass generation nor radiative cLFV. This indicates that $\mathcal{F}_{1 2} \geq \mathcal{F}_{i i}$ with $i = 1, 2$, as verified by our numerical analysis presented in section~\ref{sec:cLFVmain}. In conclusion, when $\mu_{1 2} \rightarrow 0$ and/or $\lambda_5 \rightarrow 0$, the loop-functions $\mathcal{F}_{i i}$ tend to zero faster than $\mathcal{F}_{1 2}$, and the ratio $\mathcal{R}$ defined in eq.~\eqref{eq:Rratio} obeys $0 \leq \mathcal{R} \leq 1$. 

\pagebreak

\section{Dark matter (co)annihilation diagrams}
\label{sec:DMdiag}
\begin{figure}[ht!]
    \centering
    \includegraphics[scale=0.94,trim={3.0cm 14.7cm 3.0cm 1.3cm},clip]{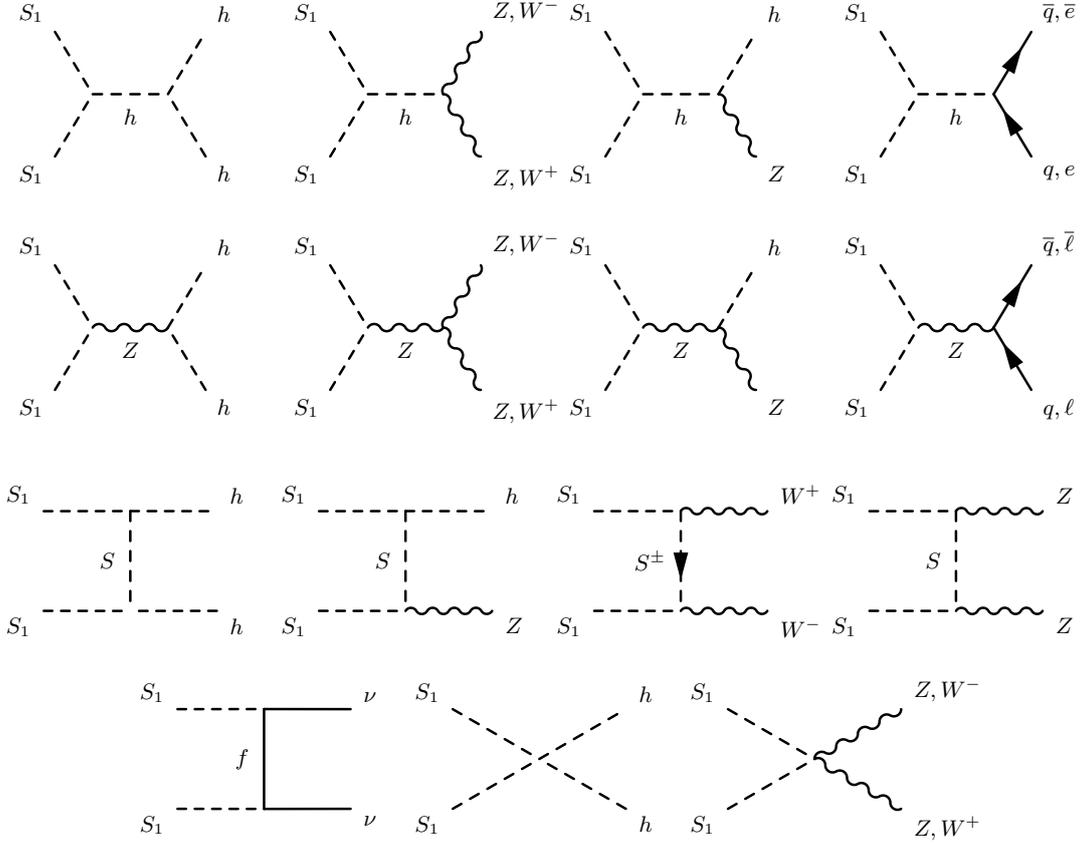}
    \caption{Annihilation diagrams for scalar DM. These coincide with the ones contributing to $S_1$ and $S_{i}$ ($i=2,3,4$) co-annihilation, after making the replacement of one initial $S_1$ state by $S_i$.}
    \label{fig:DiagAnnScalarDM}
\end{figure}

\begin{figure}[ht!]
    \centering
    \includegraphics[scale=0.94,trim={3.5cm 20.7cm 3.0cm 1.3cm},clip]{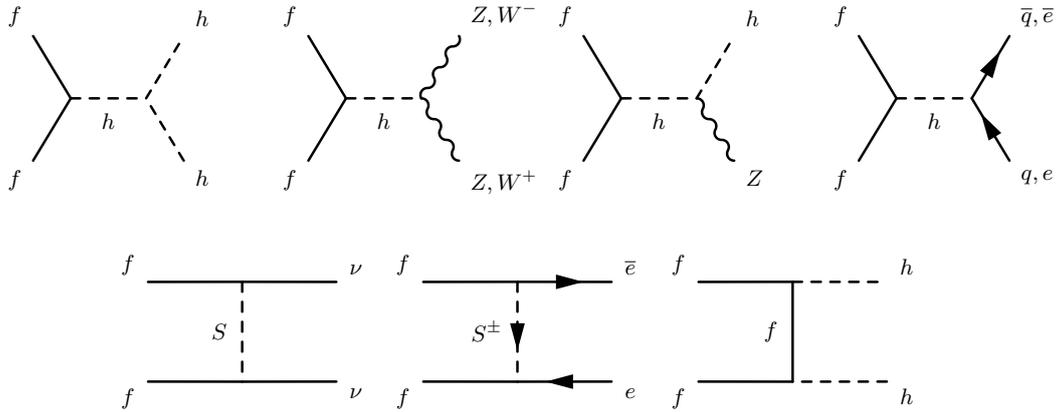}
    \caption{Annihilation diagrams for fermionic DM.}
    \label{fig:DiagAnnFermDM}
\end{figure}

\begin{figure}[ht!]
    \centering
    \includegraphics[scale=0.975,trim={2.8cm 20.7cm 3.0cm 1.3cm},clip]{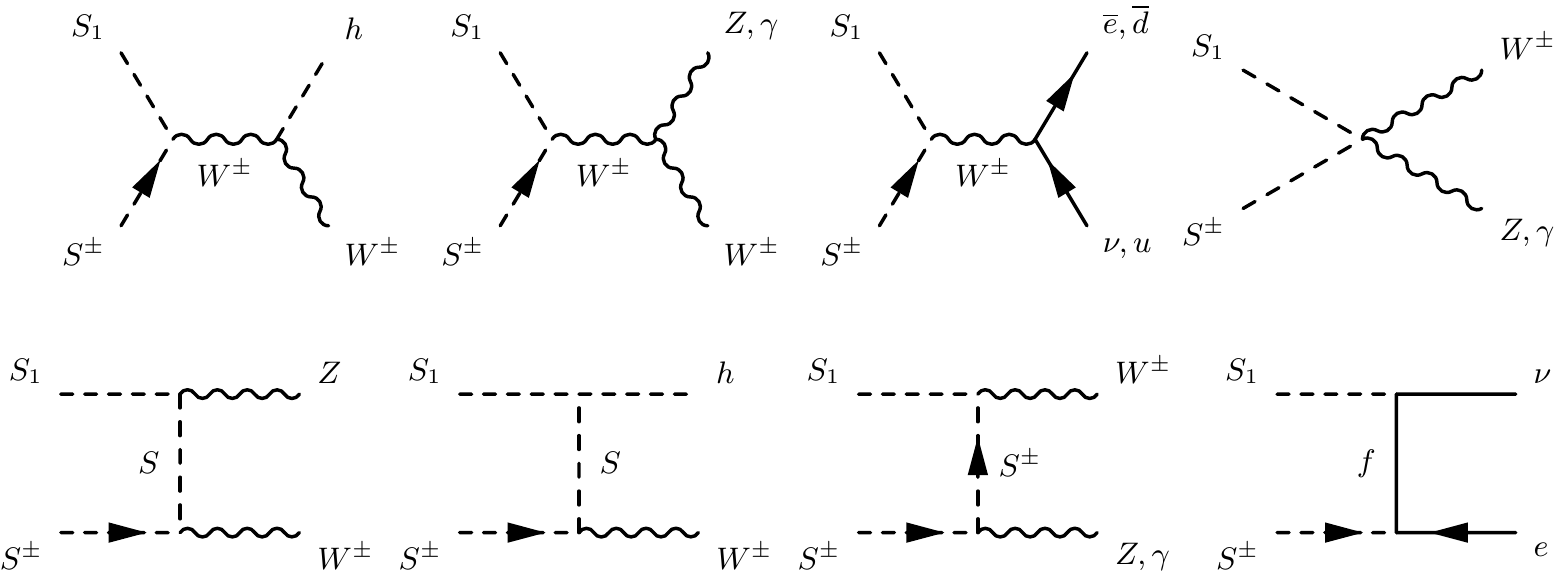}
    \caption{Diagrams contributing to co-annihilation of $S_1$ with $S^\pm$.}
    \label{fig:DiagCoAnn1}
\end{figure}

\begin{figure}[ht!]
    \centering
    \includegraphics[scale=0.975,trim={3.5cm 20.7cm 3.0cm 1.3cm},clip]{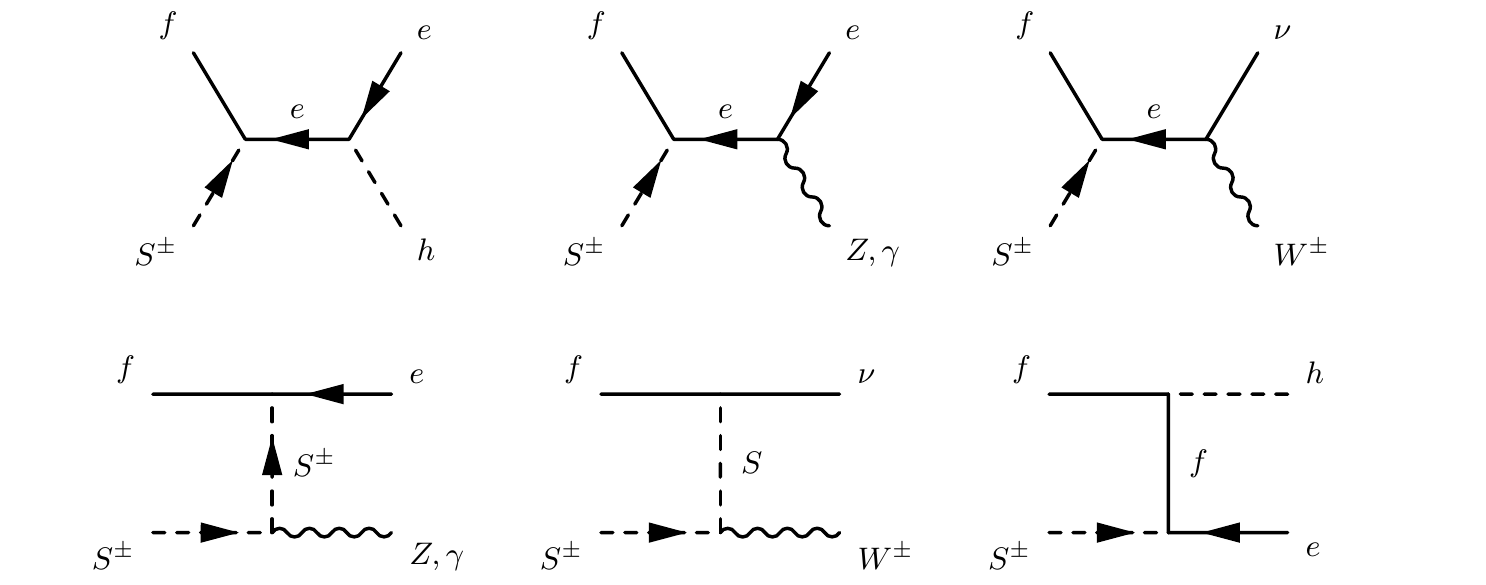}
    \caption{Diagrams contributing to co-annihilation of $f$ with $S^\pm$.}
    \label{fig:DiagCoAnn2}
\end{figure}

\begin{figure}[ht!]
    \centering
    \includegraphics[scale=0.95,trim={2.0cm 20.5cm 2.0cm 1.3cm},clip]{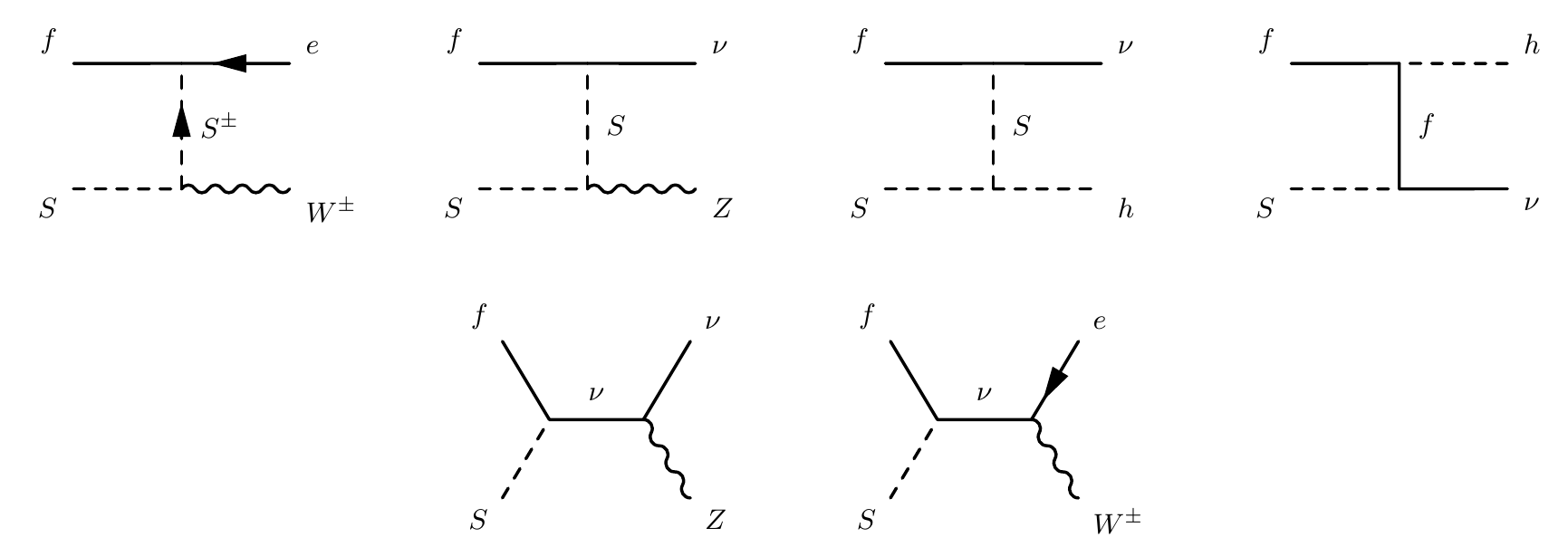}
    \caption{Diagrams contributing to co-annihilation of $f$ with $S$.}
    \label{fig:DiagCoAnn3}
\end{figure}

In this appendix we present the scalar and fermion DM (co)-annihilation diagrams in figs.~\ref{fig:DiagAnnScalarDM}-\ref{fig:DiagCoAnn3}.

\pagebreak

\end{document}